\documentclass[twocolumn]{aastex701}

\let\longtable*\relax

\usepackage{xspace}
\usepackage{hyperref}
\usepackage{amsmath}
\usepackage[font=small,skip=3pt]{caption}

\newcommand{\ROMANtext}{{\it Nancy Grace Roman Space Telescope}}

\newcommand{\HLTDStext}{High Latitude Time Domain Survey}
\newcommand{\HLTDS}{HLTDS}
\newcommand{\committee}{HLTDS definition committee}

\newcommand{\OpenU}{OpenUniverse2024}

\newcommand{\ELASTICC}{{\sc ELAsTiCC}} 
\newcommand{\ELASTICCtext}{Extended LSST Astrometric Time-series Classification Challenge}

\newcommand{\specy}{spectroscopically}
\newcommand{\spec}{spectroscopic}

\newcommand{\SNANA}{{\tt SNANA}}

\newcommand{\SCONE}{{\tt SCONE}}
\newcommand{\DESVYR}{DES-SN5YR}
\newcommand{\LOGMASS}{{\tt LOGMASS}}
\newcommand{\LOGSFR}{{\tt LOGSFR}}
\newcommand{\DIFFSKY}{{\tt Diffsky}}
\newcommand{\HOSTLIB}{{\tt HOSTLIB}}
\newcommand{\WGTMAP}{{\tt WGTMAP}}
\newcommand{\HOSTLIBsize}{400,000}

\newcommand{\ztarget}{z_{\mathrm{target}}}      
\newcommand{\SNRtarget}{{\rm S/N}_{\rm target}} 
\newcommand{\Texpose}{T_{\rm expose}}
\newcommand{\Tread}{T_{\rm read}}
\newcommand{\SNRSUM}{{\tt SNRSUM}} 
\newcommand{\eminus}{{e$^-$}}

\newcommand{\Trest}{T_{\rm rest}}

\newcommand{\MJDo}{MJD$_0$}

\newcommand{\zphot}{z_{\rm phot}}
\newcommand{\zspec}{z_{\rm spec}}
\newcommand{\ztrue}{z_{\rm true}}
\newcommand{\zres}{(\zphot - \ztrue)/(1+\ztrue)}
\newcommand{\Dz}{\Delta_z}
\newcommand{\mlimit}{m_{5\sigma}}
\newcommand{\Dmu}{\Delta\mu}
\newcommand{\mubbc}{\mu_{\rm BBC}}

\newcommand{\PIa}{P_{\rm Ia}}
\newcommand{\Plcfit}{P_{\rm SALT3}}

\newcommand{\Csyst}{{\cal C}_{\rm syst}}
\newcommand{\Cstat}{{\cal C}_{\rm stat}}
\newcommand{\Ctot}{{\cal C}_{\rm tot}}
\newcommand{\wwCDM}{$w_0w_a{\rm CDM}$}
\newcommand{\sigint}{\sigma_{\rm int}}
\newcommand{\sigmu}{\sigma_{\mu}}
\newcommand{\mSB}{m_{\rm SB}}

\newcommand{\LCDM}{$\Lambda$CDM}
\newcommand{\Om}{\Omega_M}
\newcommand{\cosparLCDM}{\{\Om,w_0,w_a \}{=}\{0.315,-1,0\}}
\newcommand{\cosparDVYR}{\{\Om,w_0,w_a \}{=}\{0.315,-0.8,-1.0\}}

\newcommand{\Ndataset}{9}
\newcommand{\Nsyst}{30}

\newcommand{\NWFDapprox}{800}  
\newcommand{\NDDFapprox}{3{,}600}  
\newcommand{\NLSSTapprox}{4{,}400}  
\newcommand{\NROMANapprox}{10{,}000}

\newcommand{\nzbinBBC}{45}                  
\newcommand{\HDsizeAVGbinned}{45}           
\newcommand{\HDsizeAVGrebin}{919}           
\newcommand{\HDsizeRebinDefine}{1440}       
\newcommand{\HDsizeAVGunbin}{14585}         
\newcommand{\NREBINx}{4}                    
\newcommand{\NREBINc}{8}                    

\newcommand{\BBClossAll}{6.5}  

\newcommand{\zmaxWFD}{0.09}
\newcommand{\zmaxDDF}{0.50}

\newcommand{\AreaWideCCS}{18.27}      
\newcommand{\AreaDeepCCS}{6.47}       
\newcommand{\AreaWideSNANA}{18.61}      
\newcommand{\AreaDeepSNANA}{6.58}       

\newcommand{\TexposeMIN}{60}     
\newcommand{\zspecmax}{0.3}

\newcommand{\NPVcadence}{4}
\newcommand{\NPVtot}{8}
\newcommand{\NPVtemplate}{2}
\newcommand{\NPVlcfit}{6}
\newcommand{\NIaPV}{470}   
\newcommand{\NnonIaPV}{22} 
\newcommand{\contamPV}{4.5} 

\newcommand{\EVincrease}{8}  

\newcommand{\FOMALLSYSzbin}{364}            
\newcommand{\FOMALLSYSrebin}{479}           
\newcommand{\FOMALLSYSunbin}{507}           
\newcommand{\FOMALLSYSnoDDFunbin}{408}      
\newcommand{\FOMRATIOrebinTOzbin}{32}       
\newcommand{\FOMRATIOunbinTOrebin}{6}       
\newcommand{\FOMALLSYSrequire}{326}         

\newcommand{\URLROMAN}{\url{https://roman.gsfc.nasa.gov}}
\newcommand{\URLSNANA}{\url{https://github.com/RickKessler/SNANA}}
\newcommand{\URLSNANAMANUAL}{\url{https://github.com/RickKessler/SNANA/blob/master/doc/snana_manual.pdf}}
\newcommand{\URLwfit}{\url{https://github.com/RickKessler/SNANA/blob/master/src/wfit.c}}
\newcommand{\URLPIP}{\url{https://github.com/dessn/Pippin}}
\newcommand{\URLEAZY}{\url{http://www.astro.yale.edu/eazy/?download}}
\newcommand{\URLSCONE}{\url{https://github.com/helenqu/scone}}

\newcommand{\URLOPSIM}{\url{https://www.lsst.org/scientists/simulations/opsim}}
\newcommand{\URLDIFFSKY}{\url{https://diffsky.readthedocs.io/en/latest}}
\newcommand{\URLLSST}{\url{https://www.lsst.org}}
\newcommand{\URLZTF}{\url{https://www.ztf.caltech.edu}}
\newcommand{\URLROMANSUBARU}{\url{https://asd.gsfc.nasa.gov/roman/wps_2023/files/004_Harikane_HLWAS.pdf}}

\newcommand{\URLEUCLID}{\url{https://www.esa.int/Science_Exploration/Space_Science/Euclid}}

\begin{document}

\title{Cosmology Constraints from Type Ia Supernova Simulations of the \ROMANtext\ 
Strategy Recommended by the {\HLTDStext} Definition Committee
}


\newcommand{\KICP}{Kavli Institute for Cosmological Physics, University of Chicago, Chicago, IL 60637, USA}
\newcommand{\UCAA}{Department of Astronomy and Astrophysics, University of Chicago, Chicago, IL 60637, USA}
\newcommand{\UMary}{University of Maryland, Baltimore County, 1000 Hilltop Cir, Baltimore, MD 21250}
\newcommand{\Goddard}{NASA Goddard Space Flight Center, 8800 Greenbelt Rd, Greenbelt, MD 20771}
\newcommand{\JHU}{Physics and Astronomy Department, Johns Hopkins University, Baltimore, MD 21218, USA}
\newcommand{\LBL}{Lawrence Berkeley National Laboratory, 1 Cyclotron Rd, Berkeley, CA, 94720, USA}
\newcommand{\UHawaii}{Department of Physics and Astronomy, 
  University of Hawai'i at Manoa, Honolulu, Hawai'i 96822}
\newcommand{\StonyBrook}{C. N. Yang Institute for Theoretical Physics, 
Stony Brook University, Stony Brook, NY 11794, USA}
\newcommand{\Duke}{Department of Physics, Duke University Durham, NC 27708, USA}
\newcommand{\UPenn}{Department of Physics and Astronomy, University of Pennsylvania, 
 209 South 33rd Street, Philadelphia, PA 1910}
\newcommand{\Baylor}{Department of Physics and Astronomy, 
   Baylor University, One Bear Place \#97316, Waco, TX 76798-7316, USA}

\author{Richard~Kessler} \affiliation{\KICP} \affiliation{\UCAA} \email{rkessler@uchicago.edu}

\author{Rebekah~Hounsell} \affiliation{\UMary} \affiliation{\Goddard} \email{}

\author{Bhavin~Joshi}  \affiliation{\JHU} \email{}

\author{David~Rubin} \affiliation{\UHawaii} \affiliation{\LBL}    \email{}

\author{Masao~Sako}  \affiliation{\UPenn} \email{}

\author{Rebecca~Chen}  \affiliation{\Duke} \email{}

\author{Vivian Miranda} \affiliation{\StonyBrook} \email{}

\author[0000-0002-1873-8973]{Benjamin.~M.~Rose} \affiliation{\Baylor} \email{}

\begin{abstract}
Within the next few years, the upcoming \ROMANtext\ will be gathering data for the
\HLTDStext\ (\HLTDS) that will be used to significantly improve the
Type~Ia supernova measurement of the dark energy equation of state parameters $w_0$ and $w_a$.
Here we generate a catalog-level simulation of the {\it in-guide} strategy recommended by the 
\HLTDS\ definition committee, and determine dark energy parameter constraints 
using a detailed analysis that includes 
light curve fitting, 
photometric redshifts and classification,
BEAMS formalism,
systematic uncertainties,
and cosmology fitting.
After analysis and selection requirements, the sample includes
${\sim}\NROMANapprox$ Roman SNe~Ia that we combine with ${\sim}\NLSSTapprox$ events from LSST.
The resulting dark energy figure of merit is well above the NASA mission requirement
of $\FOMALLSYSrequire$, with the caveat that SN~Ia model training systematics
have not been included. 
\end{abstract}


\section{Introduction} 
\label{sec:intro}

Using Type Ia Supernovae (SN~Ia), the accelerated expansion of the universe was discovered more than 25 years ago 
\citep{Riess1998,Perlmutter1999}. Numerous ground-based SN surveys and re-analyses of public data 
have since improved the statistical and systematic precision on parameters describing a mysterious {\it dark energy} 
that potentially permeates the universe. The results have been mostly consistent with a cosmological constant
described by the dark energy equation of state parameter $w=-1$ 
\citep{Astier2006, Kessler2009, Suzuki2012, Betoule2014, Scolnic2018, Alam2021}.
More recent results using higher statistics and more sophisticated methods for evaluating systematic uncertainties
have shown 2-3$\sigma$ evidence for evolving dark energy
\citep{Brout2022, Rubin2025_UNITY, DESCollaboration2024}.
In addition to the SNIa-based evidence, recent results  using baryon acoustic oscillations (BAO)
have also shown a similar deviation from a cosmological constant \citep{DESI_2024,DESI_2025}.

The 2010 Decadal Survey \citep{Decadal2010} recommended a space-based mission to study 
dark energy and exoplanets,
and the resulting \ROMANtext\footnote{\URLROMAN} is nearly built and planned to launch in late 2026. 
The Roman telescope includes a wide field instrument (WFI) with 
18 detectors covering a 0.281~deg$^2$ field of view,
and $RZYJHFK$ broadband filters 
with central wavelengths 6340, 8719, 10595, 12936, 15791, 18418, 21255~\AA, respectively.
The 5-year mission includes roughly 6 months of observing time dedicated to the \HLTDStext\ (\HLTDS).

An \committee\ was formed and tasked to work with the project infrastructure teams and
Roman community to propose an \HLTDS\ survey strategy that optimizes SN~Ia cosmology constraints 
and other time-domain science that includes searches for rare transients at high redshift. 
The committee charge included optimizing cosmological constraints
for simulated SN data combined with cosmic microwave background (CMB) sensitivity
corresponding to \citet{Planck2018}. Constraints are based on the \wwCDM\ model 
in which the dark energy equation of state parameter evolves
with redshift as $w=w_0 + w_a(1-a)$ and $a=(1+z)^{-1}$. 
The committee optimization is based on the figure of merit 
${\rm FoM} = \sqrt{\det[{\rm COV}(w_0,w_a)]}$,
which is proportional to the inverse area of the contour constraining
$w_0$ and $w_a$ \citep{Albrecht2006,Wang2008_FoM}.

Our Roman Supernova Project Infrastructure Team (SNPIT) contributed several types of simulations and
cosmology analyses to the \committee. 
The first SNPIT effort generated a set of ${\sim}1000$ survey strategies to probe a 
wide range of filter combinations, exposure times and cadence, 
and each strategy is analyzed with a Fisher matrix approach to estimate FoM.
\citep[hereafter R25]{Rubin2025_Optimize}. 
The second effort was a focused FoM study on a single strategy using a more detailed simulation 
and a more realistic analysis using photometric classification methods similar to those used in the \DESVYR\ analysis 
\citep{DESCollaboration2024,Vincenzi2024}, and photometric SN+host redshifts based on 
\citet{Kessler2010_zphot,Mitra2023,Chen2025}.
A third effort estimated yields of rare transients \citep{Rose2025_hourglass}.
In this paper we present cosmology constraints from the second effort after adjusting the survey
parameters to match the {\it in-guide} recommendation reported in \citet{CCS_report}. 
Results presented here supersede \SNANA\ based results presented
in the committee report.\footnote{The \SNANA-based results in \citet{CCS_report} 
are based on a preliminary strategy that has higher SN statistics compared to 
the published in-guide strategy.}

The following software packages are used in this analysis.
The simulation and analysis use \SNANA\footnote{\URLSNANA} \citep{SNANA},
and pippin\footnote{\URLPIP} \citep{Pippin} is used for the pipeline orchestration.
\SCONE\footnote{\URLSCONE} \citep{SCONE} is used for photometric classification
and {\tt MINUIT} \citep{MINUIT} is used for $\chi^2$-minimization for
light curve fitting and for 
``Beams with Bias Correction'' (BBC; \citet{BBC}).

The outline of this paper is as follows.
An overview of the recommended \HLTDS\ strategy is given in Sec.~\ref{sec:overview}.
The simulation method is described in Sec.~\ref{sec:snana_sim},
the analysis is described in Sec.~\ref{sec:snana_ana},
and FoM results are presented in Sec.~\ref{sec:snana_results}.
The impact of the extended survey components are discussed in Sec.~\ref{sec:noncore_visits}.

\section{Overview of Recommended {\HLTDS} }
\label{sec:overview}

\newcommand{\TobsTotal}{180}
\newcommand{\TobsTD}{157.8}
\newcommand{\TobsWFI}{122.5}
\newcommand{\TobsPRISM}{35.3}

\begin{figure}[hb]
    \includegraphics[width=0.95\linewidth]{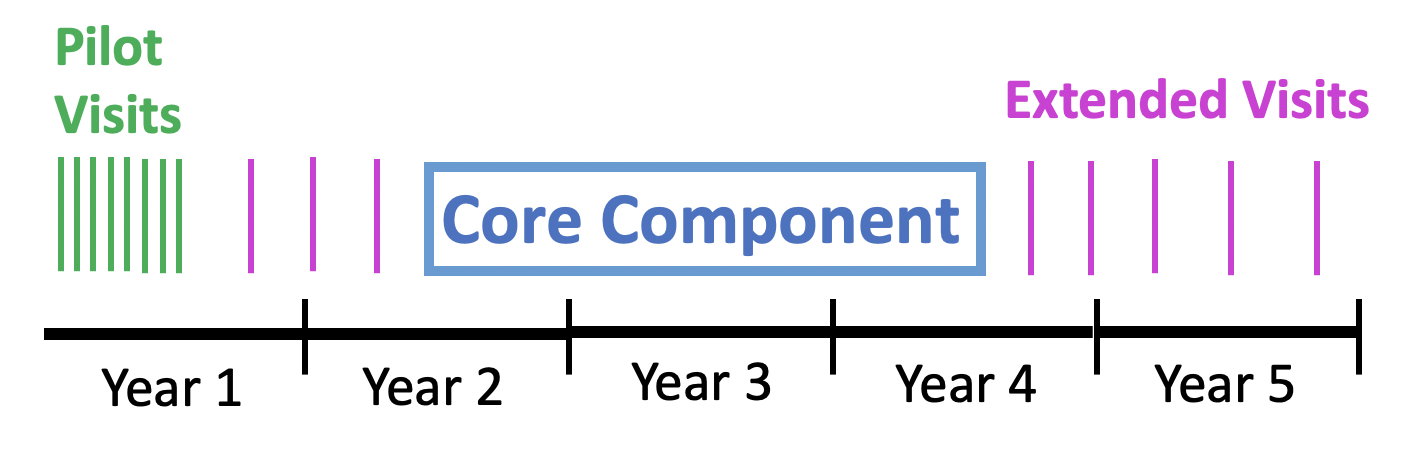}
    \vspace{-0.3cm}
     \caption{Illustration of three survey components recommended 
     in \citet{CCS_report}.
     }  
    \label{fig:ccs_survey}
\end{figure}

The \committee\ was constrained to use \TobsTotal\ observing days over the 5-year mission.
As shown in Fig.~\ref{fig:ccs_survey}, they defined 3 survey components as follows: 
\begin{itemize}
    \itemsep0em 
   \item a 2-year long core component to measure high-quality SN~Ia light curves for cosmology analysis.
      \TobsWFI\ observing days are used for the WFI, including a WIDE and DEEP tier, 
      and \TobsPRISM\ observing days are for the prism
      (\TobsTD\ total days);
    \item 8 extended visits (EV) before and after the core component to extend light curves
          for high-redshift events (DEEP tier only);
   \item \NPVtot\ Pilot visits (PV) for both tiers in year 1 to 
    (i) construct templates for real-time difference imaging during
        the core component, and 
        (ii) acquire a preliminary high-redshift sample that is much
        larger than currently available for $z>1$.
\end{itemize}

The proposed strategy for Roman has two tiers (WIDE and DEEP) and
key parameters are shown in Table~\ref{tb:ccs_strategy}.
The comma-separated list of filters and cadence times indicates {\it interlaced}
observations. 
Defining \MJDo\ as a WIDE tier observation time for $RZJ$, 
$RYH$ is observed at \MJDo$+5$, 
$RZJ$ is observed again at \MJDo$+10$, etc. 
Thus $R$-band is observed every 5 days, while each of the $ZJYH$ bands is observed every 10 days.
For the DEEP tier, $Z$-band is observed every 5 days, while each of the $YHJF$ bands
is observed every 10 days.

\begin{table}[ht!]
\begin{center}
\caption{Roman Survey Parameters\tablenotemark{a} }
\begin{tabular}{ l r c c c  }
\hline
Roman     & Area        &                            &           &  cadence      \\
Tier      &  (deg$^2$)  & $\ztarget$\tablenotemark{b} & filters  &  (days) \\
\hline
WIDE &  $\AreaWideCCS$  &     0.9  & $RZJ,RYH$   & 5,5     \\
DEEP &  $\AreaDeepCCS$  &     1.7  & $ZYH,ZJF$   & 5,5    \\
\hline 
\end{tabular}
\tablenotetext{a}{In-guide recommendation from \citet{CCS_report}.} \vspace{-1.6ex}
\tablenotetext{b}{See definition in Eq.~\ref{eq:SNRtarget}.}
\label{tb:ccs_strategy}
\end{center}
\vspace{-0.3cm}
\end{table}

In the analysis presented here we use only the WFI and only the 2-year core component,
and we combine the following samples in a joint analysis:
Roman WIDE and DEEP tiers (Table~\ref{tb:ccs_strategy}),
${\sim}\NWFDapprox$ low redshift ($z<\zmaxWFD$) events from the proposed LSST\footnote{\URLLSST} wide-fast-deep (WFD), and
${\sim}\NDDFapprox$ $z<\zmaxDDF$ events from the LSST deep drilling fields (DDF).
In addition to the nominal analysis using all four subsamples (two Roman and two LSST),
this analysis includes constraints for subsets in which 
(i) DDF is excluded,
(ii) Roman-WIDE is excluded, and
(iii) Roman-DEEP is excluded.
While we do not propose excluding any of these subsamples, 
these extra constraints may help to better understand
the \HLTDS\ sensitivity to dark energy parameters.

The simulated LSST samples (WFD and DDF) are assumed to be \specy\ confirmed
(pure SN~Ia samples) with accurate \spec\ redshifts ($\zspec$).
The simulated Roman samples include ``non-SNIa'' contamination from 
(i) core collapse SNe (SNCC), which includes SNII, SNIb, SNIc,
and (ii) peculiar SNe~Ia, which includes 91bg and Iax.

While we anticipate that \spec\ resources 
from the Roman prism \citep{Prism2023_whitepaper,Prism2025} and 
Subaru Prime Focus Spectrograph\footnote{\URLROMANSUBARU}
will result in a
subset of \spec\ classifications and redshifts for Roman, here we perform a 
conservative analysis and assume that no events are \specy\ confirmed,
and use accurate $\zspec$ only for redshifts $\ztrue{<}\zspecmax$.
The Roman data analysis therefore relies on photometric classification and 
photometric SN+host redshifts ($\zphot$). 
In contrast to this conservative photometric analysis, we implicitly assume
perfectly known SED models for light curve fitting and simulating bias corrections.

\section{Simulation}
\label{sec:snana_sim}

The simulation is used to generate \Ndataset\ 
statistically independent data sets that include Roman and LSST events. 
This choice of \Ndataset\  data sets is a compromise between CPU resources and obtaining a 
useful estimate of the dispersion in FoM values.
The simulation is also used to generate a single large
sample for classifier training (Sec.~\ref{subsec:scone}) 
and for correcting SN~Ia distance biases 
(Sec.~\ref{subsec:snana_bbc}); 
the same large sample is used for all \Ndataset\  data sets.

\begin{figure}[hb]
    \includegraphics[width=0.9\linewidth]{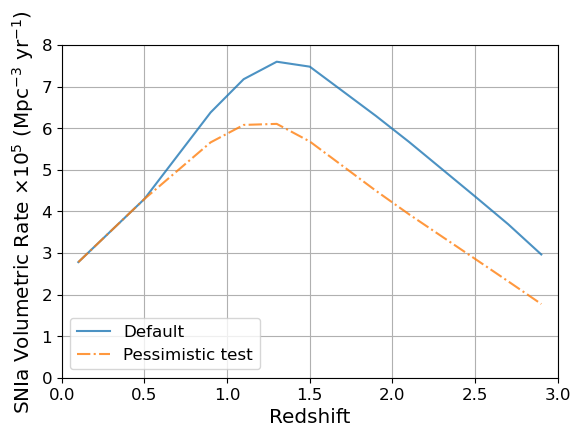} \hfill
     \caption{Volumetric SNIa rate vs. redshift used in the {\SNANA} simulation.
     	Based on \citet{Rodney2014}, and $H_0=70$~km/s/Mpc.
     }
    \label{fig:zrate}
\end{figure}

The simulated LSST and Roman data are generated using the {\SNANA} software package,
which has been used for distance bias corrections in many previous SNIa-cosmology analyses.
Accurate data/sim comparisons have been demonstrated for 
DES-SN5YR (Fig.~3 in \citet{Vincenzi2024}),
a combined spectroscopically confirmed compilation in Pantheon+ (Fig.~2 in \citet{Brout2022}),
the original Pantheon compilation (Fig.~7 in \citet{Scolnic2018}), and
PanSTARRS1 (Fig.~1 in {\cite{Jones2018_PS1}).

Compared to simulating real data, forecasting future surveys is more challenging because
it requires predicting instrumental performance. 
{\SNANA}  was used in the first detailed Roman (WFIRST) simulation and analysis with systematics 
\citep{Hounsell2018}, and more recently to define a reference survey that includes
three WFI tiers and 25\% PRISM time \citep{Rose2021}.
Here we use updated Roman instrument parameters as described in Appendix~\ref{app:instr_par}.
The LSST cadence and depth are based on OpSim v3.4.\footnote{\URLOPSIM}

The \SNANA\ simulation models three main components that are illustrated in 
Fig.~1 of \citet{Kessler2019_sim}: (i) source model, (ii) noise model, and (iii) trigger model.
These components are described below. 

\medskip
\noindent {\bf Source Model:} \\
For the SNIa light curve model we use the SALT3 SED time-series model
\citep{K21_SALT3} with updated training to include NIR observations \citep{Pierel2022}. 
The intrinsic brightness scatter is based on the dust model in \citet{BS21},
in which host-galaxy dust parameters $R_V$ and mean $A_V$ depend on the host-galaxy mass.
We use the DES-SN5YR ``dust map" describing the populations of 
$\{x_1, c, R_V, A_V, \alpha, \beta \}$
as determined by the {\tt Dust2Dust} fitting code \citep{Dust2dust2023}}.
The SNIa volumetric rate model vs. redshift (Fig.~\ref{fig:zrate})
is based on \citet{Rodney2014}. Since the high-$z$ rate uncertainty is large,
the orange dashed curve in Fig.~\ref{fig:zrate} shows the rate model used
for a pessimistic test in which the high-$z$ rate is well below the current estimate,
but still within measurement uncertainty.

SNCC (II/Ib/Ic) are generated using the SED templates from \citet{Vincenzi2019},
and the wavelength range has been extended to 25,000~\AA\ using the 
methods in \citet{Pierel2018}. 
The SNCC volumetric rate vs. redshift is from \citet{Strolger2015} (green curve in Fig.~6);
the rate fractions for II/Ib/Ic are 0.70/0.15/0.15.

Peculiar SNe Iax and 91bg are generated with the SED models in \citet{Kessler2019_plasticc}, 
and include improvements from \citet{Vincenzi2021}.
The SN~Iax volumetric rate vs. redshift follows the star formation rate in \citet{MD14},
and the rate at $z=0$ is 
$R_0 = 6{\times}10^{-6}$~yr$^{-1}$Mpc$^{-3}$. 
The 91bg rate is 10\% of the SN~Ia rate.

The cosmology model is flat \LCDM\ with parameters $\cosparLCDM$.
Average weak lensing magnifications are based on the $N$-body simulation from 
{\tt MICECAT} \citep{MICECAT2015} with mock galaxies from \citet{Carretero2015_mockgal}
(see Sec.~5.4 in \citet{Kessler2019_sim}).
The rms of the asymmetric distance scatter is roughly $0.055{\times}z$,
and we assume a lensing correction (e.g., \citet{Shah2024}) 
that results in an uncertainty of half the scatter, or $0.028{\times}z$.

\newcommand{\sigsky}{\sigma_{\rm sky}}
\newcommand{\Vsky}{V_{\rm sky}}
\newcommand{\Vsrc}{V_{\rm src}}
\newcommand{\Vread}{V_{\rm read}}
\newcommand{\Vhost}{V_{\rm host}}
\newcommand{\Vtot}{V_{\rm tot}}
\newcommand{\Npe}{N_{\rm pe}}

\medskip
\noindent {\bf Noise Model:} \\
The Roman model 
for the flux uncertainty includes variances from
instrumental readout ($\Vread$),
thermal+zodiacal ($\Vsky$), 
Poisson fluctuations from the host galaxy ($\Vhost$),
and Poisson fluctuations from the source flux ($\Vsrc$).
The total variance is $\Vtot = \Vsky + \Vread + \Vhost + \Vsrc$.
The LSST noise model is similar, except that thermal and zodiacal are replaced with sky noise.
Our FoM results are sensitive to the Roman noise model because we probe 
redshifts out to the Roman detection limit. 
The LSST samples, however, are truncated in redshift below the detection limit
($z<\zmaxWFD$ for WFD, $z<\zmaxDDF$ for DDF) and therefore our FoM has minimal 
sensitivity to LSST noise modeling.

To determine $\Vsky$ and $\Vread$, the variance per pixel is summed over an 
effective noise-equivalent-area (NEA) defined by
\begin{equation}
    {\rm NEA} = \frac{ \left[ \sum_{i}({\rm PSF}_i) \right]^2}{ \sum_{i}( {\rm PSF}_i^2) }~.
    \label{eq:NEA}
\end{equation}
For Roman, the index $i$ runs over the $19\times 19$ 
point-spread function (PSF) pixel grid,
and the PSF map is based on an average from the 
Roman images in the \OpenU\ simulation in \citet{OpenUnivserse2024}. 
The PSF size is comparable to the $0.11{\arcsec}$ pixel size, 
and therefore NEA increases as the PSF center moves from a pixel center to a pixel corner;
we do not model this pixel-location dependence.
For LSST, OpSim provides the time-dependent NEA for each observation.

$\Vhost$ is described in \S\ref{subsec:snana_host}, 
and $\Vsrc$ is the number of photoelectrons from the source flux.

\newcommand{\effdetect}{\epsilon_{\rm detect}}
\medskip
\noindent {\bf Trigger Model:} \\
For both Roman and LSST, 
the trigger model requires two detections in any band and the detections
must span at least two visits.
Thus multiple detections in one visit will not trigger the event.
An LSST detection is modeled by the efficiency ($\effdetect$) vs. 
signal-to-noise ratio (SNR) curve measured
in the LSST Data Challenge 2 (DC2); see Fig.~9  in \citet{Sanchez2022}
where $\effdetect=1/2$ for SNR=5.8.
We do not yet have $\effdetect$-vs-SNR curves for Roman,
and we therefore assume that $\effdetect=1/2$ at SNR${\sim}5$ 
and apply the following model:
$\effdetect=0$ for SNR$<4.5$,
rises linearly to $\effdetect=1$ at SNR$=5.5$,
and remains at 100\% for SNR$>5.5$.
A more realistic $\effdetect$-vs-SNR curve awaits analysis of the 
\OpenU\ images.

\newcommand{\Ntemplate}{N_{\rm template}}
\medskip
\noindent {\bf Miscellaneous \HLTDS\ Assumptions:} \vspace{-0.1cm}
\begin{itemize}
    \itemsep-0.2em 
    \item Gaps and pointing pattern are approximated by randomly removing 30\% of the 
           observations from an ideal tiling that is 100\% efficient;
    \item Difference-imaging template noise is not included;\footnote{
    For $\Ntemplate$ co-added images in the templates, the noise increases by
    $\sqrt{1+1/\Ntemplate}$. With $\Ntemplate=20$, for example, the
    noise increases by 2.5\%.
    };

    \item Efficiency for acquiring $\zspec$ is 100\% for $\ztrue{<}\zspecmax$ and zero for 
        $\ztrue{>}\zspecmax$;
    \item Efficiency for \spec\ confirmation is zero;
    \item All host galaxy associations are correct (\S\ref{subsec:snana_host}).
\end{itemize}

\bigskip
\subsection{Host Galaxy}
\label{subsec:snana_host}

A host galaxy library (\HOSTLIB) is needed to model 
(i) additional Poisson noise for both Roman and LSST and 
(ii) host photo-$z$ to use as a prior in the SN+host combined
photo-$z$ fit for Roman (\S\ref{subsec:snana_lcfit}).
The recent \OpenU\ effort selected galaxies from a \DIFFSKY\ catalog 
\citep{GCR2019_forDC2}\footnote{\URLDIFFSKY}
that includes significant upgrades compared to the catalog used for the 
LSST-DC2 image simulations \citep{LSST_DC2}.

To model transient-host correlations, the relevant {\DIFFSKY} catalog information includes
host-galaxy mags, 
\LOGMASS,\footnote{{\tt LOGMASS} = log10 of host stellar mass.}  
\LOGSFR,\footnote{{\tt LOGSFR} = log10 of host star formation rate.}
and a double Sersic profile with 
$n{=}1$ (exponential) and $n{=}4$ (de Vaucouleurs) components; 
photo-$z$'s were not included.
Here we use a random subset of {\HOSTLIBsize} galaxies for the
\SNANA\ {\HOSTLIB},
and include the following additional information:
\begin{itemize}
  \itemsep0em 
  \item Noisy mags corresponding to $S/N{=}10$ at $m{=}28$ for $RZYJHF$ stacked images.
        The \SNANA\ simulation requires three bands with $m{<}28$ for detection of the host.
  \item $\zphot$ generated by fitting $RZYJHF$ mags using the 
      EAZY fitting code\footnote{\URLEAZY} and 12 EAZY SED templates.
      The $\zphot$ PDF mean and standard deviation are stored in the {\tt HOSTLIB}, 
      and these parameters are propagated to the lightcurve fit stage (\S\ref{subsec:snana_lcfit})
      and used as a Gaussian prior.
\end{itemize}

After the simulation selects a random SN redshift ($\ztrue$) based on the rate model
(Fig.~\ref{fig:zrate}), a random galaxy from the {\HOSTLIB} is selected
with a $\ztrue$ tolerance of $0.01{+}0.01\ztrue$, and the host mags are adjusted to the SN redshift.
The tolerance is needed because there is no \HOSTLIB\ galaxy with the exact SN redshift,
and the number of potential galaxies within the tolerance should be large enough to describe
a reasonable \LOGMASS\ distribution to account for SN-host correlations. 
In addition to the \HOSTLIB, a weight map (\WGTMAP)
defines a relative probability for each galaxy to host an SNIa. 
For SN~Ia, we construct a \WGTMAP\ such that the {\LOGMASS} 
distribution matches the $z<0.5$ {\LOGMASS} distribution for the \DESVYR\ simulation.
For non-SNIa contamination, we use the {\WGTMAP}s developed for 
{\ELASTICC}\footnote{\ELASTICC\ = ``\ELASTICCtext,''  and was designed to test LSST alert brokers.}
within the LSST-DESC collaboration.
These \ELASTICC\  {\WGTMAP}s, however, are not appropriate for the new Roman \HOSTLIB, 
and the result is that non-SNIa are preferentially placed on lower-mass galaxies,
resulting in slightly higher SNR for contamination.
Additional \WGTMAP-development effort is needed for accurate correlations between
non-SNIa and host galaxies.

For LSST sims we use the {\HOSTLIB}s and {\WGTMAP}s from \citet{SCOTCH2023}.

\begin{figure}
    \includegraphics[width=0.9\linewidth]{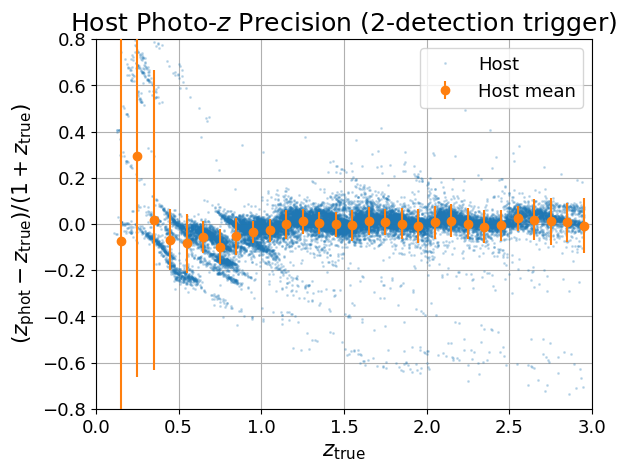} 
     \caption{Blue points show host-galaxy photo-$z$ residual $\Dz$ (Eq.~\ref{eq:Dz})
     vs. $\ztrue$ for simulated Roman events passing 2-detection trigger.
     Filled orange circles show mean bias and the error bars show 
    the standard deviation in each $\ztrue$ bin. }
    \label{fig:zphot_host} 
\end{figure}

The $\zphot$ residuals are defined as 
\begin{equation}
   \Dz \equiv \zres \label{eq:Dz}
\end{equation}
and they are shown in Fig.~\ref{fig:zphot_host}
for events passing the 2-detection trigger.
The orange error bars show the $\Dz$ standard deviation in each redshift bin. 
The residual scatter is large for $\ztrue<0.6$ because
the bluest Roman band ($R$-band) does not cover the 4000~\AA\ break. 
At higher redshifts, the average $\Dz$ scatter is ${\sim}0.07$.

The location of the SN near its host galaxy is randomly selected with a weight proportional
to the Sersic profile flux. 
The galaxy contribution to the flux uncertainty is described in Sec~4.22.2 in
the \SNANA\ manual,\footnote{\URLSNANAMANUAL}
and a key approximation is that the PSF shape is described by a Gaussian with
$\sigma=\sqrt{{\rm NEA}/(4\pi)}$, where NEA is defined in Eq.~\ref{eq:NEA}.

\begin{figure*}[ht]
    \includegraphics[width=0.32\linewidth]{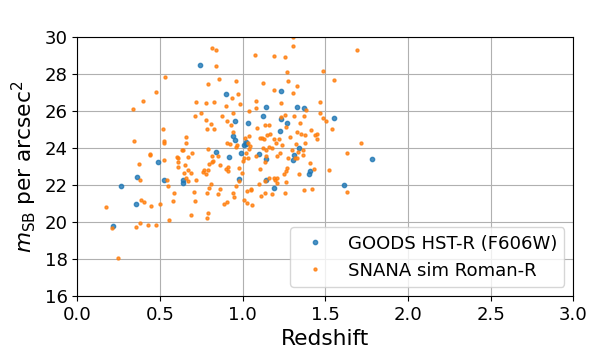}\includegraphics[width=0.32\linewidth]{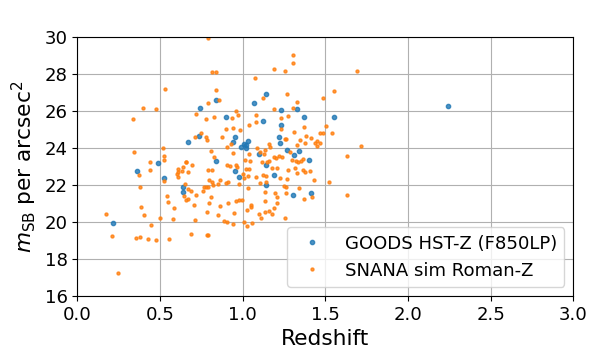}\includegraphics[width=0.32\linewidth]{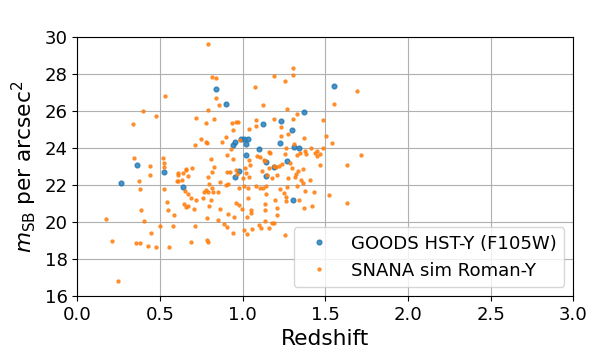}
    \includegraphics[width=0.32\linewidth]{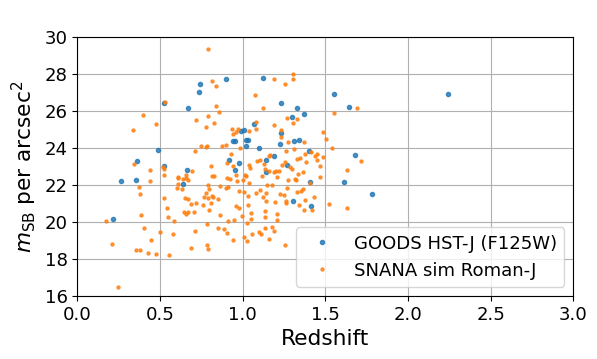}\includegraphics[width=0.32\linewidth]{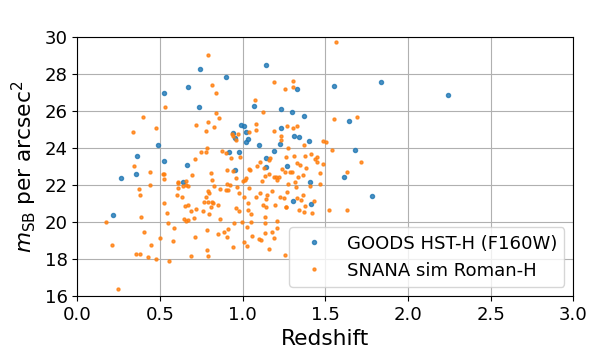}
     \caption{Distribution of local surface brightness mag ($\mSB$ per arcsec$^2$) vs. redshift for 5 bands
     that have close overlap between HST-GOODS and Roman.
     }
    \label{fig:validate_SB}
\end{figure*}

To check the Roman flux contribution from the host galaxy,
Fig.~\ref{fig:validate_SB} shows the local surface brightness (SB) magnitude, $\mSB$,
within a $0.1{\arcsec}$ radius, and compares the \SNANA\ Roman simulation to data from 
GOODS/HST.
While there is decent agreement in the $R$- and $Z$-band, the redder
bands show brighter SB in the sim compared to data. The data may be biased to fainter
SB due to selection effects (harder to detect an SN in a brighter galaxy), or the simulation
needs further tuning. We have decided to use the brighter simulated SB as a conservative
estimate.

\begin{figure}[hb]
    \includegraphics[width=0.9\linewidth]{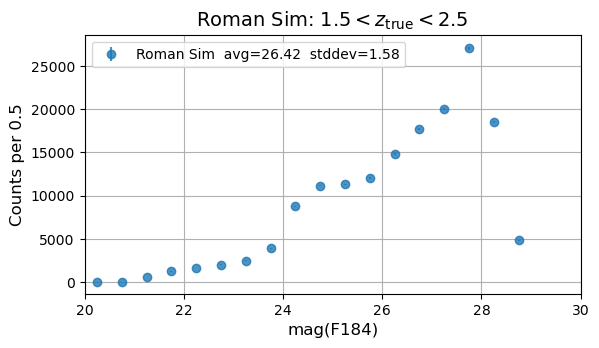}
    \includegraphics[width=0.9\linewidth]{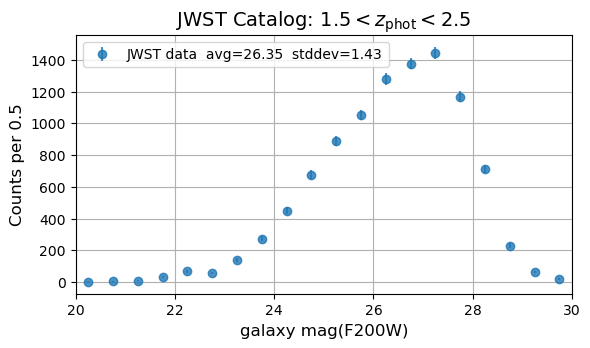}
     \caption{Distribution of galaxy mags in redshift range $1.5< z<2.5$ for 
        top: simulated Roman F184-band and 
	   bottom: JWST F200W band from the PRIMER catalog. 
	The mean and standard deviation for each distribution is shown in the legend.
     }
    \label{fig:validate_galmag}
\end{figure}

Since $F$-band has the longest exposure time, and the high-$z$ SNR in this band 
is sensitive to host galaxy Poisson noise, 
we crosscheck the simulated $F$-band galaxy mag distribution by comparing with 
real F200W galaxy mags (Fig.~\ref{fig:validate_galmag})
using the PRIMER catalog \citep{PRIMER} from the James Webb Space Telescope (JWST).
The JWST catalog likely has selection effects,
and the selected redshift range ($1.5 < z<2.5$) is based on photo-$z$'s with large
uncertainties, whereas the true redshift range is selected for the Roman simulation.
In spite of the JWST caveats, the distributions in Fig.~\ref{fig:validate_galmag}
are remarkably similar:
the means agree to better than 0.1 mag, the standard deviations agree to within 0.15 mag,
and the locations of the peak probability differ by about 0.5~mag.

Previous ground based studies have shown that the measured difference-imaging flux 
has scatter that is larger
than naive Poisson noise estimates (from sky, source, and host), and that this anomalous scatter
increases on brighter host galaxies. This effect has been shown for point sources overlaid on real 
data for the DES (Figs.~9,10 in \citet{Kessler2015_DIFFIMG}) and for sources overlaid on
simulated images for LSST (Fig.~14 in \citet{Sanchez2022}). 
This effect is ignored in the Roman sims presented here,
but will be important to investigate later using transient fluxes  measured 
from the \OpenU\ images.

\subsection{Determination of Roman Exposure Times and Survey Area}
\label{subsec:snana_texpose}

Following R25, we determine the exposure time ($\Texpose$) 
needed to obtain the the following $\SNRtarget$ at peak brightness and redshift $\ztarget$ :
\begin{equation}
 \SNRtarget  = 10 \sqrt{  \frac{\rm cadence}{2.5(1 + \ztarget)}  }~.
 \label{eq:SNRtarget}
\end{equation} 
\newcommand{\SNRtargetWIDE}{14.5}
\newcommand{\SNRtargetDEEP}{12.2}

Using the cadence and $\ztarget$ values in Table~\ref{tb:ccs_strategy},
the WIDE tier $\SNRtarget$ value is \SNRtargetWIDE\ for all bands redder than $R$-band;
we set $\SNRtarget=2$ for $R$-band to avoid excessive $\Texpose$ in the rest-frame UV region 
($<3500$~\AA) where the SN flux is much fainter compared to the optical region.
For the DEEP tier, $\SNRtarget=\SNRtargetDEEP$ for bands redder than $Z$-band;
$R$- and $Z$-band  $\SNRtarget=2$.

Since there is no analytical method to compute $\Texpose$, 
the computation uses the \SNANA\ sim to produce at-peak SNR on a grid of $\Texpose$, 
and interpolates $\Texpose$-vs-SNR to the desired SNR. 
To reduce statistical fluctuations in the $\Texpose$ computation, 
simulated SNR scatter is minimized by
fixing SALT3 parameters $c{=}0$ and $x_1{=}0$,
and removing the intrinsic scatter model. 
The only source of event-to-event variation is Poisson noise from a randomly selected
host galaxy (\S\ref{subsec:snana_host}).
To avoid excessive overhead for short exposures, 
there is a minimum exposure time requirement of $\Texpose \ge \TexposeMIN$~seconds.
A counter-intuitive artifact of this computation is that $\Texpose$ can sometimes decrease with 
increasing redshift because of rest-frame spectral features moving in and out of a filter band. 
To avoid this artifact, we follow R25 and pre-compute a table of calculated $\Texpose$ vs. $\ztarget$
and forbid simulated $\Texpose$ from decreasing with increasing $\ztarget$
(hereafter called ``monotonically increasing constraint'').

\begin{figure*}[ht!]
    \includegraphics[width=0.3\linewidth]{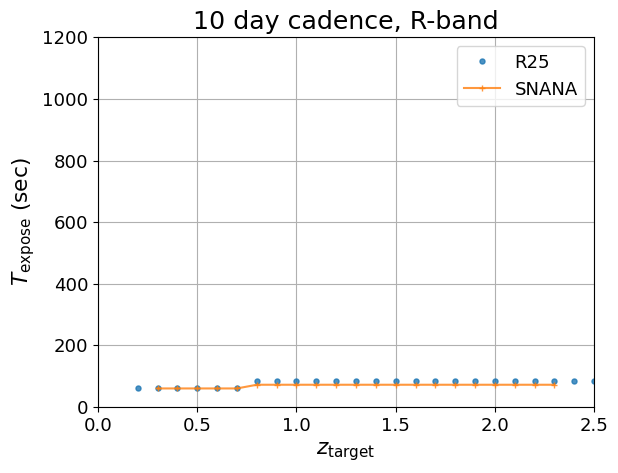}\includegraphics[width=0.3\linewidth]{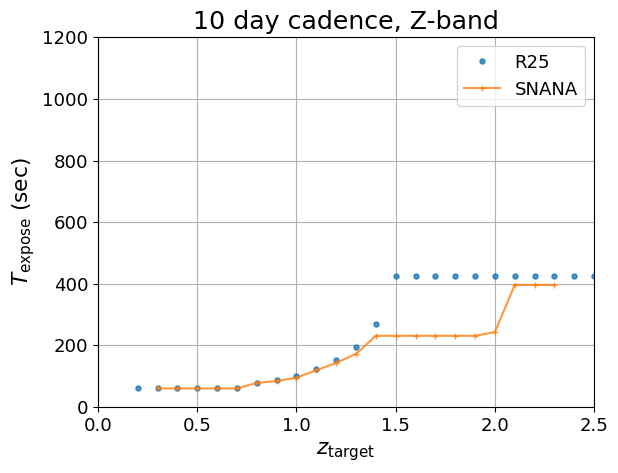}\includegraphics[width=0.3\linewidth]{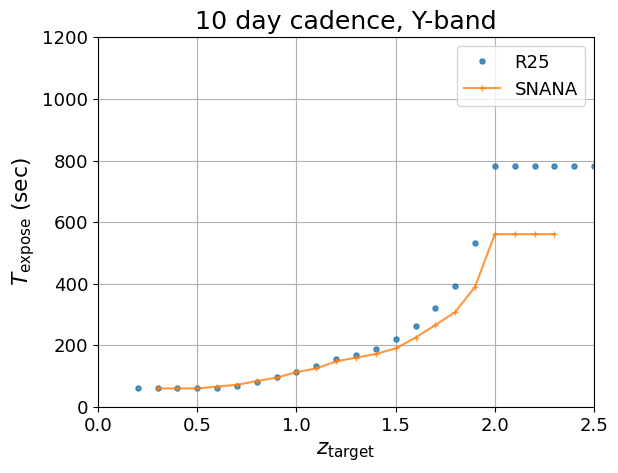}
    \includegraphics[width=0.3\linewidth]{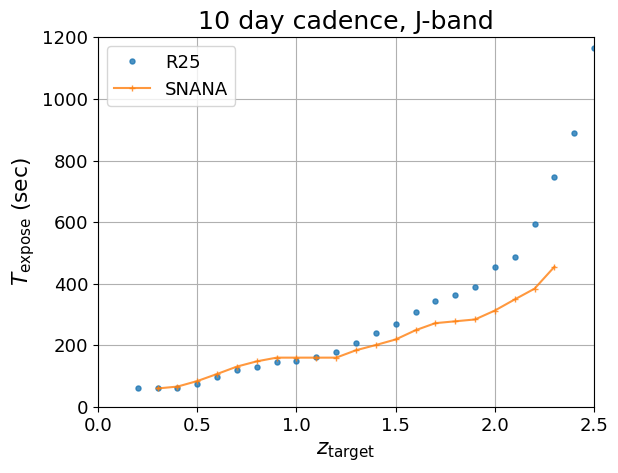}\includegraphics[width=0.3\linewidth]{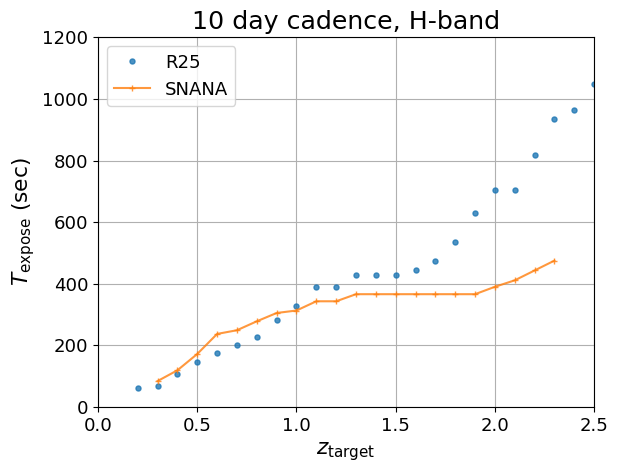}\includegraphics[width=0.3\linewidth]{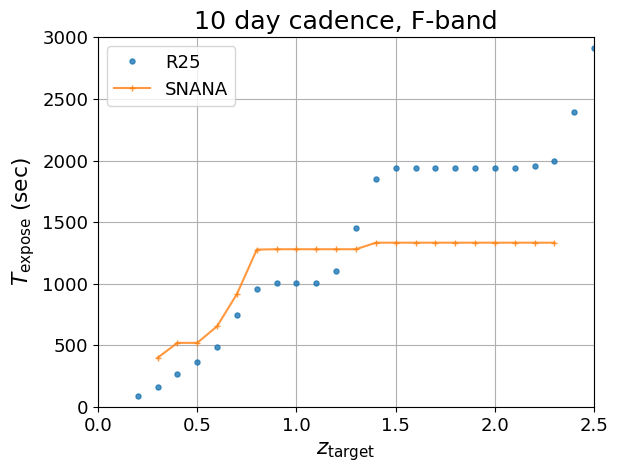}
    \caption{
     $\Texpose$ vs. $\ztarget$ with 10~day cadence for:
    R25 (blue points) and \SNANA\ (orange lines).
    Each panel shows a different band:
    the first 5 panels have the same vertical axis scale (0-1200~sec)
    while the $F$-band panel has a wider scale (0-3000~sec).
	The constraint  $\Texpose \ge \TexposeMIN$~seconds is applied to both sets of calculations.
     }
    \label{fig:texpose_vs_z}
\end{figure*}

$\Texpose$ vs. $\ztarget$ is shown in Fig.~\ref{fig:texpose_vs_z} for each band,
and compares the \SNANA-based calculation with R25,
which uses a different galaxy simulation method based on the measured SB
at SN locations in HST GOODS data.
There is good agreement in the bluest bands ($R$, $Z$, $Y$) over the entire redshift range.
In $J$-band the agreement is good up to nearly $\ztarget{\sim}1.5$, and then they become
discrepant at higher redshift. The reddest bands ($H,F$) show increasing discrepancy,
and after comparing sources of noise between \SNANA\ and R25,
we have isolated the noise discrepancy to the modeling of the host galaxy.

\newcommand{\URLDSnine}{\url{http://ds9.si.edu}}

While the \SNANA\ simulation predicts smaller $\Texpose$ than R25, 
indicating a smaller SB in the \SNANA\ simulation compared to HST GOODS data,
there is a paradox because the \SNANA\ simulation predicts brighter $\mSB$ 
than the HST GOODS data that is used in R25 (Fig.~\ref{fig:validate_SB}).

As a final test of the \SNANA\ host galaxy simulation, we selected a single galaxy
from the \OpenU\ image simulation ($z=1.7$, $m_F=23.8$~mag, and SN-host separation of $0.02\arcsec$), 
manually measured $F$-band $\mSB$ on the image using ds9\footnote{\URLDSnine}, 
and compared this result to the \SNANA\ prediction based on the Sersic profile and galaxy mag.
The sky plus read noise is in excellent agreement, 
and the \SNANA\ SB-flux prediction is about 15\% brighter than the image SB flux.
This slight discrepancy is likely within the uncertainty based on modeling of the Sersic profile,
and suggests that the \SNANA\ simulation may overestimate the SB flux rather than underestimate.
Similar image tests on HST GOODS data would be useful to help resolve the discrepancy
between the \SNANA\ and R25 host simulations.

The final exposure times per band are shown in Table~\ref{tb:texpose}.
Next, the observing areas are computed using
the total WFI observing time (\TobsWFI\ days), exposure time per band, 
cadence, and 0.281~deg$^2$ field of view. 
The WIDE/DEEP area ratio from Table~\ref{tb:ccs_strategy} is used as a constraint.
The computed areas per tier are $\AreaWideSNANA$ and $\AreaDeepSNANA$~deg$^2$
for WIDE and DEEP, respectively. These SNANA-computed areas are within
$2$\% of the \committee\ values in Table~\ref{tb:ccs_strategy}.

\begin{table}[hb!]
\begin{center}
\caption{SNANA-computed exposure time and $5\sigma$ depth per band and tier}
\begin{tabular}{ | c | c c | c c | }
\hline    
Roman   &  \multicolumn{2}{c|}{ $\Texpose$ (sec):}  &  \multicolumn{2}{c|}{ $\mlimit$ } \\
band    &    WIDE & DEEP  &  WIDE  & DEEP     \\
\hline
  $R$   &    72   &  ---  &  26.55 &  ---     \\
  $Z$   &    84   & 231   &  26.22 & 26.84     \\
  $Y$   &    95   & 266   &  26.24 & 26.86    \\
  $J$   &   160   & 272   &  26.54 & 26.86    \\
  $H$   &   305   & 366   &  26.85 & 26.95    \\
  $F$   &   ---   & 1333  &  ---   & 27.08    \\
\hline
\end{tabular} \label{tb:texpose} 
\end{center}  \vspace{-0.3cm}
\end{table}

\begin{figure*}[ht]
    \includegraphics[width=0.3\linewidth] {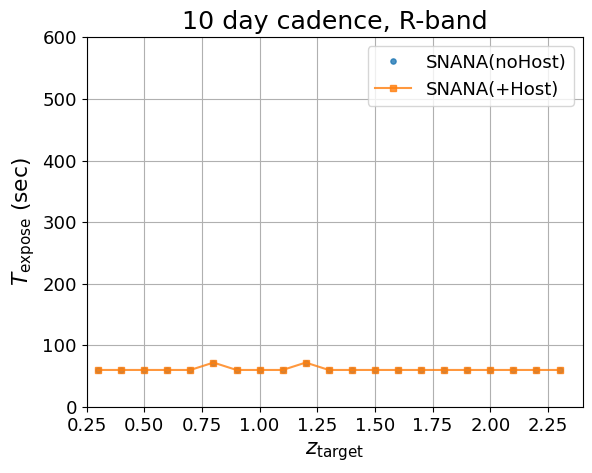}\includegraphics[width=0.3\linewidth]{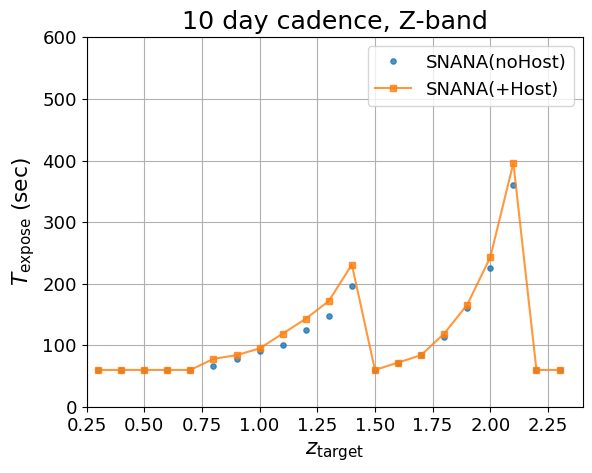}\includegraphics[width=0.3\linewidth]{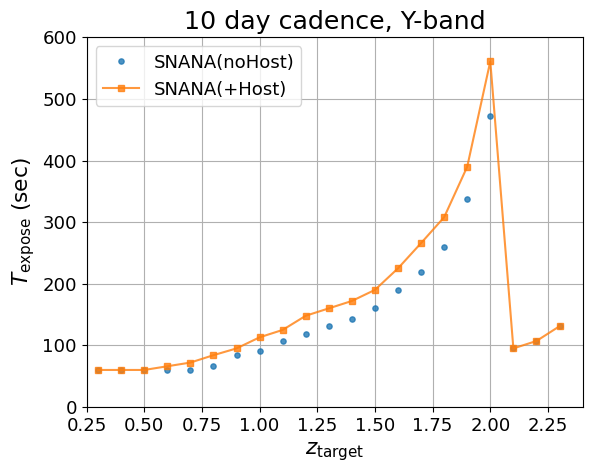}
    \includegraphics[width=0.3\linewidth]{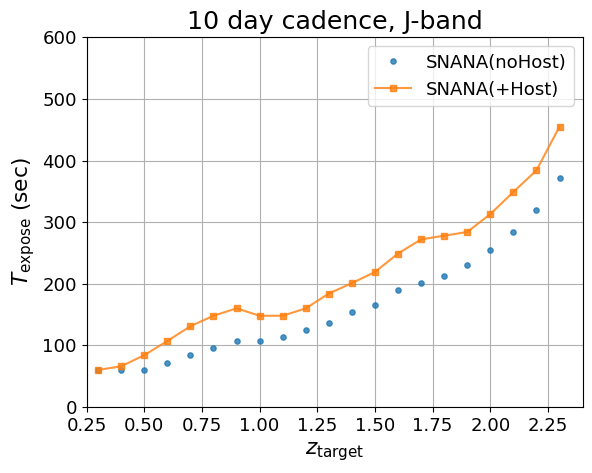}\includegraphics[width=0.3\linewidth]{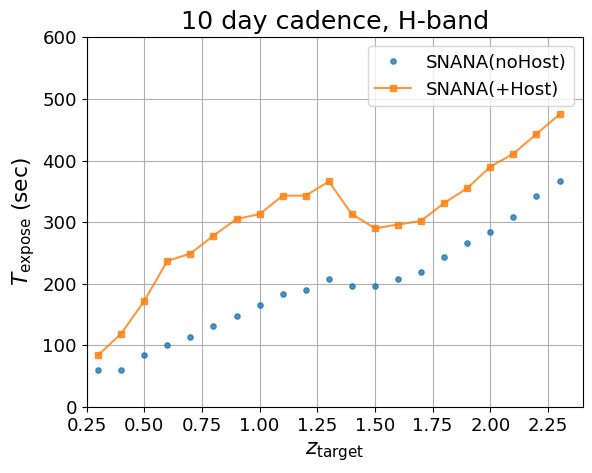}\includegraphics[width=0.3\linewidth]{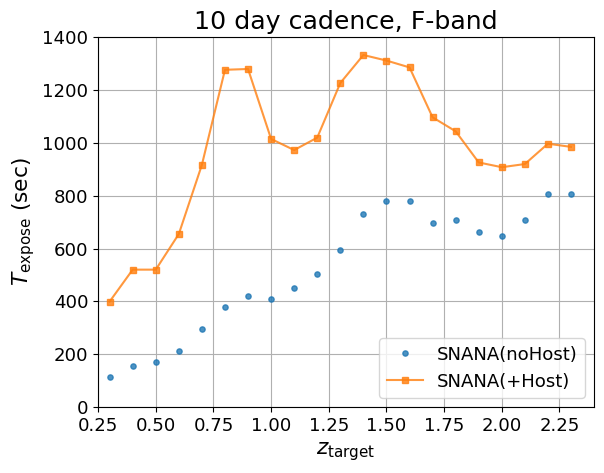} 
    \caption{\SNANA-calcuated exposure time ($\Texpose$) vs. $\ztarget$,
	with host galaxy noise (orange line) and without the host (blue points).
	Each panel shows a different band; 
    the first 5 panels have same vertical axis scale (0-600~sec)
    while the $F$-band has a wider scale (0-1400~sec).
	The constraint  $\Texpose \ge \TexposeMIN$~seconds is applied, and the
    monotonically increasing constraint has been removed.
    The orange lines here (without monotonic constraint) 
    correspond to the orange lines in Fig.~\ref{fig:texpose_vs_z}
    that impose the monotonic constraint.}
    \label{fig:texpose2_vs_z}
\end{figure*}

To illustrate the impact of host-galaxy noise, Fig.~\ref{fig:texpose2_vs_z} 
shows \SNANA-computed $\Texpose$-vs-$\ztarget$ with and without host-galaxy noise.
The differences are small in the bluer bands, and increase significantly with $\ztarget$.
Here we show a numerical illustration for $H$- and $F$-bands with 10~day cadence and $\ztarget{=}1.7$.
The calculated $\Texpose(H) = 219{\to}302$~seconds without$\to$with 
host galaxy noise; imposing the non-decreasing constraint,
$\Texpose(H) = 219{\to}366$~sec.
For the $F$-band,  $\Texpose(F) = 696\to 1097$~sec without$\to$with 
host galaxy noise; imposing the non-decreasing constraint,
$\Texpose(F) = 779\to 1333$~sec. 
Given the large impact of the host galaxy on the measurement noise, and the
discrepancy between \SNANA\ and R25 simulations (Fig.~\ref{fig:texpose_vs_z}),
the modeling of host-galaxy noise is likely the most uncertain component in the simulation.

\subsection{ Visit Depths } 

The average $5\sigma$ visit-depth band is shown in Table~\ref{tb:lsst_m5sig}
for LSST WFD and DDF. The deepest bands are $g$ and $r$, and the DDF depths
are 0.6 to 0.8 mag deeper than WFD. The corresponding Roman depths are
shown in Table~\ref{tb:texpose}, and they are 2-3 mag deeper compared to LSST.

\begin{table}[hb!]
\begin{center}
\caption{LSST Average $5\sigma$ depth for WFD and DDF}
\begin{tabular}{ | c | c c | }
\hline    
LSST   &    \multicolumn{2}{c|}{ $\mlimit$ } \\
band    &    WFD  & DDF     \\
\hline
  $u$   &    23.20 & 23.79     \\
  $g$   &    24.47 & 25.04     \\
  $r$   &    24.12 & 24.74    \\
  $i$   &    23.61 & 24.31    \\
  $z$   &    23.00 & 23.79   \\
  $Y$   &    22.20 & 22.80   \\
\hline
\end{tabular} \label{tb:lsst_m5sig} 
\end{center}  \vspace{-0.3cm}
\end{table}

\subsection{Simulation Statistics}

\newcommand{\ngen}{N_{\rm gen}}
\newcommand{\nacc}{N_{\rm trigger}}
\newcommand{\naccfracWFD}{36}

The number of events generated ($\ngen$) and passing the 2-detection trigger ($\nacc$)
is shown in Table~\ref{tb:sim_stats} for each survey/tier and each transient model.
$\ngen$ for LSST WFD and DDF were tuned to achieve a predetermined number of events
after selection cuts (Sec.~\ref{subsec:snana_lcfit}).
The $\naccfracWFD$\% trigger acceptance for WFD is defined largely by seasonal effects
in which LSST fields are not visible or the airmass is beyond observing limits.
While LSST-DDF has similar seasonal effects, it has a higher trigger acceptance 
because of the (1) time dilation induced by average higher redshift, and 
(2) more sensitive co-added depth. The two detections could occur before or long
after peak brightness and dooes not ensure observations near peak brightness.

For Roman, $\ngen{\sim}30,000$ is computed from the volumetric rate model (Fig.~\ref{fig:zrate}),
redshift range, and survey area.
The 90\% trigger acceptance is much higher (compared to LSST) 
because of the continuous viewing zone fields selected,
and therefore there are no seasonal effects.
For the non-SNIa contaminants, the trigger acceptance is lower (compared to SNIa) 
because these events are intrinsically fainter.


\begin{table*}[ht!]
\begin{center}
\caption{Simulation Statistics for each Survey and Transient Type}
\begin{tabular}{ | l c c | r r r  | }
\hline 
              &  max      & transient  & \multicolumn{3}{c|}{Number of events (fraction):\tablenotemark{a}} \\
  Survey      & redshift  & type       &  Generated\tablenotemark{b}  
                                       & pass trigger\tablenotemark{c} 
                                    & pass cuts\tablenotemark{d} \\
\hline
  LSST(WFD)  &  0.09  & Ia       &     5200 &    1877(0.361)  &    790(0.152)   \\
\hline
  LSST(DDF)  &  0.50  & Ia       &     8800 &    6263(0.712)  &   3610(0.410)   \\
\hline
  ROMAN\tablenotemark{e}  &  2.95  & Ia       &    28660 &   25781(0.900)  &   9930(0.346)   \\
  ROMAN     &  2.95  & Pec-Iax  &    20593 &    2070(0.101)  &     26(0.001)   \\
  ROMAN     &  2.95  & Pec-91bg &     6508 &    2452(0.377)  &      1(0.000)   \\
  ROMAN     &  2.95  & IIP      &    76169 &   23093(0.303)  &      0(0.000)   \\
  ROMAN     &  2.95  & IIL      &    11238 &    6902(0.614)  &     12(0.001)   \\
  ROMAN     &  2.95  & Ib       &    18730 &    9144(0.488)  &    143(0.008)   \\
  ROMAN     &  2.95  & Ic       &    18730 &   10768(0.575)  &     75(0.004)   \\
\hline
\end{tabular} \label{tb:sim_stats} 
\tablenotetext{a}{The fraction in parentheses is with respect to the number in the generated column.} \vspace{-1.6ex}
\tablenotetext{b}{Computed physical rate for Roman; artificial rate for LSST to achieve desired statistics.}  \vspace{-1.6ex}
\tablenotetext{c}{Satisfies 2-detection trigger.}  \vspace{-1.6ex}
\tablenotetext{d}{Satsifies selection cuts in \S\ref{subsec:snana_lcfit}.} \vspace{-1.6ex}
\tablenotetext{e}{Includes WIDE and DEEP tiers.}
\end{center}  \vspace{-0.3cm}
\end{table*}

\section{Analysis}
\label{sec:snana_ana}

Here we describe the analysis performed on each of the \Ndataset\ simulated data samples,
and in Sec.~\ref{sec:snana_results} the FoM results are averaged over these 9 samples.
Our analysis here largely follows the recent \DESVYR\ cosmology analysis \citep{Vincenzi2024} 
using a photometrically classified sample, 
and also follows a cosmology analysis applied to simulated LSST data
using photometric redshifts \citep{Mitra2023}.

\subsection{Selection Cuts and Lightcurve Fitting with SALT3}
\label{subsec:snana_lcfit}

The analysis begins with a multi-band light curve fit for each event using the SALT3 model,
which is needed to standardize the brightness. 
For LSST using an accurate $\zspec$, the 4 fitted parameters are
(i) time of peak brightness, $t_0$,
(ii) amplitude, $x_0$, with $m_x = -2.5\log(x_0)$,
(iii) stretch parameter, $x_1$,
(iv) color parameter, $c$.
For Roman events without $\zspec$, the redshift is a 5th fitted parameter
using the host-galaxy $\zphot$ (mean and rms of pdf) as a Gaussian prior.
The central rest-frame wavelength range for both simulation and light curve
fitting is 2800-25000\AA.

The selection requirements (cuts) are:
\vspace{-0.2cm}
\begin{itemize}
  \itemsep -0.4em 
  \item Convergent SALT3 fit.
  \item SALT3 fit probability (computed from $\chi^2$ and number of degrees of freedom)
    $\Plcfit > 0.01$
  \item At least one observation with $\Trest{<}-5$ days and another with $\Trest{>}20$~days;
  \item \SNRSUM$>40$ where \SNRSUM\ is the quadrature sum of SNR over observations
      with $-15{<}\Trest{<}+45$~days in the rest-frame. Higher-$z$ events include more
      observations in  \SNRSUM, but with lower SNR;
  \item At least 3 bands with an observation with SNR$>5$;
  \item $t_0$ is within the 2-year core component (Fig.~\ref{fig:ccs_survey}), 
  and $\sigma_{t_0}<2$~days;
  \item $|x_1| < 3$, $\sigma_{x_1}<1$, $|c|<0.3$;
  \item Valid distance bias correction (Sec.~\ref{subsec:snana_bbc}) for all \Nsyst\ systematics.
\end{itemize}
In addition to these event cuts, we only keep observations within the 2-year core component
and ignore the EV component (Fig.~\ref{fig:ccs_survey}). 
While a photometric classifier is used in the analysis (Sec.~\ref{subsec:snana_bbc}), 
there is no explicit cut on classifier probability.

\begin{figure}[ht]
    \includegraphics[width=0.89\linewidth]{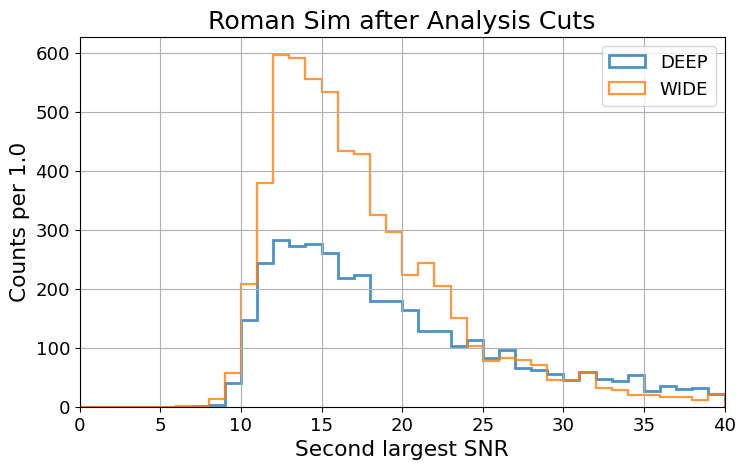} 
     \caption{Second largest SNR after analysis cuts for Roman WIDE and DEEP tiers.
     } 
    \label{fig:snrmax2}
\end{figure}

Since the $\effdetect$ vs. SNR assumption (Sec.~\ref{sec:snana_sim}) is essentially a guess
based on DES and LSST image processing,
we check the sensitivity of our analysis to the 2-detection trigger requirement
by showing the distribution of the second largest SNR in Fig.~\ref{fig:snrmax2}.
The distribution shows that our analysis is not sensitive to details of the 
efficiency model and we only assume that $\effdetect\to 1$ for SNR${>}8$.
However, if future analyses attempt to increase statistics by reducing
the $\SNRSUM$ cut, a more accurate model of $\effdetect$ would be needed.

The number of events passing cuts for each survey/tier and transient type is shown
in the last column in Table~\ref{tb:sim_stats}. The overall Roman selection efficiency for SN~Ia
is ${\sim}1/3$, which is much higher than the sub-percent efficiency for contaminants.
This efficiency difference is mainly from the SALT3 fit requirement, 
SNR cuts, and the implicit classification cut from requiring a valid bias correction.
This last cut is subtle because light curve fits on contaminants often result
in $\{z,x_1,c\}$ values that are not populated by SNe~Ia, and therefore a bias correction
cannot be determined.

To account for the fitted $\zphot$ uncertainty, one would naively add $\sigma_{zphot} \times d\mu/dz$ 
in quadrature to each distance uncertainty. However, the strong correlation between $\zphot$ and fitted
color results in a self-correction such that the naive distance uncertainty is greatly overestimated.
This effect is discussed in Sec.~\ref{sss:host_zphot_bias}, and also
see Appendix A in \citet{Chen2022} and Sec.~5.3 in \citet{Mitra2023}.

\begin{figure}[ht]
    \includegraphics[width=0.49\linewidth]{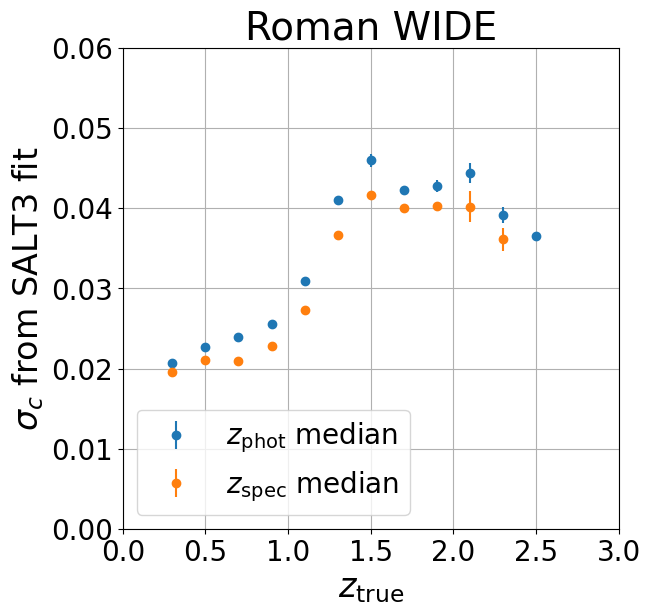} 
    \includegraphics[width=0.49\linewidth]{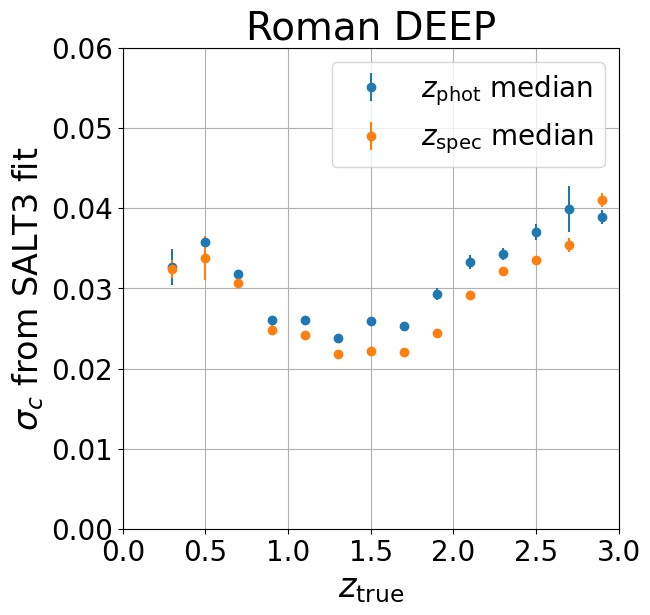}
     \caption{Median fitted color uncertainty ($\sigma_c$) vs. redshift
     for SN+host $\zphot$ (blue) and accurate $\zspec$ (orange).
     The left and right panels show Roman WIDE and DEEP tiers, respectively.
     } 
    \label{fig:zphot_cerr}
\end{figure}

Rather than adding an explicit $\zphot$-dependent uncertainty on the distance,
the impact of fitting for $\zphot$ is expressed by larger SALT3 fitted uncertainties
on the other parameters (compared to fitting with $\zspec$), which results in 
increased distance uncertainty.
Fig.~\ref{fig:zphot_cerr} shows
a comparison of the fitted color uncertainty ($\sigma_c$) between
fitting $\zphot$ and using accurate $\zspec$ 
on the same Roman sample.

\subsection[Photo-z Performance]{Photo-$z$ Performance}
\label{subsec:snana_pz}

\newcommand{\sigDz}{\sigma_{\Dz}}

\begin{figure}[hb]
    \includegraphics[width=0.49\linewidth]{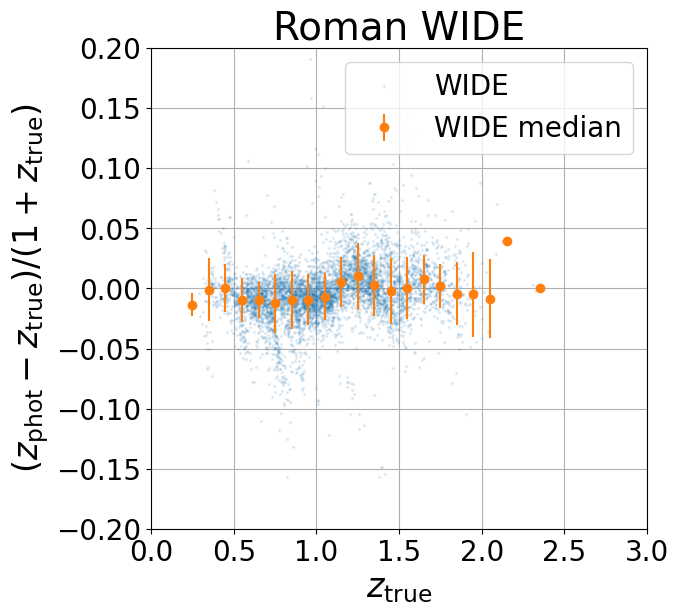}
    \includegraphics[width=0.49\linewidth]{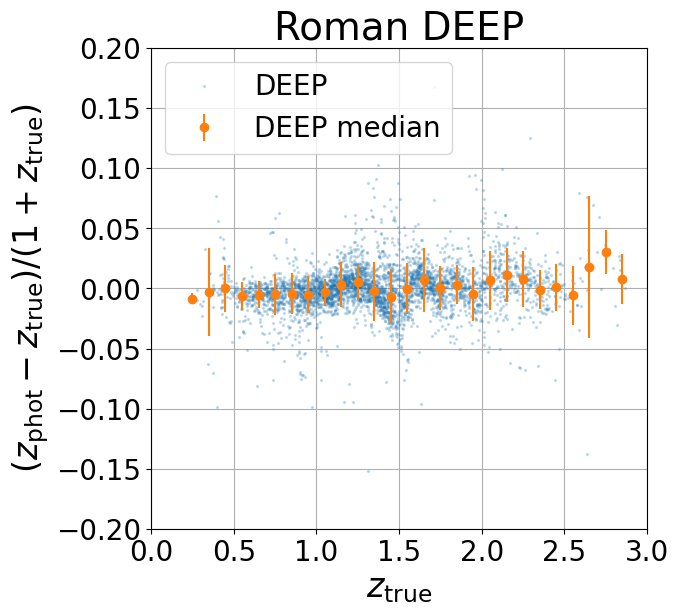}
     \caption{For true SN~Ia events after analysis cuts, 
        SALT3-fitted photo-$z$ residuals ($\zres$ vs. $\ztrue$) 
        are shown in blue for WIDE (left) and DEEP (right) tiers.
	    Filled orange circles show the mean bias, and the error bars show the standard 
    	deviation in each $\ztrue$ bin.
	    Plots exclude the $\ztrue<\zspecmax$ subset with accurate $\zspec$.
     }
    \label{fig:zphot_resid}
\end{figure}

The SALT3 (SN+host) fitted photo-$z$ performance is illustrated in 
Fig.~\ref{fig:zphot_resid} for true SN~Ia events passing cuts.
The small $\zphot$ biases in Fig.~\ref{fig:zphot_resid} are discussed in Sec.~\ref{subsec:snana_bbc},
and Table~\ref{tb:Dz_rms} quantifies the standard deviation of $\Dz$ (Eq.~\ref{eq:Dz}),
$\sigDz$, 
for different selection criteria (trigger and after analysis cuts) and for different redshift ranges
($\ztrue<0.6$ and $\ztrue>0.6$).
In the low redshift region, where the Roman filter bands do not cover the 4,000\AA\ break,
the host-only standard deviation is $\sigDz{\sim}0.3(0.4)$ for the WIDE(DEEP) tiers 
for events satisfying the 2-detection trigger.
After analysis cuts, $\sigDz$  improves to ${\sim}0.1$, but at a cost of low efficiency
as described below. The final SN+host $\sigDz$ improves further to 0.04(0.02) for WIDE(DEEP).
In the high-redshift region, the host-only $\sigDz{\sim}0.07$ for events satisfying the trigger,
and the precision is very similar after analysis cuts. 
The final SN+host $\sigDz$ is reduced by roughly a factor of 2.
While the final $\sigDz$ is comparable for both redshift ranges, the analysis efficiency
is much higher in the $\ztrue>0.6$ range.

\begin{table}[hb!]
\begin{center}
\caption{$\Dz$ Photo-$z$ Precision}
\begin{tabular}{ | l l c | c c | }
\hline    
                &  $\zphot$  & $\ztrue$ &  \multicolumn{2}{c|}{$\sigDz$ for:} \\
selection       &  source    &  range   &   WIDE       & DEEP       \\
\hline
trigger         &  host      &  0.3-0.6   &  0.290   &  0.420  \\                              
analysis + cuts &  host      &  0.3-0.6   &  0.119   &  0.090  \\ 
analysis + cuts &  host+SN   &  0.3-0.6   &  0.039   &  0.020  \\ 
\hline
trigger         &  host      &  0.6-3.0   &  0.070   &  0.072  \\ 
analysis + cuts &  host      &  0.6-3.0   &  0.075   &  0.068  \\ 
analysis + cuts &  host+SN   &  0.6-3.0   &  0.034   &  0.034  \\ 
\hline
\end{tabular} \label{tb:Dz_rms} 
\end{center}  \vspace{-0.3cm}
\end{table}

\begin{figure}[hb]
    \includegraphics[width=0.49\linewidth]{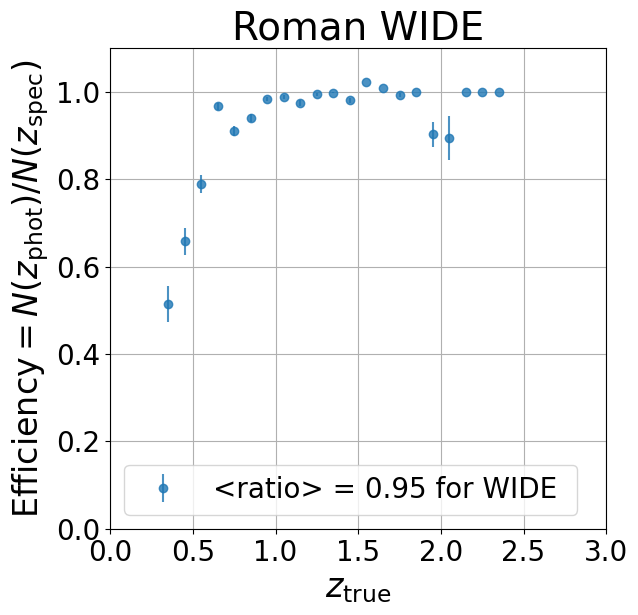} \hfill
    \includegraphics[width=0.49\linewidth]{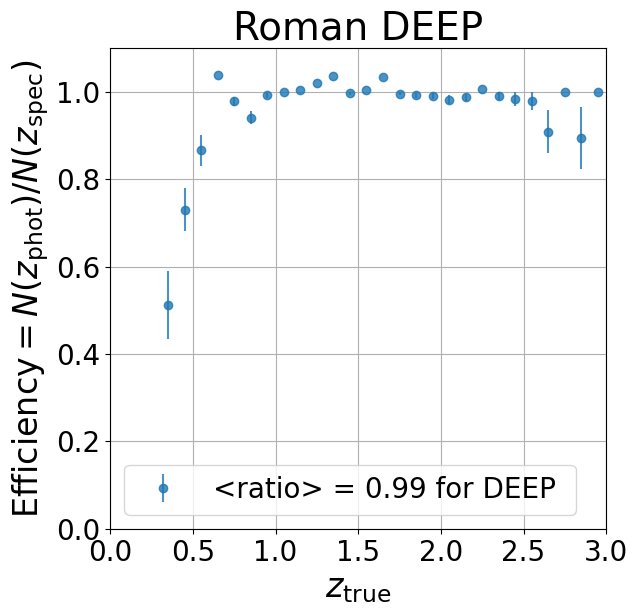} \hfill
     \caption{Roman efficiency for $\zphot$ SALT3 lightcurve fits compared to
     	lightcurve fits using accurate $\zspec$:   
        WIDE (left) and DEEP (right).    
     }
    \label{fig:eff_zphot}
\end{figure}

To estimate the loss of Roman events from a photo-$z$ fit,  Fig.~\ref{fig:eff_zphot} shows the 
$\zphot$ light curve fitting efficiency with respect to an ideal analysis using an accurate $\zspec$
for all events.
For $\ztrue  < 0.6$ the efficiency drops dramatically in both tiers.
Recall that there are no $\zspec$ for $\ztrue>\zspecmax$ in the simulation, and therefore the
$\zphot$ efficiency in Fig.~\ref{fig:eff_zphot} is a worst-case scenario.
It is somewhat fortuitous that adding host-galaxy $\zspec$ at lower redshifts has
the biggest improvement in analysis efficiency (per redshift range), 
and that this lower-redshift region is most accessible to ground-based spectroscopic resources.

The efficiency impact on SN+host $\zphot$ precision is minor
(Fig.~\ref{fig:zphot_resid}), but the $\ztrue{<}0.6$ event loss from requiring a convergent
SALT3 fit is much larger compared to higher $\ztrue$.
While it may be possible to improve the SALT3 light curve fitting code, 
a more realistic approach to recovering these lower $\ztrue$ events is to either 
(i) measure accurate host $\zspec$ out to ${\sim}0.6$ (e.g., PRISM, Subaru), 
and/or (ii) include bluer photometric bands
from external surveys (e.g., PanSTARRS1, LSST, EUCLID-VIS\footnote{\URLEUCLID}) 
to improve the host $\zphot$ precision.

\subsection{Photometric Classification}
\label{subsec:scone}

For Roman, the \SCONE\ classifier is used to train a model on a separate
training sample consisting of 30,000 SNIa and 15,000 non-SNIa,
where the simulation code and inputs are identical to those used for simulated data.
\SCONE\ uses a convolutional neural network on images constructed from light curves
using a 2D Gaussian regression to estimate flux on a grid of phase and wavelength.
Other information, such as host galaxy properties and redshift, is not used.
To have a more effective training on fewer events, 
the training is performed on events passing all selection requirements;
the resulting trained model is applied to each simulated data event
to determine $\PIa$, the probability that each event is a Type SN~Ia.

\begin{figure}[hb]
    \includegraphics[width=0.9\linewidth]{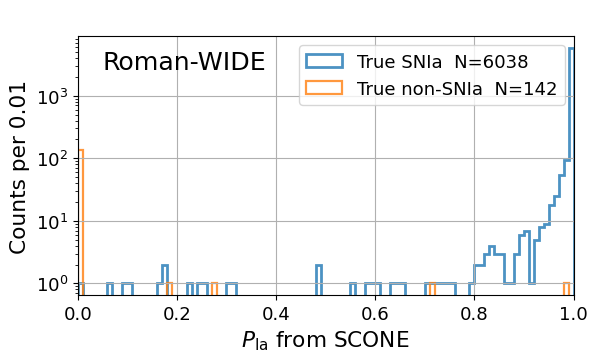}
    \includegraphics[width=0.9\linewidth]{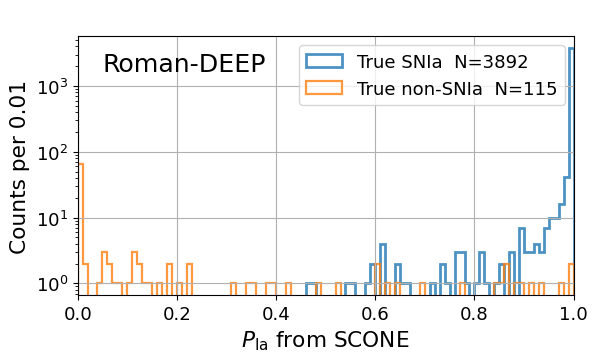}
     \caption{{\SCONE}-$\PIa$ distribution for true SNIa events and for true non-SNIa events;
     	WIDE (top) and DEEP (bottom) for 1 of the 9 data sets.
     }
    \label{fig:contam_scone_pia}
\end{figure}

\begin{figure}[hb]
    \includegraphics[width=0.47\linewidth]{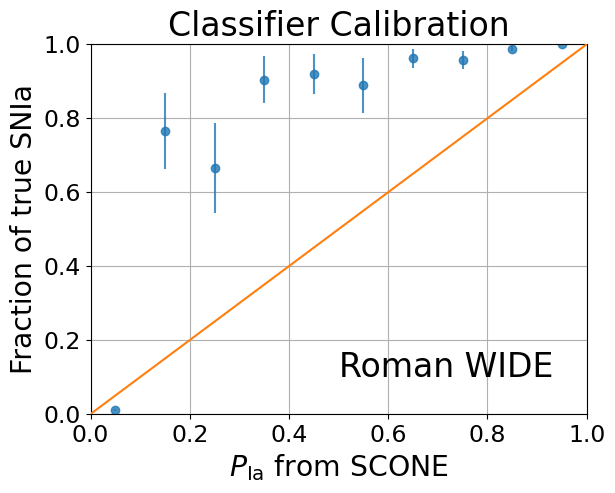}
    \includegraphics[width=0.47\linewidth]{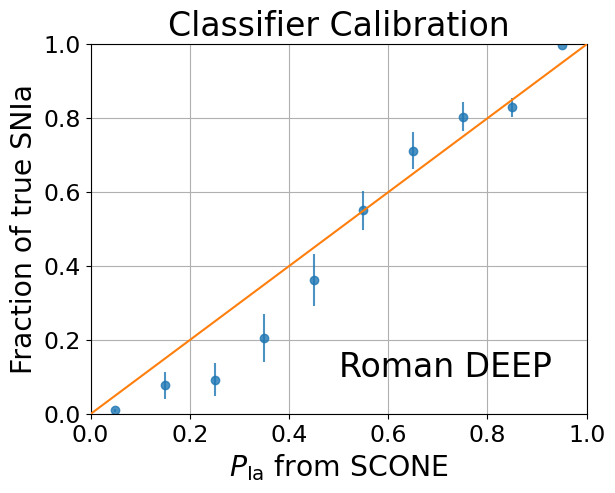}
     \caption{Fraction of true SN~Ia events vs. {\SCONE}-$\PIa$ for 
     WIDE (left) and DEEP (right) tiers;
     summed over all 9 data sets for better statistics.
     SN~Ia fractions for a well calibrated classifier should lie on the orange diagonal line.
     }
    \label{fig:calib_scone}
\end{figure}

\begin{figure}[hb]
    \includegraphics[width=0.9\linewidth]{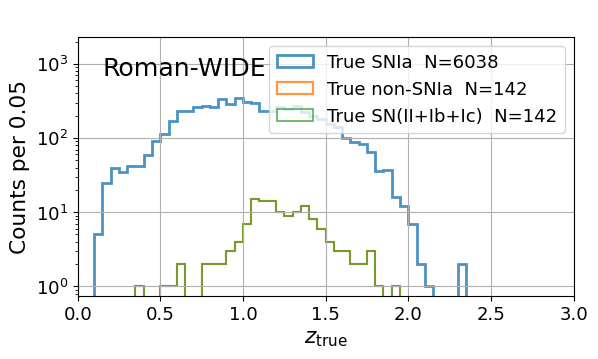}
    \includegraphics[width=0.9\linewidth]{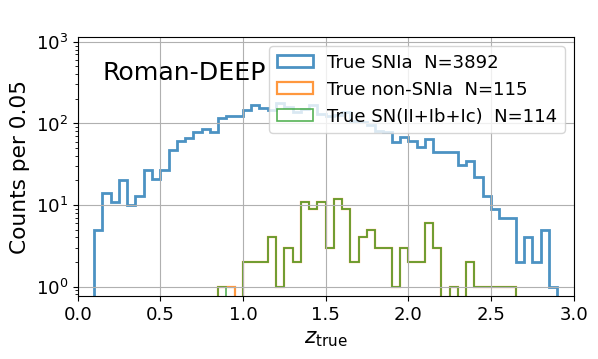}
     \caption{After analysis cuts, true redshift distribution for:
     true SNIa events (blue);
	 true non-SNIa (II+Ib+Ic+Iax+91bg) events (orange); 
	 true non-SNIa subset of SN(II+Ib+Ic) (green).
     WIDE (top) and DEEP (bottom).
     True non-SNIa events are dominated by SN(II+Ib+Ic) with only 1 peculiar SNIax;
     hence, the orange curve is not visible in WIDE, and barely visible for DEEP.}
    \label{fig:contam_ztrue}
\end{figure}

$\PIa$ distributions are shown in Fig.~\ref{fig:contam_scone_pia} for true
SNIa (blue) and for true non-SNIa (orange), and for WIDE and DEEP tiers;
visual inspection indicates that this classifier has excellent discrimination.
The \SCONE\ calibration is illustrated in Fig.~\ref{fig:calib_scone}, which shows
the fraction of true SN~Ia events vs. $\PIa$. A well calibrated classifier has
SN~Ia fractions on the diagonal; the WIDE tier shows a poor calibration while
the DEEP tier is fairly well calibrated.
True redshift distributions are shown in Fig.~\ref{fig:contam_ztrue},
and show that the contamination after selection requirements is
(i) dominated by core collapse events with minimal contribution from peculiar SNIa and
(ii) mostly in the redshift range $z>1$ where current rate measurements have large statistical uncertainties.

For LSST, we did not simulate non-SNIa contamination and thus there is no classification step;
each event is assigned $\PIa=1$. 

\bigskip
\subsection{BBC}
\label{subsec:snana_bbc}

\newcommand{\DCC}{$D_{CC}$}
\newcommand{\DCCsim}{$D_{CC}^{\rm sim}$}
\newcommand{\DCCzpoly}{$D_{CC}^{\rm zpoly}$}

The next analysis stage is ``BEAMS with Bias Corrections'' (BBC) to 
implement the BEAMS\footnote{BEAMS = Bayesian Estimation Applied to Multiple Species} 
formalism \citep{BEAMS2007,BEAMS2012}, 
and to apply distance bias corrections for the Hubble diagram (HD) as a multi-dimensional function of 
\{redshift, color, stretch\} 
\citep{BBC,Popovic2021}.
The bias correction component relies on a detailed SN~Ia simulation as described in
Sec.~\ref{sec:snana_sim}.
The BEAMS component relies on a non-SNIa prior for the mean distance residual and scatter;
see {\DCC}\footnote{Here we use \DCC\ notation to represent all non-SNIa contamination, and not
the core collapse (CC) subset suggested by the CC subscript.} 
in Eqs~6-10 in \citet{BBC}.

\citet{BEAMS2012} characterized this prior using $z$-dependent polynomials (\DCCzpoly) 
whose coefficients are additional fitted parameters. 
\citet{BBC} introduced a physically motivated prior based on simulating non-SNIa events (\DCCsim).
\DCCsim\ was used for the nominal \DESVYR\ analysis,
and \DCCzpoly\ was used as a crosscheck and resulted in good agreement 
(see Table~10 in \citet{Vincenzi2024}). 
Here we use the \DCCsim\ approach
with the caveat of assuming that Roman will eventually measure the high-$z$ rates 
with sufficient precision to make the corresponding systematic negligible.
The agreement between the two \DCC\  methods in \DESVYR\ suggests
that we could use the simpler \DCCzpoly\ approach. However, there are two reasons for
continuing to use the \DCCsim\ approach:
(1) agreement between the two \DCC\ methods adds confidence to an 
analysis relying on photometric classification, and 
(2) even if we abandon the simulation-based \DCCsim\  in BBC, 
we need the detailed non-SNIa sims anyway for training photometric classifiers.

\begin{figure}[hb]
    \vspace{-0.3cm}
    \includegraphics[width=0.9\linewidth]{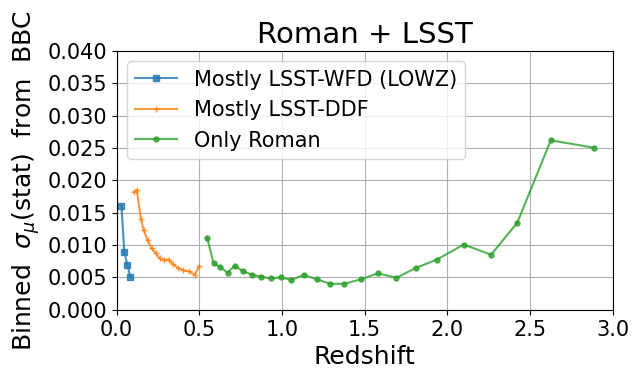}  \hfill
    \vspace{-.2cm}
     \caption{Binned distance uncertainty (stat only) from BBC.  
        The redshift-dependent structure is caused by a different
        sample dominating in each redshift region indicated by the legend.
     }
    \label{fig:muerr_vs_z}
\end{figure}

The BBC-fitted parameters are:
\begin{itemize}
  \itemsep0em  
  \item $\boldsymbol{\alpha}$: Stretch-luminosity parameter for standardization correction  $\alpha \times x_1$.
  \item $\boldsymbol{\beta}$: Color-luminosity parameter for standardization correction $\beta \times c$.
  \item $\boldsymbol{\sigint}$: Intrinsic scatter term.
  \item $\boldsymbol{z}${\bf-Binned HD:} Bias-corrected distance modulus ($\mubbc$) and uncertainty in 
           \nzbinBBC\ redshift bins with bin size proportional to $(1+z)^2$. 
           	Fig~\ref{fig:muerr_vs_z} shows the $z$-binned $\sigmu$ vs. redshift.
  \item {\bf Unbinned HD:} Bias-corrected $\mubbc$ and $\sigmu$ for each event as described in 
      \citet{Kessler2023_redemption}.
  \item {\bf Rebinned HD:} To reduce CPU resources and still benefit from unbinned self-calibration
           \citep{Binning_is_Sinning},
           a rebinned HD is created using \NREBINx\ stretch and \NREBINc\ color bins \citep{Kessler2023_redemption}.
          Combined with \nzbinBBC\ redshift bins, the rebinned HD size is expected to be 
           $\nzbinBBC{\times}\NREBINx{\times}\NREBINc{=}\HDsizeRebinDefine$. 
           After discarding empty bins, the average rebinned HD size is \HDsizeAVGrebin.
\end{itemize}

For the small $\zphot$ biases in Fig.~\ref{fig:zphot_resid},
the BBC formalism properly accounts for these biases by computing the true distance
modulus at the measured $\zphot$ instead of $\ztrue$.
An important caveat is that the host-$\zphot$ pdf must be well modeled for BBC to be reliable,
and while host-$\zphot$ modeling is always accurate for sims, the true test will be 
our ability to understand real data.

\begin{figure}[ht]
   \medskip
    \includegraphics[width=0.9\linewidth]{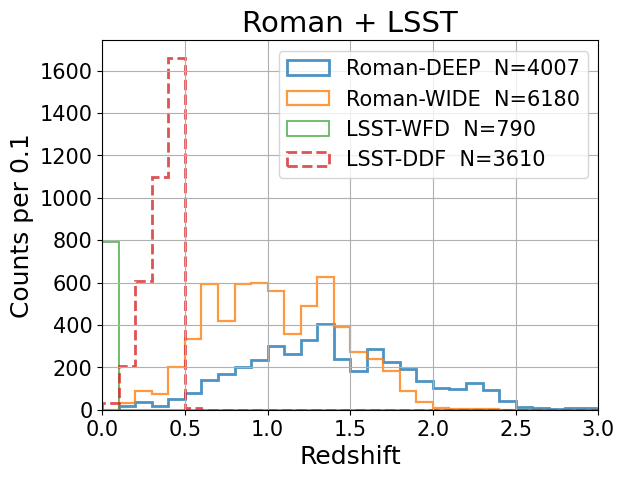}
    \vspace{-0.3cm}
     \caption{zHD distribution after BBC stage, for all four subsets as indicated in the legend.
         zHD is the Hubble diagram redshift: $\zspec$ for LSST and $\zphot$ for Roman.
     }
    \label{fig:zHD}
    \vspace{0.2cm}
\end{figure}

The BBC sample sizes are shown in Table~\ref{tb:bbc_stats}.
Statistics for 1 of the \Ndataset\ data sets is shown, 
totaling nearly 15k events among the four survey/tier subsets.
The corresponding redshift distributions are shown in Fig.~\ref{fig:zHD}.

\newcommand{\unitk}{{\times}10^{-3}}
\begin{table*}[ht]
\begin{center}
\caption{BBC Statistics\tablenotemark{a} } 
\begin{tabular}{ l | r r r | l | l  }  
\hline    
                       & \multicolumn{3}{c |}{sample size ($\unitk$) for:}        &           &         \\
                       &  data  &    SNIa for        & non-SNIa                   & biasCor   & all-syst \\ 
Survey(Tier)  & set-1  &  biasCor\tablenotemark{b}   & for prior\tablenotemark{c} & event loss & event loss\tablenotemark{d}  \\ 
%
\hline  
    LSST(WFD)  &  0.79 &    85.1 & --- & 1.8\%   & 3.1\%  \\ 
    LSST(DDF)  &  3.61 &   226.2 & --- & 0.5\%   & 0.8\%  \\ 
  ROMAN(WIDE)  &  6.18 &  2063.8 &  16.6 & 2.1\%   & 7.9\%  \\ 
  ROMAN(DEEP)  &  4.01 &  1339.7 &  11.4 & 1.6\%   & 9.7\%  \\ 
\hline  
          ALL  & 14.59 &  3714.8 &  28.0 & 1.6\%   & 6.5\%  \\ 
\hline   
\end{tabular}
\label{tb:bbc_stats}
\tablenotetext{a}{All numbers are after lightcurve fit and selection cuts.} \vspace{-1.6ex}
\tablenotetext{b}{Used to estimate distance bias corrections.} \vspace{-1.6ex}
\tablenotetext{c}{Used to estimate non-SNIa prior for BEAMS.} \vspace{-1.6ex} 
\tablenotetext{d}{Fails analysis cuts for one or more systematics, or fails valid biasCor requirement.}
\end{center}
\vspace{-0.3cm}
\end{table*}

The finite bias-correction size results in rejecting ${\sim}1$\%  of the events due to 
invalid bias correction (see ``biasCor event loss'' column in Table~\ref{tb:bbc_stats}),
which typically occurs for events in small probability regions in the space of ${z,x_1,c}$.
This loss is compounded by requiring a valid light curve fit and valid biasCor 
for all \Nsyst\ systematics,
which results in a $\BBClossAll$\% loss (see "all-syst event loss" in Table~\ref{tb:bbc_stats}).
In future analyses a larger biasCor sample can reduce the biasCor loss, but improved
photo-$z$ light curve fitting is needed to reduce the all-syst loss.

\newcommand{\contamBBC}{CONTAM$_{\rm BBC}$}
\newcommand{\contamSIM}{CONTAM$_{\rm SIM}$}
\newcommand{\contamBias}{0.14}   
\newcommand{\contamRMS}{0.10}   
\newcommand{\contamAVG}{2.5}    

\begin{figure}[h]
    \includegraphics[width=0.85\linewidth]{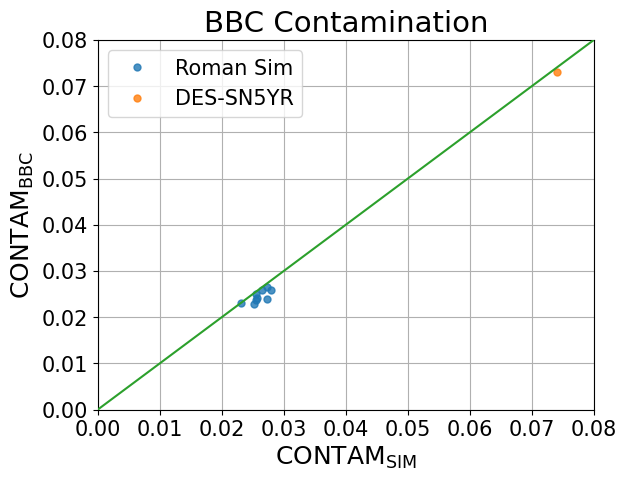} 
     \vspace{-0.3cm}
     \caption{BBC-measured contamination (\contamBBC) vs. true contamination (\contamSIM)
     	 for the \Ndataset\ simulated data sets (blue), and for the \DESVYR\ analysis (orange).
	   The green  line shows \contamBBC = \contamSIM.
     }
    \label{fig:bbc_contam}
\end{figure}

A critical crosscheck for BBC is to compare a contamination estimate from BBC (\contamBBC) with
the simulated contamination (\contamSIM). After the BBC fit, \contamBBC\ is computed from
the sum of BEAMS probabilities (Eq.~9 in \citet{Kessler2023_redemption}). 
For the \DESVYR\ analysis, \contamBBC$=0.073$ and \contamSIM$=0.074$ 
(Table 10 in \citet{Vincenzi2024}),\footnote{Here we apply a 12\% correction to account 
for a BBC reporting mistake that included 194 spec-confirmed low-$z$ events in the denominator for the 
\DESVYR\ contamination. This reporting mistake did not effect BBC fitted results.} 
adding confidence to their treatment of contamination. The corresponding comparison for our 
\Ndataset\ Roman data sets is shown in Fig.~\ref{fig:bbc_contam}. 
The average contamination is \contamAVG\%, and \contamBBC\ and \contamSIM\ 
agree on average to within \contamBias\% with an rms dispersion of \contamRMS\%.

\newcommand{\TestCutPIa}{0.01}  
\newcommand{\TestCutRejectIa}{0.01\%}  
\newcommand{\TestCutRejectCC}{79\%}    
\newcommand{\TestCutPurity}{99.5\%}    

While here we follow \DESVYR\ by not applying an explicit cut on $\PIa$ 
(Fig.~\ref{fig:contam_scone_pia}),
we note that a dramatic purity improvement is feasible. For example,
requiring $\PIa > \TestCutPIa$ results in a \TestCutRejectIa\ loss of true SNe~Ia,
while rejecting \TestCutRejectCC\ of the non-SNIa contamination.
The resulting SN~Ia purity of \TestCutPurity\ should be useful for related studies 
such as measuring rates and host-galaxy correlations.

\subsection{Stat+Syst Covariance Matrix}
\label{subsec:snana_covmat}

The following systematic uncertainties are included in this analysis:
\begin{itemize}
  \itemsep 0em
  \item {\bf Cal\_ROMAN\_ZP, Cal\_LSST\_ZP:} 
              0.005~mag zero point (ZP) error for Roman \& LSST;
  \item {\bf Cal\_ROMAN\_wave, Cal\_LSST\_wave:} 
         5~\AA\ error on mean filter-$\lambda$ for Roman \& LSST;
  \item {\bf Cal\_per\_A:} 0.0071 mag/$\mu$m in global calibration;
  \item {\bf Cal\_NONLIN:}  nonlinearity of 0.05\% over 4.5 dex (Roman only);

  \item {\bf ZPHOT\_HOST:} 0.01 coherent shift in host $\zphot$ (Roman only),
  which is comparable to the bias in Fig.~10 of \citet{Myles2021}\footnote{Their Fig.~10 shows photo-$z$ variance,
  so take square root for bias estimate.} and also comparable
  to the SN+host $\zphot$ bias in Fig.~\ref{fig:zphot_resid};
  \item {\bf ZERRSCALE:}  20\% increase in host-$\zphot$ uncertainty (Roman only);
  \item {\bf ZSHIFT:}  $4{\times}10^{-5}$ shift in $ \zspec$ (LSST only);
  \item {\bf MWEBV:}   5\% error in Galactic $E(B{-}V)$ for Roman \& LSST.
\end{itemize}
There are additional model systematics that are under active development within our SNPIT, 
but not yet included in this analysis:
\vspace{-1ex} 
\begin{itemize}
  \itemsep -0.5em
    \item  ZP calibration error in SALT3 training;
    \item  Model error in SALT3 training;
    \item  Intrinsic scatter model;
    \item  Rate-vs-redshift for non-SNIa contamination used for \DCCsim.
\end{itemize}
\vspace{-1ex} 
Based on the R25 method, these model systematics are expected to add a significant contribution.

With 12 ZP errors and 12 filter-wavelength errors, the total number of systematics is \Nsyst.
Following Eq.~6 in \citet{Conley2011}, the light curve fitting and BBC are run separately for each
systematic shift and each $\delta\mu$ contribution is evaluated for the systematics covariance, $\Csyst$. 
The statistical covariance ($\Cstat$) includes
Poisson noise, weak-lensing uncertainty of $0.028z$, $\sigint$ from BBC, and a contribution from 
peculiar velocity. The total covariance is 
\begin{equation}
   \Ctot = \Cstat + \Csyst
\end{equation}
To quantify sources of systematic uncertainty, 
many $\Ctot$ variants are constructed with subsets of the above systematics.
These $\Ctot$ variants are made for binned and rebinned HDs.
Because of the large CPU and memory resources needed for an unbinned HD, 
only the nominal unbinned $\Ctot$ is computed (no variants).

\subsubsection[Sensitivity to Host zphot Bias]{Sensitivity to Host $\zphot$ Bias}
\label{sss:host_zphot_bias}

\newcommand{\dmudz}{d\mu/dz}
\newcommand{\dmudzSYST}{d\mu/dz_{\rm syst}}
\newcommand{\dmudzLCDM}{d\mu/dz_{\Lambda\rm CDM}}

\begin{figure}[ht]
    \includegraphics[width=0.8\linewidth]{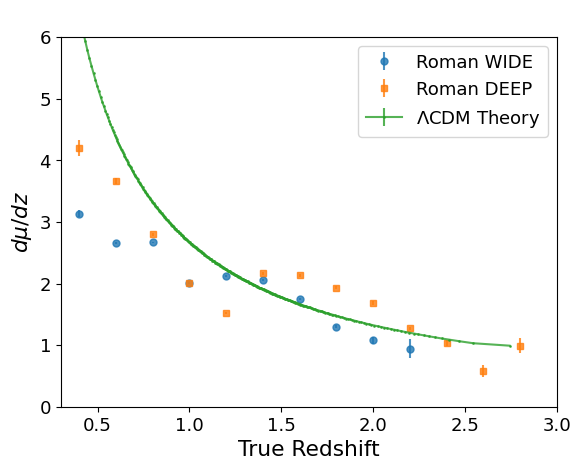} 
     \caption{$\dmudz$ vs. true redshift for each Roman tier,
        and a \LCDM\ theory curve in green.
     } 
    \label{fig:dmudz}
\end{figure}

In Sec.~5.3 of \citet{Mitra2023}, they show that the anticorrelation between the
SALT3-fitted $\zphot$ error and color error results in self-correcting the distance moduli,
and the resulting uncertainty is much smaller compared to adding a naive $\sigma_z \times d\mu/dz$ term.
Using the systematic with 0.01 shift in host $\zphot$,
the self-correction is illustrated by computing the systematic distance change with respect 
to the change in $\zphot$:
\begin{equation}
    \dmudzSYST\ \equiv \Delta\mu_{\rm syst} / \Delta{\zphot}_{\rm ,syst}
\end{equation}
and comparing to the theoretical \LCDM\ expectation, $\dmudzLCDM$.
An exact self-correction results if $\dmudzSYST = \dmudzLCDM$,
and their Fig.~9 shows moderate agreement for $0.5 < \ztrue < 1.2$.
For our Roman sims, Fig.~\ref{fig:dmudz} shows the same comparison over
a wider redshift range, and we find qualitative agreement with \citet{Mitra2023}.

\subsection{Cosmology Fitting}
\label{subsec:snana_wfit}

\newcommand{\sigwo}{\sigma_{w_0}}
\newcommand{\sigwa}{\sigma_{w_a}}
\newcommand{\chisqprior}{\chi^2_{\rm prior}}

We fit each data set for cosmology parameters \{$\Om,w_0,w_a$\} 
using  the \wwCDM\ model and minimizing
\begin{equation}  
    \chi^2  =  \Dmu^T \Ctot^{-1} \Dmu  +  \chisqprior~,  
\end{equation} 
where $\Dmu \equiv \mubbc - \mu_{\rm theory}(z,\Om,w_0,w_a)$,
and $\chisqprior$ incorporates a CMB
prior based on the $R$-shift parameter with uncertainty $\sigma_{R} = 0.0044$. 
To avoid bias from the CMB constraint, $R$ is computed from the same cosmology parameters
used to generate the simulation.
We use a relatively fast grid-search program in \SNANA\footnote{\URLwfit}
that evaluates the likelihood by looping over a 3D ($\Om,w_0,w_a$) grid,
and reporting marginalized values and uncertainties.
After the fit has completed, FoM is computed as
\begin{equation}
   {\rm FoM} = \left[ \sigwo \sigwa \sqrt{(1-\rho^2)} \right]^{-1}
   \label{eq:FoM}
\end{equation}
where $\sigwo$ and $\sigwa$ are the marginalized uncertainties on $w_0$ and $w_a$,
and $\rho$ is the reduced covariance between these two parameters.

\section{FoM Results}
\label{sec:snana_results}

\begin{figure}[ht]
    \vspace{-0.2cm}
    \includegraphics[width=1.03\linewidth]{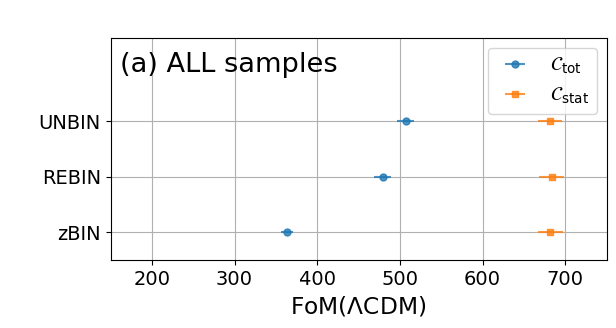}
    \includegraphics[width=1.03\linewidth]{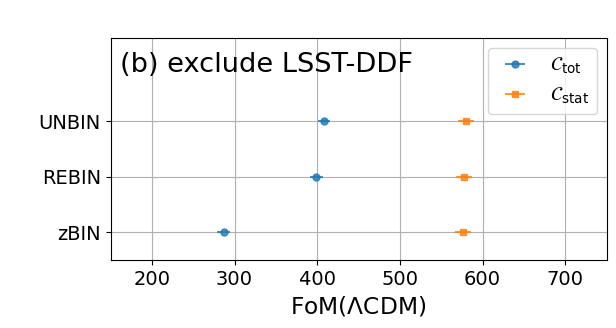}
    \includegraphics[width=1.03\linewidth]{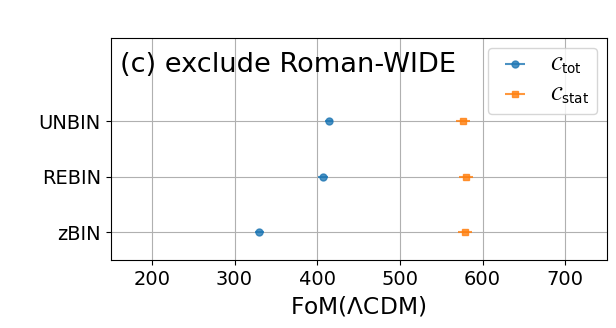}
    \includegraphics[width=1.03\linewidth]{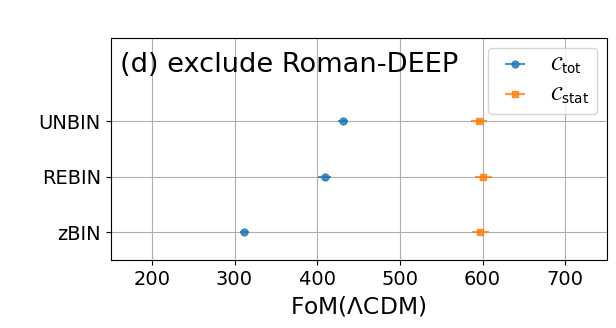}
     \caption{
        For analysis of simulations using \LCDM\ cosmological parameters ($\cosparLCDM$),
        FoM (horizontal axis) vs. Hubble diagram binning method (vertical axis); 
     	$\Ctot$ in blue circle and $\Cstat$ in orange square.
     	Error bar shows standard deviation among the \Ndataset\ data sets.
        Panel (a) includes all four tiers.
        The remaining panels exclude one subsample indicated on the plot.
     }
    \label{fig:fom_lcdm_by_sample}
\end{figure}

The FoM results are shown in Fig.~\ref{fig:fom_lcdm_by_sample} for all systematics ($\Ctot$)
and for stat-only ($\Cstat$). 
Panel (a) shows the nominal FoM with all four samples (Roman WIDE+DEEP \& LSST WFD+DDF),
and for the three different binning methods:
$z$-binned, rebinned, and unbinned with average HD sizes of 
$\HDsizeAVGbinned$, $\HDsizeAVGrebin$, and $\HDsizeAVGunbin$, respectively.
Using $\Cstat$, the FoMs are the same for all binning
methods as expected. Using $\Ctot$, 
the $z$-binned FoM$=\FOMALLSYSzbin$, and switching to rebinning
increases FoM to \FOMALLSYSrebin. 
The optimal unbinned results improve FoM to \FOMALLSYSunbin, 
but at a significant cost in computing time. 

\newcommand{\CovCPUratioExpect}{4,500}  
\newcommand{\CovCPUratioActual}{2,500}  
\newcommand{\wfitCPUratioActual}{30}   

The $\Ctot$ matrix inversion time scales as the cube of the HD size, 
resulting in an expected unbin/rebin CPU-ratio of \CovCPUratioExpect;
we observe a CPU ratio of \CovCPUratioActual, close to the expected ratio. 
The ratio of cosmology fitting times is about a factor of \wfitCPUratioActual. 
The rebin method was designed to benefit from self calibration using only a small fraction of the 
computing time compared to an unbinned HD,
and it does indeed deliver most of the benefit with a \FOMRATIOrebinTOzbin\% FoM improvement
over a redshift-binned HD.
However, there is an additional \FOMRATIOunbinTOrebin\% FoM improvement using the optimal unbinned HD.
Further work is needed to either improve the rebin method or to prepare for adequate computing 
resources to process the optimal unbinned HD for the many systematic variants,
and with a potentially much larger LSST sample.

Next we exclude LSST-DDF (panel (b) in Fig.~\ref{fig:fom_lcdm_by_sample}), 
combining Roman WIDE, Roman DEEP and LSST-WFD;
this results in a ${\sim}20$\% drop in FoM.
Excluding Roman-WIDE and restoring LSST-DDF (panel (c)) results in a 
slightly smaller FoM drop (compared with all samples in panel (a)). 
Finally, excluding Roman-DEEP results in similar FoM compared to excluding Roman-WIDE.

\begin{figure}[h]
    \includegraphics[width=0.98\linewidth]{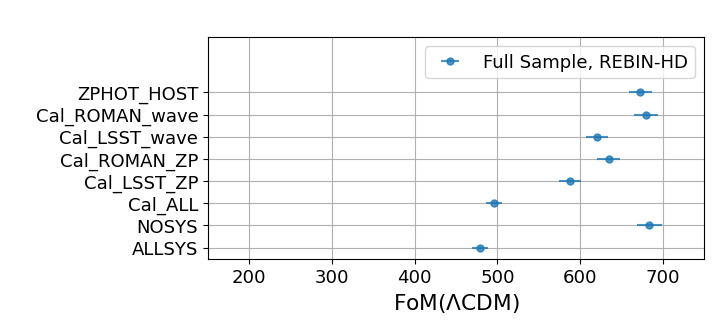} 
     \caption{FoM (horizontal axis) vs. systematics (vertical axis).
     	The error bar shows the standard deviation among the \Ndataset\ data sets. 
     }
    \label{fig:fom_vs_syst}
\end{figure}

Fig.~\ref{fig:fom_vs_syst} shows FoM for many combinations of systematics using the full sample.
Here we use the rebin method to limit computing resources.
Calibration accounts for most of the FoM degradation.
The 5~millimag zeropoint uncertainties result in more degradation than the
5~\AA\ filter-wavelength uncertainties, 
and LSST calibration uncertainties degrade FoM more than the Roman uncertainties.

\begin{figure}[h]
    \includegraphics[width=0.98\linewidth]{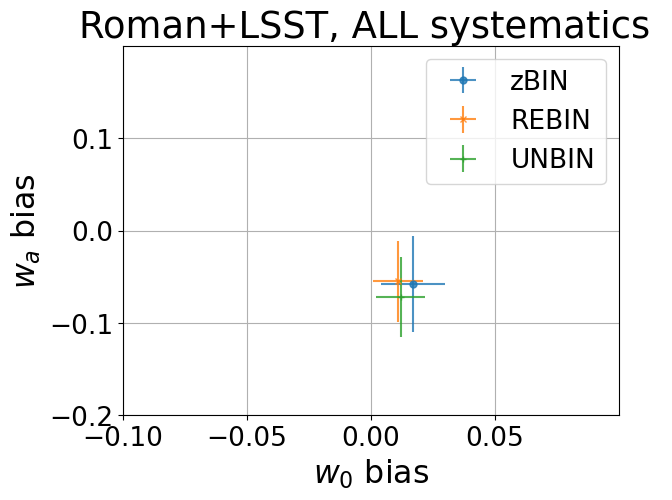} 
     \caption{Average bias on $w_a$ vs. average bias on $w_0$ for each Hubble diagram binning method. 
     The error bars show the uncertainty on the mean bias. 
     Ideally, the bias values should be consistent with zero.
     }
    \label{fig:w0wa_bias}
\end{figure}

Fig.~\ref{fig:w0wa_bias} shows the full-sample bias on $w_0$ and $w_a$ from the cosmology fitting stage, 
and averaged over the \Ndataset\ samples.
There is a small but significant bias for all binning methods,
indicating that further development is needed for the analysis.
The poor classifier calibration in the WIDE tier (Fig.~\ref{fig:calib_scone})
may contribute to this bias.

\subsection{FoM Results with non-standard Cosmology}

While our main results are based on simulations with the 
standard assumption of a flat \LCDM\ model  ($\cosparLCDM$),
recent results from 
\citet{DESCollaboration2024, DESI_2024,DESI_2025}
suggest a non-standard model with evolving dark energy. 
Here we briefly explore the FoM sensitivity by repeating the simulation 
and analysis with cosmology parameters
that are similar to recent results: $\cosparDVYR$.

\begin{figure}[ht]
    \vspace{-0.2cm}
    \includegraphics[width=1.03\linewidth]{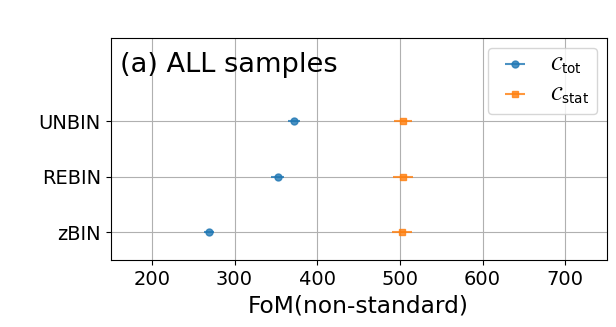}
    \includegraphics[width=1.03\linewidth]{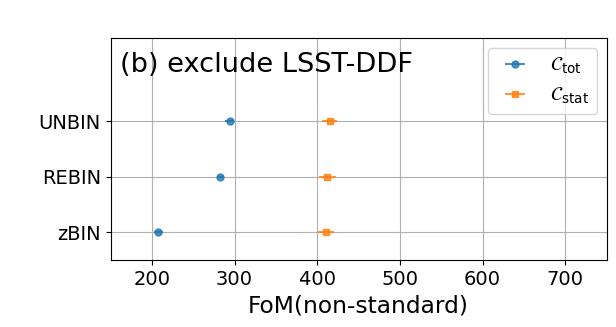}
    \includegraphics[width=1.03\linewidth]{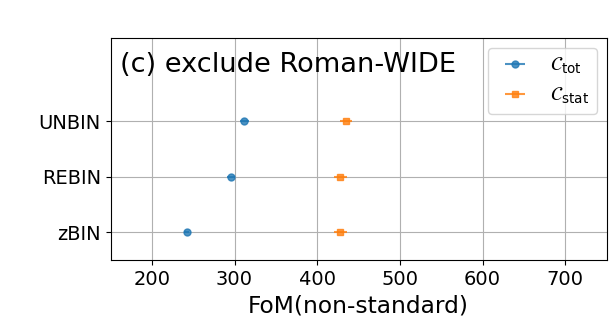}
    \includegraphics[width=1.03\linewidth]{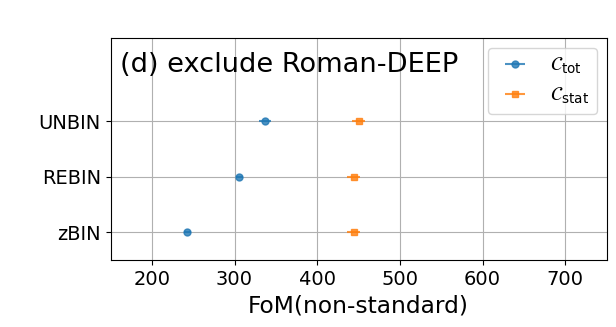}
     \caption{
        For simulations using non-standard cosmological parameters ($\cosparDVYR$),
         FoM (horizontal axis) vs. Hubble diagram binning method (vertical axis); 
     	$\Ctot$ in blue circle and $\Cstat$ in orange square.
     	The error bar shows the standard deviation among the \Ndataset\ data sets.
        Panel (a) includes all four tiers.
        The remaining panels exclude one subsample indicated on the plot.
     } 
    \label{fig:fom_d5yr_by_sample}
\end{figure}

\begin{figure}[ht]
    \vspace{-0.2cm}
    \includegraphics[width=1.03\linewidth]{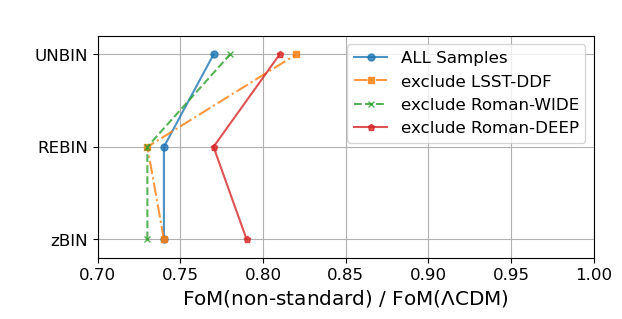}
     \caption{
        For analysis using $\Ctot$,
        horizontal axis shows FoM ratio between 
        using non-standard cosmological parameters ($\cosparDVYR$) and
        using \LCDM\ parameters ($\cosparLCDM$);
        vertical axis shows Hubble diagram binning method. 
        Each set of points connected by lines corresponds to a sample indicated
        in the legend.
     }
    \label{fig:fom_ratios_d5yr}
\end{figure}

The non-standard FoMs are shown in Fig.~\ref{fig:fom_d5yr_by_sample},
and FoM ratios between simulating non-standard and standard (\LCDM) cosmologies 
are shown in Fig.~\ref{fig:fom_ratios_d5yr}.
For the nominal sample, the non-standard FoMs are nearly 25\% smaller 
(blue points in Fig.~\ref{fig:fom_ratios_d5yr})
compared to the standard FoMs in Fig.~\ref{fig:fom_lcdm_by_sample}.
Excluding Roman-WIDE (green points in Fig.~\ref{fig:fom_ratios_d5yr}) results in 
roughly 5\% more FoM degradation compared to excluding Roman-DEEP (red points).
This comparison is not surprising because the non-standard cosmology
has much less dark energy at early times (high redshift) and thus
the DEEP tier is not quite as sensitive as the WIDE tier. 

There is a binning method artifact in which the $z$-binned and rebinning methods
result in slightly more FoM degradation for the non-standard cosmology 
(by a few percent) compared to the unbinned method.
This artifact is not understood.

\begin{figure}[h]
    \includegraphics[width=0.98\linewidth]{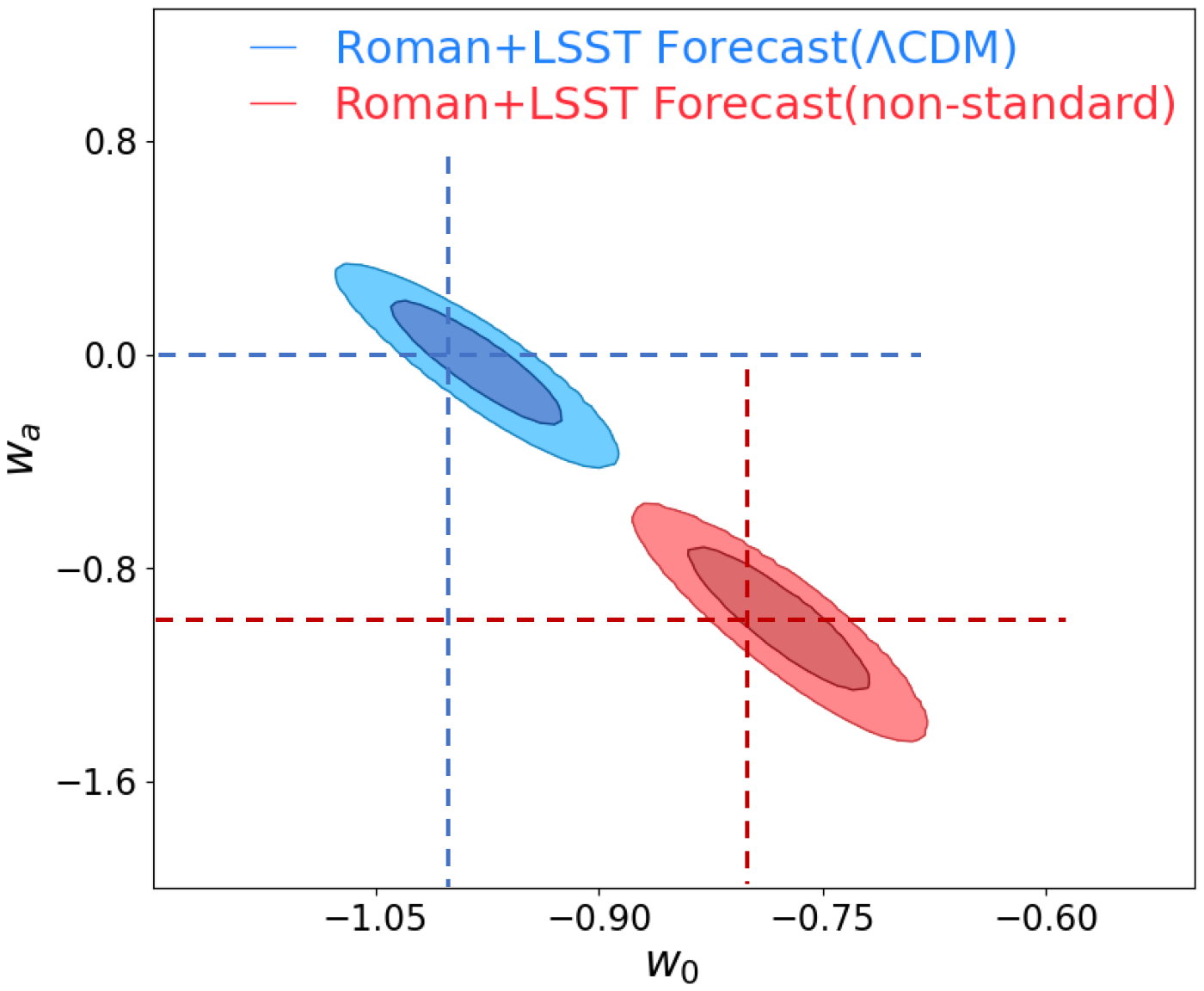} 
     \caption{$w_0$-$w_a$ contour forecasts (68\% and 95\% confidence) from analyzing
     simulated data generated with the 
     standard \LCDM\ cosmology in blue ($\cosparLCDM$) and the
     non-standard cosmology in red ($\cosparDVYR$).
     To roughly compensate for missing systematics, we show results from 
     the $z$-binned Hubble diagram that degrades FoM by nearly 40\% compared to
     an unbinned Hubble diagram.
     Among the $\Ndataset$ simulated samples, we show a result for which
     the best-fit cosmology parameters are close to the true values
     indicated by the dashed lines.
     }
    \label{fig:w0wa-contours}
\end{figure}

Fig.~\ref{fig:w0wa-contours} shows $w_0$-$w_a$ contour forecasts for the
\LCDM\ (blue) and non-standard cosmologies (red).
To roughly compensate for missing systematics (Sec.~\ref{subsec:snana_covmat}),
we show the $z$-binned Hubble diagram results for which the FoM is degraded by nearly 40\%
compared to the optimal unbinned method.

\subsection{FoM Results with Pessimistic Volumetric Rate}
\label{subsec:low_zrate}

\begin{figure}[h]
    \includegraphics[width=0.98\linewidth]{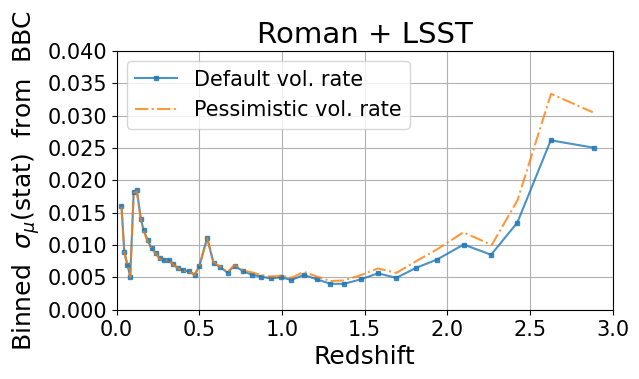} 
    \includegraphics[width=0.98\linewidth]{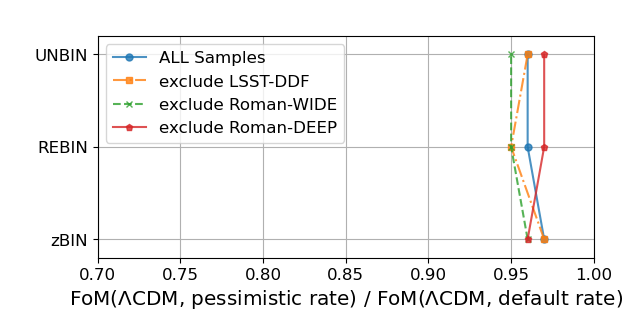} 
     \caption{Top panel: BBC fitted distance uncertainty in redshift bins
     for default SN~Ia volumetric rate (blue) and for pessimistic rate (orange);
     see Fig.~\ref{fig:zrate} for volumetric rates vs. redshift.
     Bottom panel: FoM ratio (all systematics) between pessimistic and default rate,
     for all sample combinations (legend) and for the 3 Hubble diagram binning methods (vertical axis).
     }  
    \label{fig:pessimistic_zrate}
\end{figure}

The main results presented above are based on the default SN~Ia volumetric rate 
vs. redshift shown by the blue curve in Fig.~\ref{fig:zrate}. 
Here we show sensitivity changes based on a lower (pessimistic) SN~Ia rate estimate 
shown by the orange curve in Fig.~\ref{fig:zrate}. 
The pessimistic rate is the same as the default rate for $z<0.5$,
15\% lower at $z=1$, 30\% lower at $z=2$, and 35\% lower at $z=2.5$.
The resulting change in BBC-fitted distance precision is shown in 
the top panel of Fig.~\ref{fig:pessimistic_zrate}.
The corresponding FoM ratio between the two rate models
is shown in the bottom panel of Fig.~\ref{fig:zrate}; 
the pessimistic FoM is only a few percent smaller compared
to using the nominal rate, and this result holds for all 
sample combinations and all binning methods.

\section{Extended and Pilot Visits}
\label{sec:noncore_visits}

For Roman, roughly 22 of the 180 observing days have been allocated to time outside
the core component. Here we present a preliminary evaluation of the
DEEP-tier extended visits (EV), and pilot visits (PV) in both tiers.
The simulation for the main analysis (Sec.~\ref{sec:snana_sim})
does not include these extra visits, and the core component time
has been reduced to account for EV and PV.

\subsection{Extended Visits}
\label{subsec:extended_visits}

The analysis cuts in Sec.~\ref{subsec:snana_lcfit} include light curve sampling cuts
that require $t_0$ to be within the 2-year core component (Fig.~\ref{fig:ccs_survey}),
and also require observations before and after $t_0$. 
Without the EV component, $t_0$ must be well within the 2-year core component
time window in order to satisfy the early and late observation cuts.
Since \citet{CCS_report} did not specify EV times, here we implement the following
8 additional visits as a test:
$-180$, $-80$, $-20$ days with respect to the start of the core component, and
20, 60, 140, 240, 420 days with respect to the end of the core component.
         
\begin{figure}[hb]
    \includegraphics[width=0.47\linewidth]{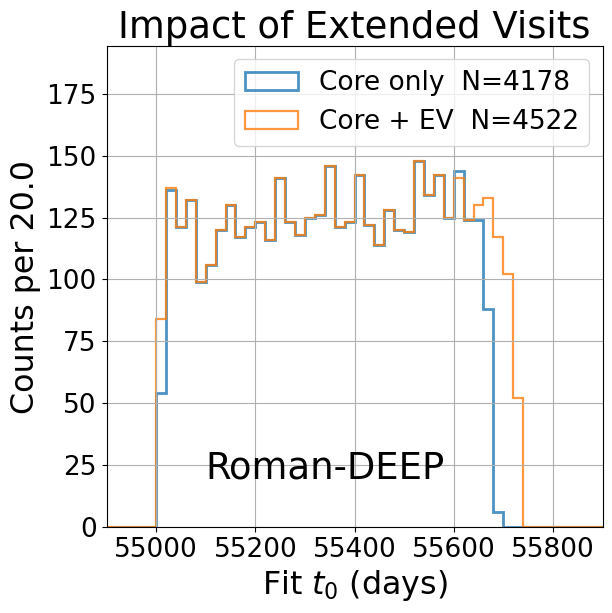}
    \includegraphics[width=0.49\linewidth]{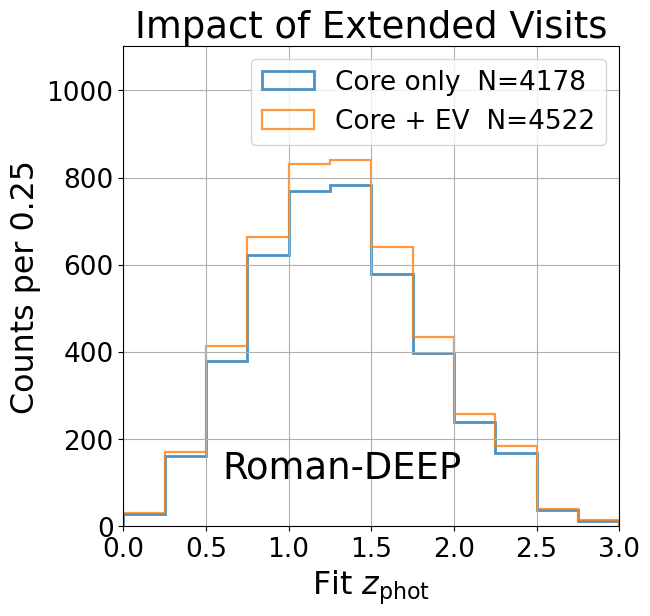}
     \caption{Distribution of fitted SALT3 parameters 
     without extended visits (blue) and with extended visits (orange);
     for $t_0$ (left) and $\zphot$(right).
     The core component MJD window is 55000 to 55725.
     }
    \label{fig:core+EV}
\end{figure}

The impact of adding the DEEP-tier EV component is shown in
Fig.~\ref{fig:core+EV}. There is an \EVincrease\% increase in DEEP-tier events after 
light curve fitting and cuts, and as expected, these extra events have a 
fitted time of peak brightness ($t_0$) near the end of the 
core component at MJD=55725.
The redshift region with the largest increase is $1 < z < 1.75$.
There is a negligible increase at the highest redshifts ($z>2.5$) 
because the EV only adds 1 or 2 late epochs, which is enough to satisfy the
$\Trest > 20$ day cut, but is not enough to satisfy the \SNRSUM\ cut.
While this \EVincrease\% increase appears somewhat useful, these extra events have a much 
lower quality cadence in the EV time range, and thus they may have
limited impact if detailed late-time features are needed to 
improve standardization of the brightness.
For example, \citet{SALT3+x2} used ZTF\footnote{\URLZTF} data \citep{ZTF_DR2_OVERVIEW}
to show that including a second stretch component ($x_2$) in the SALT3 light curve fit results in 
improved standardization; it is not clear if this improvement can be realized for the EV-recovered Roman events.

\subsection{Pilot Visits}
\label{subsec:pilot_visits}

Our analysis implicitly assumes that deep co-added imaging 
is used to construct templates for difference imaging.
To acquire these templates, \citet{CCS_report} allocates \NPVtot\ first-year PVs
in both the WIDE and DEEP tiers (Fig.~\ref{fig:ccs_survey}).
In addition to providing templates, the PV component is intended to provide an
early sample of quality high-redshift SNe that can potentially inform
the core component strategy that starts a year later. 
Useful PV information would include a better $z>1$ rate estimate
for both SN~Ia and SNCC, and limits on evolution of the light curve model.
Carrying out these preliminary analyses and updating the core component
strategy on a 1 year time scale will be very challenging.

Here we present a preliminary evaluation of a PV sample using \NPVlcfit\ of the \NPVtot\ PVs for light curves,
and using the first \NPVtemplate\ PVs for templates.
To account for the limited template depth, the zodiacal and thermal noise are 
increased by a factor of  $\sqrt{1+1/2} = 1.22$.
We use the same WIDE and DEEP tiers as in the core component, 
and use the same filters and $\Texpose$ values in Table~\ref{tb:texpose}.
Since \citet{CCS_report} does not suggest a specific cadence, we explore
\NPVcadence\ cadences: 10, 15, 20, and 30 days between visits.
We apply the cuts in Sec.~\ref{subsec:snana_lcfit}, and relax the $\Trest$ cut to be
$>10$~days (instead of 20 days) to include more events in the limited PV window.
We did not rerun the galaxy photo-$z$ fits using galaxy mags with larger
uncertainties, and therefore the host-galaxy photo-$z$ prior is overly optimistic.

\begin{figure}[hb]
    \includegraphics[width=0.8\linewidth]{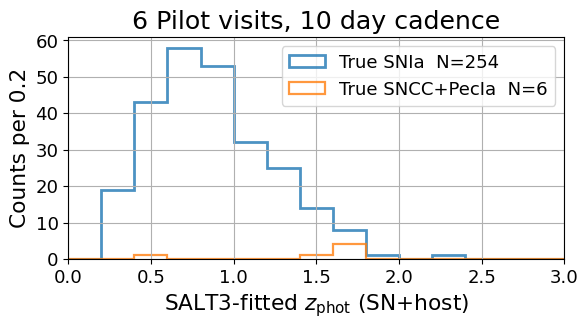}
    \includegraphics[width=0.8\linewidth]{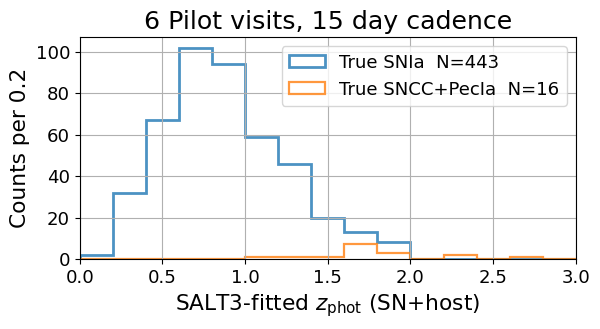}
    \includegraphics[width=0.8\linewidth]{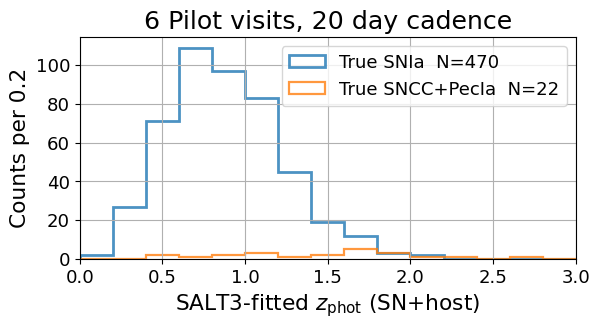}
    \includegraphics[width=0.8\linewidth]{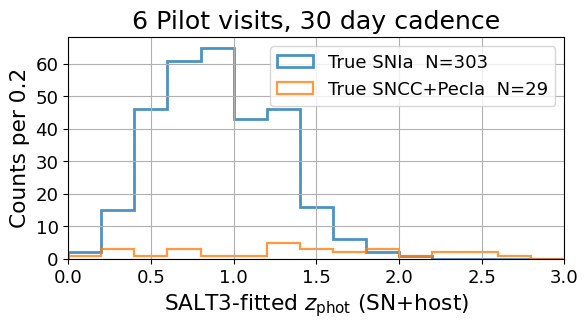}
     \caption{$\zphot$ distribution for true SNe~Ia (blue) and true non-SNIa contamination (orange).
           The 6 visit cadence is shown above each panel.
     }
    \label{fig:PV_6visits}
\end{figure}

After light curve fitting, Fig.~\ref{fig:PV_6visits} shows the resulting $\zphot$ distributions 
for true SNe~Ia and true contamination, and for the \NPVcadence\ cadences.
The 20 day cadence is optimal for the number of SN~Ia events,
resulting in \NIaPV\ true SNe~Ia and \NnonIaPV\ contaminants.
The {\contamPV}\% contamination is somewhat larger compared to the
the core component (see Fig.~\ref{fig:bbc_contam}), but an explicit
{\SCONE}-$\PIa$ cut may reduce this contamination.
About 1/3 of these events are at high-$z$ ($\zphot >1$), which is a significantly larger high-$z$ sample
than what is currently available.

\section{Conclusions} 
\label{sec:Conclusion}

This work is part of a Roman Supernova Project Infrastructure Team effort to 
provide input for the {\it in-guide} strategy reported in \citet{CCS_report}, 
and summarized here in Table~\ref{tb:ccs_strategy}.
We have generated simulations for four subsamples (Roman WIDE \& DEEP, LSST WFD \& DDF)
using a flat \LCDM\ model with $\cosparLCDM$. 
The analysis is based on that used in \DESVYR, 
and we have expanded their analysis to replace \spec\ redshifts with photometric redshifts;
this is therefore the first rigorous analysis with systematics that uses 
both photometric classification and photometric redshifts.

After applying our analysis and selection requirements,
the resulting Hubble diagram contains \HDsizeAVGunbin\ events (Table~\ref{tb:bbc_stats})
and FoM$=\FOMALLSYSunbin$ (Fig.\ref{fig:fom_lcdm_by_sample}a)
using the optimal unbinned Hubble diagram.
Excluding LSST-DDF results in FoM$=\FOMALLSYSnoDDFunbin$ (Fig.\ref{fig:fom_lcdm_by_sample}b).
Excluding WIDE or DEEP results in a comparable FoM degradation to ${\sim}420$,
and interestingly this degradation is slightly smaller compared to excluding LSST-DDF.
While these FoM values are well above the mission requirement of \FOMALLSYSrequire, 
several important systematics have been ignored (Sec.~\ref{subsec:snana_covmat}),
primarily those related to the SALT3 lightcurve training.

Motivated by evolving dark energy evidence from 
\DESVYR\ \citep{DESCollaboration2024} and
DESI \citep{DESI_2024,DESI_2025},
we analyzed simulations with a non-standard cosmology ($\cosparDVYR$);
the resulting FoMs are ${\sim}25$\% smaller compared to using the \LCDM\ model.

The galaxy $\zphot$ are notably degraded for redshifts $z{<}0.6$ (Fig.~\ref{fig:zphot_host})
because the bluest Roman band ($R$) is not blue enough to cover the 4,000~\AA\ break at lower redshifts.
Increasing S/N for the $R$ and $Z$ bands is unlikely to help; 
instead, the galaxy $\zphot$ should be determined by combining 
Roman data with ground-based optical photometry, such as from Pan-STARRS, DES, and LSST.

The simulated data sets include non-SNIa contamination from SNII/Ib/Ic, and from peculiar 91bg and Iax.
In the analysis we use the BEAMS formalism, as implemented by BBC, to account for contamination.
After all selection requirements, our prediction for contamination is $\contamAVG$\% 
(Figures~\ref{fig:contam_scone_pia} and \ref{fig:bbc_contam}),
which is nearly a factor of 3 smaller compared to the recent \DESVYR\ analysis.
While this small contamination is encouraging, most of the contamination occurs
for true redshifts $z{>}1$ where current rate measurements have significant statistical uncertainties.
Therefore SNCC and peculiar-SNIa rate-vs-redshift measurements 
need further improvement at high redshifts in order to improve the
crosscheck between measured and predicted contamination (Fig.~\ref{fig:bbc_contam}).

To significantly reduce CPU time for inverting $\Ctot$ (Sec.~\ref{subsec:snana_covmat}) 
and for cosmology fitting (Sec.~\ref{subsec:snana_wfit}),
we have shown that the HD ``rebinning''  method works fairly well, 
resulting in a FoM that is only \FOMRATIOunbinTOrebin\% smaller compared to using an unbinned HD
that requires $10^2$-$10^4$ more CPU.
Combining a more realistic LSST sample size with Roman, 
processing numerous unbinned HDs may become computational unfeasible. 
To obtain optimal dark energy constraints, effort is need to either 
(i) improve the rebinning method,
(ii) identify adequate computing resources to process the largest expected HD within a specified wall-time, 
and/or 
(iii) develop new systematics methods such as forward modeling.

\section{Acknowledgements}
Support for this work was provided by NASA under contract 80NSSC24M0023 
through the Roman Supernova Project Infrastructure Team.
We acknowledge the University of Chicago's Research Computing Center for their support of this work.
RH was supported by NASA under award number 80GSFC24M0006.

\medskip
{\bf Author Contributions} \\
RK: led simulation, analysis, writing.
RH: coordination of simulation inputs, editing.
BJ: computed galaxy photo-$z$ using EAZY code.
DR: Fisher matrix analysis input to select in-guide strategy.
MS: co-chaired \HLTDS\ committee that determined in-guide strategy.
RC: editing.
VM: aided in discussion of phantom DE affecting FoM.
BR: editing and validation.

\clearpage
\appendix
\counterwithin{figure}{section}

\newcommand{\epersecperpix}{{\eminus}/sec/pix}
\newcommand{\eperpix}{{\eminus}/pix}

\newcommand{\sigread}{\sigma_{\rm read}}
\newcommand{\ZPimg}{{\rm ZP}_{\rm img}}
\newcommand{\ZPe}{{\rm ZP}_{e}}

\newcommand{\NAB}{N_{\rm AB}}
\newcommand{\Tlam}{T_{\lambda}}

\newcommand{\COVread}{C_{\rm read}}
\newcommand{\COVthermal}{C_{\rm thermal}}
\newcommand{\COVzodi}{C_{\rm zodi}}

\newcommand{\Nread}{N_{\rm read}}
\newcommand{\Nthermal}{N_{\rm thermal}}
\newcommand{\Nzodi}{N_{\rm zodi}}

\newcommand{\URLZODI}{\url{https://www.osti.gov/biblio/842543}}  
\newcommand{\URLFILTERTRANS}{\url{https://github.com/GalSim-developers/GalSim/blob/releases/2.5/share/roman/Roman_effarea_20210614.txt}}
\newcommand{\URLWEBBPSF}{\url{https://webbpsf.readthedocs.io/en/latest}}

\section{Instrumental Parameters}
\label{app:instr_par}

The band-dependent instrumental parameters are shown in Table~\ref{tb:instr_par},
and all gains are assumed to be 1{\eminus}/ADU.
The thermal and zodiacal covariances ($\COVthermal$ and $\COVzodi$) are equal to the
number of electrons per second per pixel.
The read noise covariance (\eperpix) is approximately given by $\COVread = 25 + 3072/R_t$,
where $R_t = \Texpose/\Tread$, 
$\Texpose$ is the exposure time, and
$\Tread = 3.04$~sec is the readout time.
The \SNANA\ simulation does not read the Roman instrument parameters (except for NEA), 
but instead it reads a more general set of parameters that can be
computed for arbitrary surveys:
\begin{eqnarray}
  \sigsky^2  & = & (\COVthermal + \COVzodi ) \times \Texpose \\
  \sigread^2 & = & \COVread \\
  \ZPimg    & = & {\ZPe} + 2.5\log_{10}(\Texpose) 
\end{eqnarray}

Here is a numerical example for DEEP $F$-band using $\Texpose=1333$ from Table~\ref{tb:texpose}:
\begin{itemize}
    \itemsep0em 
   \item $\COVthermal = 0.155 \times 1333 = 206.615$, $\COVzodi = 0.194 \times 1333 = 258.602$; 
           $\sigsky = \sqrt{465.217} = 21.57$~.
   \item  $\sigread = \sqrt{25 + 3072/(1333/3.04)} = 5.65$~.
   \item  $\ZPimg =  25.913 + 2.5\log_{10}(1333) = 33.725$~.
\end{itemize}

\begin{table}[ht!]
\begin{center}
\caption{Roman Instrument Parameters used in the \SNANA\ Simulation}
\begin{tabular}{ | c c c c c | c c c | }
\hline
     &      &     &     &   &                                \multicolumn{3}{c|}{computed for {\SNANA}: } \\
     & $\COVthermal$\tablenotemark{a} & $\COVzodi$\tablenotemark{b}      
     & $\ZPe$\tablenotemark{c}  & NEA\tablenotemark{d}      & $\sigsky$    & $\sigread$  &  $\ZPimg$  \\
band & {\epersecperpix} & {\epersecperpix}  & {\eminus}/sec  & (pixels) & (WIDE/DEEP)  & (WIDE/DEEP) &  (WIDE/DEEP) \\ 
\hline
$R$  & 0.003      & 0.315       &  26.619     & 6.74   & ~4.78 / ------  & 12.01 / ------ & 31.262 / ------  \\
$Z$  & 0.003      & 0.316       &  26.303     & 8.50   & ~5.18 / ~8.58  & 11.33 / 8.02 & 31.114 / 32.212  \\
$Y$  & 0.003      & 0.357       &  26.356     & 9.55   & ~5.85 / ~9.79  & 10.83 / 7.70 & 31.300 / 32.418  \\
$J$  & 0.003      & 0.359       &  26.354     & 9.79   & ~7.61 / ~9.92  & ~9.01 / 7.65 & 31.864 / 32.440  \\
$H$  & 0.048      & 0.339       &  26.377     & 11.2   & 10.86 / 11.90  & ~7.42 / 7.08 & 32.588 / 32.786  \\
$F$  & 0.155      & 0.194       &  25.913     & 16.3   & ------/ 21.57  & ------ / 5.65 & ------ / 33.725  \\
\hline 
\end{tabular}
\setlength{\tabcolsep}{0.5pt} 
\tablenotetext{a}{Based on temperature $T=264$~K} 
    \vspace{-1.4ex}
\tablenotetext{b}{Based on \citet{Aldering2002}  }
    \vspace{-1.4ex}
\tablenotetext{c}{Zeropoint $\ZPe = 2.5\log_{10} [\int {\NAB} \lambda\Tlam d\lambda]$~; 
    $\NAB$ is AB spectrum;
    $\Tlam$ is filter transmission.\footnote{\URLFILTERTRANS} }  
        \vspace{-1.4ex}
\tablenotetext{d}{See Eq.~\ref{eq:NEA} for $RZYJ$; use WEBBPSF\footnote{\URLWEBBPSF} for $HF$.}
\label{tb:instr_par} 
\end{center} \vspace{-0.3cm} 
\end{table}


\clearpage
\bibliography{main}

\begin{thebibliography}{}
\expandafter\ifx\csname natexlab\endcsname\relax\def\natexlab#1{#1}\fi
\providecommand{\url}[1]{\href{#1}{#1}}
\providecommand{\dodoi}[1]{doi:~\href{http://doi.org/#1}{\nolinkurl{#1}}}
\providecommand{\doeprint}[1]{\href{http://ascl.net/#1}{\nolinkurl{http://ascl.net/#1}}}
\providecommand{\doarXiv}[1]{\href{https://arxiv.org/abs/#1}{\nolinkurl{https://arxiv.org/abs/#1}}}

\bibitem[{{Adame} {et~al.}(2025){Adame}, {Aguilar}, {Ahlen}, {Alam},
  {Alexander}, {Alvarez}, {Alves}, {Anand}, {Andrade}, {Armengaud}, {Avila},
  {Aviles}, {Awan}, {Bahr-Kalus}, {Bailey}, {Baltay}, {Bault}, {Behera},
  {BenZvi}, {Bera}, {Beutler}, {Bianchi}, {Blake}, {Blum}, {Brieden},
  {Brodzeller}, {Brooks}, {Buckley-Geer}, {Burtin}, {Calderon}, {Canning},
  {Carnero Rosell}, {Cereskaite}, {Cervantes-Cota}, {Chabanier}, {Chaussidon},
  {Chaves-Montero}, {Chen}, {Chen}, {Claybaugh}, {Cole}, {Cuceu}, {Davis},
  {Dawson}, {de la Macorra}, {de Mattia}, {Deiosso}, {Dey}, {Dey}, {Ding},
  {Doel}, {Edelstein}, {Eftekharzadeh}, {Eisenstein}, {Elliott}, {Fagrelius},
  {Fanning}, {Ferraro}, {Ereza}, {Findlay}, {Flaugher}, {Font-Ribera},
  {Forero-S{\'a}nchez}, {Forero-Romero}, {Frenk}, {Garcia-Quintero},
  {Gazta{\~n}aga}, {Gil-Mar{\'\i}n}, {Gontcho a Gontcho}, {Gonzalez-Morales},
  {Gonzalez-Perez}, {Gordon}, {Green}, {Gruen}, {Gsponer}, {Gutierrez}, {Guy},
  {Hadzhiyska}, {Hahn}, {Hanif}, {Herrera-Alcantar}, {Honscheid}, {Howlett},
  {Huterer}, {Ir{\v{s}}i{\v{c}}}, {Ishak}, {Juneau}, {Kara{\c{c}}ayl{\i}},
  {Kehoe}, {Kent}, {Kirkby}, {Kremin}, {Krolewski}, {Lai}, {Lan}, {Landriau},
  {Lang}, {Lasker}, {Le Goff}, {Le Guillou}, {Leauthaud}, {Levi}, {Li},
  {Linder}, {Lodha}, {Magneville}, {Manera}, {Margala}, {Martini}, {Maus},
  {McDonald}, {Medina-Varela}, {Meisner}, {Mena-Fern{\'a}ndez}, {Miquel},
  {Moon}, {Moore}, {Moustakas}, {Mueller}, {Mu{\~n}oz-Guti{\'e}rrez}, {Myers},
  {Nadathur}, {Napolitano}, {Neveux}, {Newman}, {Nguyen}, {Nie}, {Niz},
  {Noriega}, {Padmanabhan}, {Paillas}, {Palanque-Delabrouille}, {Pan},
  {Penmetsa}, {Percival}, {Pieri}, {Pinon}, {Poppett}, {Porredon}, {Prada},
  {P{\'e}rez-Fern{\'a}ndez}, {P{\'e}rez-R{\`a}fols}, {Rabinowitz}, {Raichoor},
  {Ram{\'\i}rez-P{\'e}rez}, {Ramirez-Solano}, {Rashkovetskyi}, {Ravoux},
  {Rezaie}, {Rich}, {Rocher}, {Rockosi}, {Roe}, {Rosado-Marin}, {Ross},
  {Rossi}, {Ruggeri}, {Ruhlmann-Kleider}, {Samushia}, {Sanchez}, {Saulder},
  {Schlafly}, {Schlegel}, {Schubnell}, {Seo}, {Shafieloo}, {Sharples},
  {Silber}, {Slosar}, {Smith}, {Sprayberry}, {Tan}, {Tarl{\'e}}, {Taylor},
  {Trusov}, {Ure{\~n}a-L{\'o}pez}, {Vaisakh}, {Valcin}, {Valdes},
  {Vargas-Maga{\~n}a}, {Verde}, {Walther}, {Wang}, {Wang}, {Weaver},
  {Weaverdyck}, {Wechsler}, {Weinberg}, {White}, {Yu}, {Yu}, {Yuan},
  {Y{\`e}che}, {Zaborowski}, {Zarrouk}, {Zhang}, {Zhao}, {Zhao}, {Zhou}, \&
  {Zhuang}}]{DESI_2024}
{Adame}, A.~G., {Aguilar}, J., {Ahlen}, S., {et~al.} 2025, \jcap, 2025, 021,
  \dodoi{10.1088/1475-7516/2025/02/021}

\bibitem[{{Alam} {et~al.}(2021){Alam}, {Aubert}, {Avila}, {Balland},
  {Bautista}, {Bershady}, {Bizyaev}, {Blanton}, {Bolton}, {Bovy}, {Brinkmann},
  {Brownstein}, {Burtin}, {Chabanier}, {Chapman}, {Choi}, {Chuang}, {Comparat},
  {Cousinou}, {Cuceu}, {Dawson}, {de la Torre}, {de Mattia}, {Agathe}, {des
  Bourboux}, {Escoffier}, {Etourneau}, {Farr}, {Font-Ribera}, {Frinchaboy},
  {Fromenteau}, {Gil-Mar{\'\i}n}, {Le Goff}, {Gonzalez-Morales},
  {Gonzalez-Perez}, {Grabowski}, {Guy}, {Hawken}, {Hou}, {Kong}, {Parker},
  {Klaene}, {Kneib}, {Lin}, {Long}, {Lyke}, {de la Macorra}, {Martini},
  {Masters}, {Mohammad}, {Moon}, {Mueller}, {Mu{\~n}oz-Guti{\'e}rrez}, {Myers},
  {Nadathur}, {Neveux}, {Newman}, {Noterdaeme}, {Oravetz}, {Oravetz},
  {Palanque-Delabrouille}, {Pan}, {Paviot}, {Percival}, {P{\'e}rez-R{\`a}fols},
  {Petitjean}, {Pieri}, {Prakash}, {Raichoor}, {Ravoux}, {Rezaie}, {Rich},
  {Ross}, {Rossi}, {Ruggeri}, {Ruhlmann-Kleider}, {S{\'a}nchez}, {S{\'a}nchez},
  {S{\'a}nchez-Gallego}, {Sayres}, {Schneider}, {Seo}, {Shafieloo}, {Slosar},
  {Smith}, {Stermer}, {Tamone}, {Tinker}, {Tojeiro}, {Vargas-Maga{\~n}a},
  {Variu}, {Wang}, {Weaver}, {Weijmans}, {Y{\`e}che}, {Zarrouk}, {Zhao},
  {Zhao}, \& {Zheng}}]{Alam2021}
{Alam}, S., {Aubert}, M., {Avila}, S., {et~al.} 2021, \prd, 103, 083533,
  \dodoi{10.1103/PhysRevD.103.083533}

\bibitem[{{Albrecht} {et~al.}(2006){Albrecht}, {Bernstein}, {Cahn}, {Freedman},
  {Hewitt}, {Hu}, {Huth}, {Kamionkowski}, {Kolb}, {Knox}, {Mather}, {Staggs},
  \& {Suntzeff}}]{Albrecht2006}
{Albrecht}, A., {Bernstein}, G., {Cahn}, R., {et~al.} 2006, arXiv e-prints,
  astro, \dodoi{10.48550/arXiv.astro-ph/0609591}

\bibitem[{Aldering(2002)}]{Aldering2002}
Aldering, G. 2002, SNAP sky background at the north ecliptic pole, Tech. Rep.
  LBNL-51157, Lawrence Berkeley National Lab. (LBNL), Berkeley, CA (United
  States), \dodoi{10.2172/842543}

\bibitem[{{Aldering} {et~al.}(2023){Aldering}, {Rubin}, {Rose}, {Hounsell},
  {Perlmutter}, \& {Deustua}}]{Prism2023_whitepaper}
{Aldering}, G., {Rubin}, D., {Rose}, B., {et~al.} 2023, arXiv e-prints,
  arXiv:2306.17219, \dodoi{10.48550/arXiv.2306.17219}

\bibitem[{{Astier} {et~al.}(2006){Astier}, {Guy}, {Regnault}, {Pain},
  {Aubourg}, {Balam}, {Basa}, {Carlberg}, {Fabbro}, {Fouchez}, {Hook},
  {Howell}, {Lafoux}, {Neill}, {Palanque-Delabrouille}, {Perrett}, {Pritchet},
  {Rich}, {Sullivan}, {Taillet}, {Aldering}, {Antilogus}, {Arsenijevic},
  {Balland}, {Baumont}, {Bronder}, {Courtois}, {Ellis}, {Filiol},
  {Gon{\c{c}}alves}, {Goobar}, {Guide}, {Hardin}, {Lusset}, {Lidman},
  {McMahon}, {Mouchet}, {Mourao}, {Perlmutter}, {Ripoche}, {Tao}, \&
  {Walton}}]{Astier2006}
{Astier}, P., {Guy}, J., {Regnault}, N., {et~al.} 2006, \aap, 447, 31,
  \dodoi{10.1051/0004-6361:20054185}

\bibitem[{{Betoule} {et~al.}(2014){Betoule}, {Kessler}, {Guy}, {Mosher},
  {Hardin}, {Biswas}, {Astier}, {El-Hage}, {Konig}, {Kuhlmann}, {Marriner},
  {Pain}, {Regnault}, {Balland}, {Bassett}, {Brown}, {Campbell}, {Carlberg},
  {Cellier-Holzem}, {Cinabro}, {Conley}, {D'Andrea}, {DePoy}, {Doi}, {Ellis},
  {Fabbro}, {Filippenko}, {Foley}, {Frieman}, {Fouchez}, {Galbany}, {Goobar},
  {Gupta}, {Hill}, {Hlozek}, {Hogan}, {Hook}, {Howell}, {Jha}, {Le Guillou},
  {Leloudas}, {Lidman}, {Marshall}, {M{\"o}ller}, {Mour{\~a}o}, {Neveu},
  {Nichol}, {Olmstead}, {Palanque-Delabrouille}, {Perlmutter}, {Prieto},
  {Pritchet}, {Richmond}, {Riess}, {Ruhlmann-Kleider}, {Sako}, {Schahmaneche},
  {Schneider}, {Smith}, {Sollerman}, {Sullivan}, {Walton}, \&
  {Wheeler}}]{Betoule2014}
{Betoule}, M., {Kessler}, R., {Guy}, J., {et~al.} 2014, \aap, 568, A22,
  \dodoi{10.1051/0004-6361/201423413}

\bibitem[{{Brout} {et~al.}(2021){Brout}, {Hinton}, \&
  {Scolnic}}]{Binning_is_Sinning}
{Brout}, D., {Hinton}, S.~R., \& {Scolnic}, D. 2021, \apjl, 912, L26,
  \dodoi{10.3847/2041-8213/abf4db}

\bibitem[{{Brout} \& {Scolnic}(2021)}]{BS21}
{Brout}, D., \& {Scolnic}, D. 2021, \apj, 909, 26,
  \dodoi{10.3847/1538-4357/abd69b}

\bibitem[{{Brout} {et~al.}(2022){Brout}, {Scolnic}, {Popovic}, {Riess}, {Carr},
  {Zuntz}, {Kessler}, {Davis}, {Hinton}, {Jones}, {Kenworthy}, {Peterson},
  {Said}, {Taylor}, {Ali}, {Armstrong}, {Charvu}, {Dwomoh}, {Meldorf},
  {Palmese}, {Qu}, {Rose}, {Sanchez}, {Stubbs}, {Vincenzi}, {Wood}, {Brown},
  {Chen}, {Chambers}, {Coulter}, {Dai}, {Dimitriadis}, {Filippenko}, {Foley},
  {Jha}, {Kelsey}, {Kirshner}, {M{\"o}ller}, {Muir}, {Nadathur}, {Pan}, {Rest},
  {Rojas-Bravo}, {Sako}, {Siebert}, {Smith}, {Stahl}, \& {Wiseman}}]{Brout2022}
{Brout}, D., {Scolnic}, D., {Popovic}, B., {et~al.} 2022, \apj, 938, 110,
  \dodoi{10.3847/1538-4357/ac8e04}

\bibitem[{{Carretero} {et~al.}(2015){Carretero}, {Castander}, {Gazta{\~n}aga},
  {Crocce}, \& {Fosalba}}]{Carretero2015_mockgal}
{Carretero}, J., {Castander}, F.~J., {Gazta{\~n}aga}, E., {Crocce}, M., \&
  {Fosalba}, P. 2015, \mnras, 447, 646, \dodoi{10.1093/mnras/stu2402}

\bibitem[{{Chen} {et~al.}(2022){Chen}, {Scolnic}, {Rozo}, {Rykoff}, {Popovic},
  {Kessler}, {Vincenzi}, {Davis}, {Armstrong}, {Brout}, {Galbany}, {Kelsey},
  {Lidman}, {M{\"o}ller}, {Rose}, {Sako}, {Sullivan}, {Taylor}, {Wiseman},
  {Asorey}, {Carr}, {Conselice}, {Kuehn}, {Lewis}, {Macaulay},
  {Rodriguez-Monroy}, {Tucker}, {Abbott}, {Aguena}, {Allam},
  {Andrade-Oliveira}, {Annis}, {Bacon}, {Bertin}, {Bocquet}, {Brooks}, {Burke},
  {Carnero Rosell}, {Carrasco Kind}, {Carretero}, {Cawthon}, {Costanzi}, {da
  Costa}, {Pereira}, {Desai}, {Diehl}, {Doel}, {Everett}, {Ferrero},
  {Flaugher}, {Friedel}, {Frieman}, {Garc{\'\i}a-Bellido}, {Gatti},
  {Gaztanaga}, {Gruen}, {Hinton}, {Hollowood}, {Honscheid}, {James}, {Lahav},
  {Lima}, {March}, {Menanteau}, {Miquel}, {Morgan}, {Palmese},
  {Paz-Chinch{\'o}n}, {Pieres}, {Plazas Malag{\'o}n}, {Prat}, {Romer},
  {Roodman}, {Sanchez}, {Schubnell}, {Serrano}, {Sevilla-Noarbe}, {Smith},
  {Soares-Santos}, {Suchyta}, {Tarle}, {Thomas}, {To}, {Tucker}, \&
  {Varga}}]{Chen2022}
{Chen}, R., {Scolnic}, D., {Rozo}, E., {et~al.} 2022, \apj, 938, 62,
  \dodoi{10.3847/1538-4357/ac8b82}

\bibitem[{{Chen} {et~al.}(2025){Chen}, {Scolnic}, {Vincenzi}, {Rykoff},
  {Myles}, {Kessler}, {Popovic}, {Sako}, {Smith}, {Armstrong}, {Brout},
  {Davis}, {Galbany}, {Lee}, {Lidman}, {M{\"o}ller}, {S{\'a}nchez}, {Sullivan},
  {Qu}, {Wiseman}, {Abbott}, {Aguena}, {Allam}, {Alves}, {Andrade-Oliveira},
  {Annis}, {Bacon}, {Brooks}, {Carnero Rosell}, {Carretero}, {Choi},
  {Conselice}, {da Costa}, {Pereira}, {Diehl}, {Doel}, {Everett}, {Ferrero},
  {Flaugher}, {Frieman}, {Garc{\'\i}a-Bellido}, {Gatti}, {Gaztanaga},
  {Giannini}, {Gruen}, {Gruendl}, {Gutierrez}, {Herner}, {Hinton}, {Hollowood},
  {Honscheid}, {Huterer}, {James}, {Kuehn}, {Lewis}, {Lima}, {Marshall},
  {Mena-Fern{\'a}ndez}, {Menanteau}, {Miquel}, {Ogando}, {Palmese}, {Pieres},
  {Plazas Malag{\'o}n}, {Roodman}, {Samuroff}, {Sanchez}, {Sanchez Cid},
  {Sevilla-Noarbe}, {Suchyta}, {Swanson}, {Tarle}, {To}, {Tucker}, {Vikram},
  {Weaverdyck}, {Weller}, \& {DES Collaboration}}]{Chen2025}
{Chen}, R.~C., {Scolnic}, D., {Vincenzi}, M., {et~al.} 2025, \mnras, 536, 1948,
  \dodoi{10.1093/mnras/stae2703}

\bibitem[{{Conley} {et~al.}(2011){Conley}, {Guy}, {Sullivan}, {Regnault},
  {Astier}, {Balland}, {Basa}, {Carlberg}, {Fouchez}, {Hardin}, {Hook},
  {Howell}, {Pain}, {Palanque-Delabrouille}, {Perrett}, {Pritchet}, {Rich},
  {Ruhlmann-Kleider}, {Balam}, {Baumont}, {Ellis}, {Fabbro}, {Fakhouri},
  {Fourmanoit}, {Gonz{\'a}lez-Gait{\'a}n}, {Graham}, {Hudson}, {Hsiao},
  {Kronborg}, {Lidman}, {Mourao}, {Neill}, {Perlmutter}, {Ripoche}, {Suzuki},
  \& {Walker}}]{Conley2011}
{Conley}, A., {Guy}, J., {Sullivan}, M., {et~al.} 2011, \apjs, 192, 1,
  \dodoi{10.1088/0067-0049/192/1/1}

\bibitem[{{Crocce} {et~al.}(2015){Crocce}, {Castander}, {Gazta{\~n}aga},
  {Fosalba}, \& {Carretero}}]{MICECAT2015}
{Crocce}, M., {Castander}, F.~J., {Gazta{\~n}aga}, E., {Fosalba}, P., \&
  {Carretero}, J. 2015, \mnras, 453, 1513, \dodoi{10.1093/mnras/stv1708}

\bibitem[{{DES Collaboration} {et~al.}(2024){DES Collaboration}, {Abbott},
  {Acevedo}, {Aguena}, {Alarcon}, {Allam}, {Alves}, {Amon}, {Andrade-Oliveira},
  {Annis}, {Armstrong}, {Asorey}, {Avila}, {Bacon}, {Bassett}, {Bechtol},
  {Bernardinelli}, {Bernstein}, {Bertin}, {Blazek}, {Bocquet}, {Brooks},
  {Brout}, {Buckley-Geer}, {Burke}, {Camacho}, {Camilleri}, {Campos}, {Carnero
  Rosell}, {Carollo}, {Carr}, {Carretero}, {Castander}, {Cawthon}, {Chang},
  {Chen}, {Choi}, {Conselice}, {Costanzi}, {da Costa}, {Crocce}, {Davis},
  {DePoy}, {Desai}, {Diehl}, {Dixon}, {Dodelson}, {Doel}, {Doux},
  {Drlica-Wagner}, {Elvin-Poole}, {Everett}, {Ferrero}, {Fert{\'e}},
  {Flaugher}, {Foley}, {Fosalba}, {Friedel}, {Frieman}, {Frohmaier}, {Galbany},
  {Garc{\'\i}a-Bellido}, {Gatti}, {Gaztanaga}, {Giannini}, {Glazebrook},
  {Graur}, {Gruen}, {Gruendl}, {Gutierrez}, {Hartley}, {Herner}, {Hinton},
  {Hollowood}, {Honscheid}, {Huterer}, {Jain}, {James}, {Jeffrey}, {Kasai},
  {Kelsey}, {Kent}, {Kessler}, {Kim}, {Kirshner}, {Kovacs}, {Kuehn}, {Lahav},
  {Lee}, {Lee}, {Lewis}, {Li}, {Lidman}, {Lin}, {Malik}, {Marshall}, {Martini},
  {Mena-Fern{\'a}ndez}, {Menanteau}, {Miquel}, {Mohr}, {Mould}, {Muir},
  {M{\"o}ller}, {Neilsen}, {Nichol}, {Nugent}, {Ogando}, {Palmese}, {Pan},
  {Paterno}, {Percival}, {Pereira}, {Pieres}, {Malag{\'o}n}, {Popovic},
  {Porredon}, {Prat}, {Qu}, {Raveri}, {Rodr{\'\i}guez-Monroy}, {Romer},
  {Roodman}, {Rose}, {Sako}, {Sanchez}, {Sanchez Cid}, {Schubnell}, {Scolnic},
  {Sevilla-Noarbe}, {Shah}, {Smith}, {Smith}, {Soares-Santos}, {Suchyta},
  {Sullivan}, {Suntzeff}, {Swanson}, {S{\'a}nchez}, {Tarle}, {Taylor},
  {Thomas}, {To}, {Toy}, {Troxel}, {Tucker}, {Tucker}, {Uddin}, {Vincenzi},
  {Walker}, {Weaverdyck}, {Wechsler}, {Weller}, {Wester}, {Wiseman},
  {Yamamoto}, {Yuan}, {Zhang}, \& {Zhang}}]{DESCollaboration2024}
{DES Collaboration}, {Abbott}, T.~M.~C., {Acevedo}, M., {et~al.} 2024, \apjl,
  973, L14, \dodoi{10.3847/2041-8213/ad6f9f}

\bibitem[{{DESI Collaboration} {et~al.}(2025){DESI Collaboration},
  {Abdul-Karim}, {Aguilar}, {Ahlen}, {Alam}, {Allen}, {Allende Prieto},
  {Alves}, {Anand}, {Andrade}, {Armengaud}, {Aviles}, {Bailey}, {Baltay},
  {Bansal}, {Bault}, {Behera}, {BenZvi}, {Bianchi}, {Blake}, {Brieden},
  {Brodzeller}, {Brooks}, {Buckley-Geer}, {Burtin}, {Calderon}, {Canning},
  {Carnero Rosell}, {Carrilho}, {Casas}, {Castander}, {Cereskaite}, {Charles},
  {Chaussidon}, {Chaves-Montero}, {Chebat}, {Chen}, {Claybaugh}, {Cole},
  {Cooper}, {Cuceu}, {Dawson}, {de la Macorra}, {de Mattia}, {Deiosso}, {Della
  Costa}, {Demina}, {Dey}, {Dey}, {Ding}, {Doel}, {Edelstein}, {Eisenstein},
  {Elbers}, {Fagrelius}, {Fanning}, {Fernandez-Garcia}, {Ferraro},
  {Font-Ribera}, {Forero-Romero}, {Frenk}, {Garcia-Quintero}, {Garrison},
  {Gaztanaga}, {Gil-Mar]in}, {Gontcho}, {Gonzalez}, {Gonzalez-Morales},
  {Gordon}, {Green}, {Gutierrez}, {Guy}, {Hadzhiyska}, {Hahn}, {He}, {Herbold},
  {Herrera-Alcantar}, {Ho}, {Honscheid}, {Howlett}, {Huterer}, {Ishak},
  {Juneau}, {Kamble}, {Karac\{c\}ayl\{\}}, {Kehoe}, {Kent}, {Kim}, {Kirkby},
  {Kisner}, {Koposov}, {Kremin}, {Krolewski}, {Lahav}, {Lamman}, {Landriau},
  {Lang}, {Lasker}, {Le Goff}, {Le Guillou}, {Leauthaud}, {Levi}, {Li}, {Li},
  {Lodha}, {Lokken}, {Lozano-Rodriguez}, {Magneville}, {Manera}, {Martini},
  {Matthewson}, {Meisner}, {Mena-Fern{\'}andez}, {Menegas}, {Mergulhao},
  {Miquel}, {Moustakas}, {Mutildenoz-Gutierrez}, {Munoz-Santos}, {Myers},
  {Nadathur}, {Naidoo}, {Napolitano}, {Newman}, {Niz}, {Noriega}, {Paillas},
  {Palanque-Delabrouille}, {Pan}, {Peacock}, {Pellejero Ibanez}, {Percival},
  {Perez-Fernandez}, {Perez-Rafols}, {Pieri}, {Poppett}, {Prada}, {Rabinowitz},
  {Raichoor}, {Ramirez-Perez}, {Rashkovetskyi}, {Ravoux}, {Rich}, {Rocher},
  {Rockosi}, {Rohlf}, {Roman-Herrera}, {Ross}, {Rossi}, {Ruggeri},
  {Ruhlmann-Kleider}, {Samushia}, {Sanchez}, {Sanders}, {Schlegel},
  {Schubnell}, {Seo}, {Shafieloo}, {Sharples}, {Silber}, {Sinigaglia},
  {Sprayberry}, {Tan}, {Tarle}, {Taylor}, {Turner}, {Uretildena-Lopez},
  {Vaisakh}, {Valdes}, {Valogiannis}, {Vargas-Magana}, {Verde}, {Walther},
  {Weaver}, {Weinberg}, {White}, {Wolfson}, {Yeche}, {Yu}, {Zaborowski},
  {Zarrouk}, {Zhai}, {Zhang}, {Zhao}, {Zhao}, {Zhou}, \& {Zou}}]{DESI_2025}
{DESI Collaboration}, {Abdul-Karim}, M., {Aguilar}, J., {et~al.} 2025, arXiv
  e-prints, arXiv:2503.14738, \dodoi{10.48550/arXiv.2503.14738}

\bibitem[{{Dunlop} {et~al.}(2021){Dunlop}, {Abraham}, {Ashby}, {Bagley},
  {Best}, {Bongiorno}, {Bouwens}, {Bowler}, {Brammer}, {Bremer}, {Calabro'},
  {Carnall}, {Castellano}, {Cirasuolo}, {Conselice}, {Cullen}, {Dave}, {Dayal},
  {Dekel}, {Dickinson}, {Duncan}, {Elbaz}, {Ellis}, {Ferguson}, {Ferrara},
  {Finkelstein}, {Fontana}, {Furlanetto}, {Fynbo}, {Gallerani}, {Gardner},
  {Giavalisco}, {Grazian}, {Grogin}, {Harikane}, {Hopkins}, {Ilbert},
  {Illingworth}, {Juneau}, {Jung}, {Kartaltepe}, {Kassin}, {Kauffmann},
  {Khochfar}, {Kirkpatrick}, {Kocevski}, {Koekemoer}, {Labbe}, {Laporte},
  {Larson}, {Lucas}, {Magee}, {Mason}, {McCracken}, {McLeod}, {McLure},
  {Merlin}, {Mesinger}, {Milvang-Jensen}, {Newman}, {Oesch}, {Ouchi},
  {Pacifici}, {Papovich}, {Peacock}, {Peeples}, {Pentericci}, {Perez-Gonzalez},
  {Pirzkal}, {Pope}, {Pye}, {Reddy}, {Robertson}, {Salvato}, {Santini},
  {Schaerer}, {Shapley}, {Simons}, {Smit}, {Smith}, {Snyder}, {Somerville},
  {Stanway}, {Stefanon}, {Tasca}, {Tikkanen}, {Tresse}, {Trump}, {Whitaker},
  {Wilkins}, {Wright}, {Wyithe}, {van Dokkum}, \& {van der Werf}}]{PRIMER}
{Dunlop}, J.~S., {Abraham}, R.~G., {Ashby}, M. L.~N., {et~al.} 2021, {PRIMER:
  Public Release IMaging for Extragalactic Research}, JWST Proposal. Cycle 1,
  ID. \#1837

\bibitem[{{Eegholm} {et~al.}(2025){Eegholm}, {Marx}, {Chambers}, {Chu}, {Gao},
  {Dominguez}, {Lehan}, {Hagopian}, {Gong}, {Seide}, {Bray}, {He}, {Patel},
  {Quijada}, {Malumuth}, {Morey}, {Atanassova}, {Jepsen}, {Krom}, {Kittle},
  {Choi}, {Salem}, {Kolos}, {Peabody}, {Lyons}, {Corsetti}, {Seals},
  {Pasquale}, {Rizzo}, {Rhoads}, {Schlieder}, {Malhotra}, {Kruk}, {Whipple}, \&
  {Content}}]{Prism2025}
{Eegholm}, B.~H., {Marx}, C.~T., {Chambers}, V.~J., {et~al.} 2025, Journal of
  Astronomical Telescopes, Instruments, and Systems, 11, 025001,
  \dodoi{10.1117/1.JATIS.11.2.025001}

\bibitem[{{Hinton} \& {Brout}(2020)}]{Pippin}
{Hinton}, S., \& {Brout}, D. 2020, JOSS, 5, 2122, \dodoi{10.21105/joss.02122}

\bibitem[{{Hlozek} {et~al.}(2012){Hlozek}, {Kunz}, {Bassett}, {Smith},
  {Newling}, {et~al.}}]{BEAMS2012}
{Hlozek}, R., {Kunz}, M., {Bassett}, B., {et~al.} 2012, \apj, 752, 79,
  \dodoi{10.1088/0004-637X/752/2/79}

\bibitem[{{HLTDS~Definition~Committee}(2025)}]{CCS_report}
{HLTDS~Definition~Committee}. 2025, Findings of the High Lattitude Time Domain
  Survey Definition Committee, Tech. rep.
\newblock
  \url{https://asd.gsfc.nasa.gov/roman/comm_forum/forum_17/Core_Community_Survey_Reports-rev03-compressed.pdf}

\bibitem[{{Hounsell} {et~al.}(2018){Hounsell}, {Scolnic}, {Foley}, {Kessler},
  {Miranda}, {Avelino}, {Bohlin}, {Filippenko}, {Frieman}, {Jha}, {Kelly},
  {Kirshner}, {Mandel}, {Rest}, {Riess}, {Rodney}, \&
  {Strolger}}]{Hounsell2018}
{Hounsell}, R., {Scolnic}, D., {Foley}, R.~J., {et~al.} 2018, \apj, 867, 23,
  \dodoi{10.3847/1538-4357/aac08b}

\bibitem[{James(1994)}]{MINUIT}
James, F. 1994

\bibitem[{{Jones} {et~al.}(2018){Jones}, {Scolnic}, {Riess}, {Rest},
  {Kirshner}, {et~al.}}]{Jones2018_PS1}
{Jones}, D.~O., {Scolnic}, D.~M., {Riess}, A.~G., {et~al.} 2018, \apj, 857, 51,
  \dodoi{10.3847/1538-4357/aab6b1}

\bibitem[{{Kenworthy} {et~al.}(2021){Kenworthy}, {Jones}, {Dai}, {Kessler},
  {Scolnic}, {Brout}, {Siebert}, {Pierel}, {Dettman}, {Dimitriadis}, {Foley},
  {Jha}, {Pan}, {Riess}, {Rodney}, \& {Rojas-Bravo}}]{K21_SALT3}
{Kenworthy}, W.~D., {Jones}, D.~O., {Dai}, M., {et~al.} 2021, \apj, 923, 265,
  \dodoi{10.3847/1538-4357/ac30d8}

\bibitem[{{Kenworthy} {et~al.}(2025){Kenworthy}, {Goobar}, {Jones},
  {Johansson}, {Thorp}, {Kessler}, {Burgaz}, {Dhawan}, {Dimitriadis},
  {Galbany}, {Ginolin}, {Kim}, {Maguire}, {M{\"u}ller-Bravo}, {Nugent},
  {Nordin}, {Popovic}, {Pessi}, {Rigault}, {Rosnet}, {Sollerman}, {Terwel},
  {Townsend}, {Laher}, {Purdum}, {Rosselli}, \& {Rusholme}}]{SALT3+x2}
{Kenworthy}, W.~D., {Goobar}, A., {Jones}, D.~O., {et~al.} 2025, \aap, 697,
  A125, \dodoi{10.1051/0004-6361/202452578}

\bibitem[{{Kessler} {et~al.}(2019{\natexlab{a}}){Kessler}, {Brout}, {Crawford},
  {et~al.}}]{Kessler2019_sim}
{Kessler}, R., {Brout}, D., {Crawford}, S., {et~al.} 2019{\natexlab{a}},
  \mnras, 485, 1171, \dodoi{10.1093/mnras/stz463}

\bibitem[{{Kessler} \& {Scolnic}(2017)}]{BBC}
{Kessler}, R., \& {Scolnic}, D. 2017, \apj, 836, 56,
  \dodoi{10.3847/1538-4357/836/1/56}

\bibitem[{{Kessler} {et~al.}(2023){Kessler}, {Vincenzi}, \&
  {Armstrong}}]{Kessler2023_redemption}
{Kessler}, R., {Vincenzi}, M., \& {Armstrong}, P. 2023, \apjl, 952, L8,
  \dodoi{10.3847/2041-8213/ace34d}

\bibitem[{{Kessler} {et~al.}(2009{\natexlab{a}}){Kessler}, {Becker}, {Cinabro},
  {Vanderplas}, {Frieman}, {Marriner}, {Davis}, {Dilday}, {Holtzman}, {Jha},
  {Lampeitl}, {Sako}, {Smith}, {Zheng}, {Nichol}, {Bassett}, {Bender}, {Depoy},
  {Doi}, {Elson}, {Filippenko}, {Foley}, {Garnavich}, {Hopp}, {Ihara},
  {Ketzeback}, {Kollatschny}, {Konishi}, {Marshall}, {McMillan}, {Miknaitis},
  {Morokuma}, {M{\"o}rtsell}, {Pan}, {Prieto}, {Richmond}, {Riess}, {Romani},
  {Schneider}, {Sollerman}, {Takanashi}, {Tokita}, {van der Heyden}, {Wheeler},
  {Yasuda}, \& {York}}]{Kessler2009}
{Kessler}, R., {Becker}, A.~C., {Cinabro}, D., {et~al.} 2009{\natexlab{a}},
  \apjs, 185, 32, \dodoi{10.1088/0067-0049/185/1/32}

\bibitem[{{Kessler} {et~al.}(2009{\natexlab{b}}){Kessler}, {Bernstein},
  {Cinabro}, {Dilday}, {Frieman}, {Jha}, {Kuhlmann}, {Miknaitis}, {Sako},
  {Taylor}, \& {Vanderplas}}]{SNANA}
{Kessler}, R., {Bernstein}, J.~P., {Cinabro}, D., {et~al.} 2009{\natexlab{b}},
  \pasp, 121, 1028, \dodoi{10.1086/605984}

\bibitem[{{Kessler} {et~al.}(2010){Kessler}, {Cinabro}, {Bassett}, {Dilday},
  {Frieman}, {Garnavich}, {Jha}, {Marriner}, {Nichol}, {Sako}, {Smith},
  {Bernstein}, {Bizyaev}, {Goobar}, {Kuhlmann}, {Schneider}, \&
  {Stritzinger}}]{Kessler2010_zphot}
{Kessler}, R., {Cinabro}, D., {Bassett}, B., {et~al.} 2010, \apj, 717, 40,
  \dodoi{10.1088/0004-637X/717/1/40}

\bibitem[{{Kessler} {et~al.}(2015){Kessler}, {Marriner}, {Childress},
  {Covarrubias}, {D'Andrea}, {Finley}, {Fischer}, {Foley}, {Goldstein},
  {Gupta}, {Kuehn}, {Marcha}, {Nichol}, {Papadopoulos}, {Sako}, {Scolnic},
  {Smith}, {Sullivan}, {Wester}, {Yuan}, {Abbott}, {Abdalla}, {Allam},
  {Benoit-L{\'e}vy}, {Bernstein}, {Bertin}, {Brooks}, {Carnero Rosell},
  {Carrasco Kind}, {Castander}, {Crocce}, {da Costa}, {Desai}, {Diehl},
  {Eifler}, {Fausti Neto}, {Flaugher}, {Frieman}, {Gerdes}, {Gruen}, {Gruendl},
  {Honscheid}, {James}, {Kuropatkin}, {Li}, {Maia}, {Marshall}, {Martini},
  {Miller}, {Miquel}, {Nord}, {Ogando}, {Plazas}, {Reil}, {Romer}, {Roodman},
  {Sanchez}, {Sevilla-Noarbe}, {Smith}, {Soares-Santos}, {Sobreira}, {Tarle},
  {Thaler}, {Thomas}, {Tucker}, {Walker}, \& {DES
  Collaboration}}]{Kessler2015_DIFFIMG}
{Kessler}, R., {Marriner}, J., {Childress}, M., {et~al.} 2015, \aj, 150, 172,
  \dodoi{10.1088/0004-6256/150/6/172}

\bibitem[{{Kessler} {et~al.}(2019{\natexlab{b}}){Kessler}, {Narayan},
  {Avelino}, {Bachelet}, {Biswas}, {Brown}, {Chernoff}, {Connolly}, {Dai},
  {Daniel}, {Di Stefano}, {Drout}, {Galbany}, {Gonz{\'a}lez-Gait{\'a}n},
  {Graham}, {Hlo{\v{z}}ek}, {Ishida}, {Guillochon}, {Jha}, {Jones}, {Mandel},
  {Muthukrishna}, {O'Grady}, {Peters}, {Pierel}, {Ponder}, {Pr{\v{s}}a},
  {Rodney}, {Villar}, {LSST Dark Energy Science Collaboration}, \& {Transient
  and Variable Stars Science Collaboration}}]{Kessler2019_plasticc}
{Kessler}, R., {Narayan}, G., {Avelino}, A., {et~al.} 2019{\natexlab{b}},
  \pasp, 131, 094501, \dodoi{10.1088/1538-3873/ab26f1}

\bibitem[{{Korytov} {et~al.}(2019){Korytov}, {Hearin}, {Kovacs}, {Larsen},
  {Rangel}, {Hollowed}, {Benson}, {Heitmann}, {Mao}, {Bahmanyar}, {Chang},
  {Campbell}, {DeRose}, {Finkel}, {Frontiere}, {Gawiser}, {Habib}, {Joachimi},
  {Lanusse}, {Li}, {Mandelbaum}, {Morrison}, {Newman}, {Pope}, {Rykoff},
  {Simet}, {To}, {Vikraman}, {Wechsler}, {White}, \& {(The LSST Dark Energy
  Science Collaboration}}]{GCR2019_forDC2}
{Korytov}, D., {Hearin}, A., {Kovacs}, E., {et~al.} 2019, \apjs, 245, 26,
  \dodoi{10.3847/1538-4365/ab510c}

\bibitem[{{Kunz} {et~al.}(2007){Kunz}, {Bassett}, \& {Hlozek}}]{BEAMS2007}
{Kunz}, M., {Bassett}, B.~A., \& {Hlozek}, R.~A. 2007, \prd, 75, 103508,
  \dodoi{10.1103/PhysRevD.75.103508}

\bibitem[{{Lokken} {et~al.}(2023){Lokken}, {Gagliano}, {Narayan},
  {Hlo{\v{z}}ek}, {Kessler}, {Crenshaw}, {Salo}, {Alves}, {Chatterjee},
  {Vincenzi}, {Malz}, \& {LSST Dark Energy Science Collaboration}}]{SCOTCH2023}
{Lokken}, M., {Gagliano}, A., {Narayan}, G., {et~al.} 2023, \mnras, 520, 2887,
  \dodoi{10.1093/mnras/stad302}

\bibitem[{{LSST Dark Energy Science Collaboration (LSST DESC)}
  {et~al.}(2021){LSST Dark Energy Science Collaboration (LSST DESC)},
  {Abolfathi}, {Alonso}, {Armstrong}, {Aubourg}, {Awan}, {Babuji}, {Bauer},
  {Bean}, {Beckett}, {Biswas}, {Bogart}, {Boutigny}, {Chard}, {Chiang},
  {Claver}, {Cohen-Tanugi}, {Combet}, {Connolly}, {Daniel}, {Digel},
  {Drlica-Wagner}, {Dubois}, {Gangler}, {Gawiser}, {Glanzman}, {Gris}, {Habib},
  {Hearin}, {Heitmann}, {Hernandez}, {Hlo{\v{z}}ek}, {Hollowed}, {Ishak},
  {Ivezi{\'c}}, {Jarvis}, {Jha}, {Kahn}, {Kalmbach}, {Kelly}, {Kovacs},
  {Korytov}, {Krughoff}, {Lage}, {Lanusse}, {Larsen}, {Le Guillou}, {Li},
  {Longley}, {Lupton}, {Mandelbaum}, {Mao}, {Marshall}, {Meyers}, {Moniez},
  {Morrison}, {Nomerotski}, {O'Connor}, {Park}, {Park}, {Peloton}, {Perrefort},
  {Perry}, {Plaszczynski}, {Pope}, {Rasmussen}, {Reil}, {Roodman}, {Rykoff},
  {S{\'a}nchez}, {Schmidt}, {Scolnic}, {Stubbs}, {Tyson}, {Uram}, {Villarreal},
  {Walter}, {Wiesner}, {Wood-Vasey}, \& {Zuntz}}]{LSST_DC2}
{LSST Dark Energy Science Collaboration (LSST DESC)}, {Abolfathi}, B.,
  {Alonso}, D., {et~al.} 2021, \apjs, 253, 31, \dodoi{10.3847/1538-4365/abd62c}

\bibitem[{{Madau} \& {Dickinson}(2014)}]{MD14}
{Madau}, P., \& {Dickinson}, M. 2014, \araa, 52, 415,
  \dodoi{10.1146/annurev-astro-081811-125615}

\bibitem[{{Mitra} {et~al.}(2023){Mitra}, {Kessler}, {More}, {Hlozek}, \& {LSST
  Dark Energy Science Collaboration}}]{Mitra2023}
{Mitra}, A., {Kessler}, R., {More}, S., {Hlozek}, R., \& {LSST Dark Energy
  Science Collaboration}. 2023, \apj, 944, 212,
  \dodoi{10.3847/1538-4357/acb057}

\bibitem[{{Myles} {et~al.}(2021){Myles}, {Alarcon}, {Amon}, {S{\'a}nchez},
  {Everett}, {DeRose}, {McCullough}, {Gruen}, {Bernstein}, {Troxel},
  {Dodelson}, {Campos}, {MacCrann}, {Yin}, {Raveri}, {Amara}, {Becker}, {Choi},
  {Cordero}, {Eckert}, {Gatti}, {Giannini}, {Gschwend}, {Gruendl}, {Harrison},
  {Hartley}, {Huff}, {Kuropatkin}, {Lin}, {Masters}, {Miquel}, {Prat},
  {Roodman}, {Rykoff}, {Sevilla-Noarbe}, {Sheldon}, {Wechsler}, {Yanny},
  {Abbott}, {Aguena}, {Allam}, {Annis}, {Bacon}, {Bertin}, {Bhargava},
  {Bridle}, {Brooks}, {Burke}, {Carnero Rosell}, {Carrasco Kind}, {Carretero},
  {Castander}, {Conselice}, {Costanzi}, {Crocce}, {da Costa}, {Pereira},
  {Desai}, {Diehl}, {Eifler}, {Elvin-Poole}, {Evrard}, {Ferrero}, {Fert{\'e}},
  {Flaugher}, {Fosalba}, {Frieman}, {Garc{\'\i}a-Bellido}, {Gaztanaga},
  {Giannantonio}, {Hinton}, {Hollowood}, {Honscheid}, {Hoyle}, {Huterer},
  {James}, {Krause}, {Kuehn}, {Lahav}, {Lima}, {Maia}, {Marshall}, {Martini},
  {Melchior}, {Menanteau}, {Mohr}, {Morgan}, {Muir}, {Ogando}, {Palmese},
  {Paz-Chinch{\'o}n}, {Plazas}, {Rodriguez-Monroy}, {Samuroff}, {Sanchez},
  {Scarpine}, {Secco}, {Serrano}, {Smith}, {Soares-Santos}, {Suchyta},
  {Swanson}, {Tarle}, {Thomas}, {To}, {Varga}, {Weller}, \&
  {Wester}}]{Myles2021}
{Myles}, J., {Alarcon}, A., {Amon}, A., {et~al.} 2021, \mnras, 505, 4249,
  \dodoi{10.1093/mnras/stab1515}

\bibitem[{National\;Research\;Council(2010)}]{Decadal2010}
National\;Research\;Council. 2010, New Worlds, New Horizons in Astronomy and
  Astrophysics (Washington, DC: The National Academies Press),
  \dodoi{10.17226/12951}

\bibitem[{{OpenUniverse} {et~al.}(2025){OpenUniverse}, {The LSST Dark Energy
  Science Collaboration}, {The Roman HLIS Project Infrastructure Team}, {The
  Roman RAPID Project Infrastructure Team}, {The Roman Supernova Cosmology
  Project Infrastructure Team}, {Alarcon}, {Aldoroty}, {Beltz-Mohrmann},
  {Bera}, {Blazek}, {Bogart}, {Braeunlich}, {Broughton}, {Cao}, {Chiang},
  {Chisari}, {Desai}, {Fang}, {Galbany}, {Hearin}, {Heitmann}, {Hirata},
  {Hounsell}, {Jain}, {Jarvis}, {Jencson}, {Kannawadi}, {Kasliwal}, {Kessler},
  {Kiessling}, {Knop}, {Kovacs}, {Laher}, {Laliotis}, {Lin}, {Lopes},
  {Mahabal}, {Mandelbaum}, {Masiero}, {Mau}, {Meehan}, {Meyers}, {Moraes},
  {Paladini}, {Pearl}, {Plazas Malagon}, {Rose}, {Rubin}, {Rusholme}, {Santos},
  {{\v{S}}ar{\v{c}}evi{\'c}}, {Scolnic}, {Troxel}, {Van Alfen}, {Van Dyke},
  {Walter}, {Wu}, {Yamamoto}, {Yan}, \& {Zhang}}]{OpenUnivserse2024}
{OpenUniverse}, {The LSST Dark Energy Science Collaboration}, {The Roman HLIS
  Project Infrastructure Team}, {et~al.} 2025, arXiv e-prints,
  arXiv:2501.05632, \dodoi{10.48550/arXiv.2501.05632}

\bibitem[{{Perlmutter} {et~al.}(1999){Perlmutter}, {Aldering}, {Goldhaber},
  {Knop}, {Nugent}, {Castro}, {Deustua}, {Fabbro}, {Goobar}, {Groom}, {Hook},
  {Kim}, {Kim}, {Lee}, {Nunes}, {Pain}, {Pennypacker}, {Quimby}, {Lidman},
  {Ellis}, {Irwin}, {McMahon}, {Ruiz-Lapuente}, {Walton}, {Schaefer}, {Boyle},
  {Filippenko}, {Matheson}, {Fruchter}, {Panagia}, {Newberg}, {Couch}, \&
  {Project}}]{Perlmutter1999}
{Perlmutter}, S., {Aldering}, G., {Goldhaber}, G., {et~al.} 1999, \apj, 517,
  565, \dodoi{10.1086/307221}

\bibitem[{{Pierel} {et~al.}(2018){Pierel}, {Rodney}, {Avelino}, {Bianco},
  {Filippenko}, {Foley}, {Friedman}, {Hicken}, {Hounsell}, {Jha}, {Kessler},
  {Kirshner}, {Mandel}, {Narayan}, {Scolnic}, \& {Strolger}}]{Pierel2018}
{Pierel}, J.~D.~R., {Rodney}, S., {Avelino}, A., {et~al.} 2018, \pasp, 130,
  114504, \dodoi{10.1088/1538-3873/aadb7a}

\bibitem[{{Pierel} {et~al.}(2022){Pierel}, {Jones}, {Kenworthy}, {Dai},
  {Kessler}, {Ashall}, {Do}, {Peterson}, {Shappee}, {Siebert}, {Barna},
  {Brink}, {Burke}, {Calamida}, {Camacho-Neves}, {de Jaeger}, {Filippenko},
  {Foley}, {Galbany}, {Fox}, {Gomez}, {Hiramatsu}, {Hounsell}, {Howell}, {Jha},
  {Kwok}, {P{\'e}rez-Fournon}, {Poidevin}, {Rest}, {Rubin}, {Scolnic},
  {Shirley}, {Strolger}, {Tinyanont}, \& {Wang}}]{Pierel2022}
{Pierel}, J.~D.~R., {Jones}, D.~O., {Kenworthy}, W.~D., {et~al.} 2022, \apj,
  939, 11, \dodoi{10.3847/1538-4357/ac93f9}

\bibitem[{{Planck Collaboration} {et~al.}(2020){Planck Collaboration},
  {Aghanim}, {Akrami}, {Ashdown}, {Aumont}, {Baccigalupi}, {Ballardini},
  {Banday}, {Barreiro}, {Bartolo}, {Basak}, {Battye}, {Benabed}, {Bernard},
  {Bersanelli}, {Bielewicz}, {Bock}, {Bond}, {Borrill}, {Bouchet}, {Boulanger},
  {Bucher}, {Burigana}, {Butler}, {Calabrese}, {Cardoso}, {Carron},
  {Challinor}, {Chiang}, {Chluba}, {Colombo}, {Combet}, {Contreras}, {Crill},
  {Cuttaia}, {de Bernardis}, {de Zotti}, {Delabrouille}, {Delouis}, {Di
  Valentino}, {Diego}, {Dor{\'e}}, {Douspis}, {Ducout}, {Dupac}, {Dusini},
  {Efstathiou}, {Elsner}, {En{\ss}lin}, {Eriksen}, {Fantaye}, {Farhang},
  {Fergusson}, {Fernandez-Cobos}, {Finelli}, {Forastieri}, {Frailis},
  {Fraisse}, {Franceschi}, {Frolov}, {Galeotta}, {Galli}, {Ganga},
  {G{\'e}nova-Santos}, {Gerbino}, {Ghosh}, {Gonz{\'a}lez-Nuevo}, {G{\'o}rski},
  {Gratton}, {Gruppuso}, {Gudmundsson}, {Hamann}, {Handley}, {Hansen},
  {Herranz}, {Hildebrandt}, {Hivon}, {Huang}, {Jaffe}, {Jones}, {Karakci},
  {Keih{\"a}nen}, {Keskitalo}, {Kiiveri}, {Kim}, {Kisner}, {Knox},
  {Krachmalnicoff}, {Kunz}, {Kurki-Suonio}, {Lagache}, {Lamarre}, {Lasenby},
  {Lattanzi}, {Lawrence}, {Le Jeune}, {Lemos}, {Lesgourgues}, {Levrier},
  {Lewis}, {Liguori}, {Lilje}, {Lilley}, {Lindholm}, {L{\'o}pez-Caniego},
  {Lubin}, {Ma}, {Mac{\'\i}as-P{\'e}rez}, {Maggio}, {Maino}, {Mandolesi},
  {Mangilli}, {Marcos-Caballero}, {Maris}, {Martin}, {Martinelli},
  {Mart{\'\i}nez-Gonz{\'a}lez}, {Matarrese}, {Mauri}, {McEwen}, {Meinhold},
  {Melchiorri}, {Mennella}, {Migliaccio}, {Millea}, {Mitra},
  {Miville-Desch{\^e}nes}, {Molinari}, {Montier}, {Morgante}, {Moss}, {Natoli},
  {N{\o}rgaard-Nielsen}, {Pagano}, {Paoletti}, {Partridge}, {Patanchon},
  {Peiris}, {Perrotta}, {Pettorino}, {Piacentini}, {Polastri}, {Polenta},
  {Puget}, {Rachen}, {Reinecke}, {Remazeilles}, {Renzi}, {Rocha}, {Rosset},
  {Roudier}, {Rubi{\~n}o-Mart{\'\i}n}, {Ruiz-Granados}, {Salvati}, {Sandri},
  {Savelainen}, {Scott}, {Shellard}, {Sirignano}, {Sirri}, {Spencer},
  {Sunyaev}, {Suur-Uski}, {Tauber}, {Tavagnacco}, {Tenti}, {Toffolatti},
  {Tomasi}, {Trombetti}, {Valenziano}, {Valiviita}, {Van Tent}, {Vibert},
  {Vielva}, {Villa}, {Vittorio}, {Wandelt}, {Wehus}, {White}, {White},
  {Zacchei}, \& {Zonca}}]{Planck2018}
{Planck Collaboration}, {Aghanim}, N., {Akrami}, Y., {et~al.} 2020, \aap, 641,
  A6, \dodoi{10.1051/0004-6361/201833910}

\bibitem[{{Popovic} {et~al.}(2023){Popovic}, {Brout}, {Kessler}, \&
  {Scolnic}}]{Dust2dust2023}
{Popovic}, B., {Brout}, D., {Kessler}, R., \& {Scolnic}, D. 2023, \apj, 945,
  84, \dodoi{10.3847/1538-4357/aca273}

\bibitem[{{Popovic} {et~al.}(2021){Popovic}, {Brout}, {Kessler}, {Scolnic}, \&
  {Lu}}]{Popovic2021}
{Popovic}, B., {Brout}, D., {Kessler}, R., {Scolnic}, D., \& {Lu}, L. 2021,
  \apj, 913, 49, \dodoi{10.3847/1538-4357/abf14f}

\bibitem[{{Qu} {et~al.}(2021){Qu}, {Sako}, {M{\"o}ller}, \& {Doux}}]{SCONE}
{Qu}, H., {Sako}, M., {M{\"o}ller}, A., \& {Doux}, C. 2021, \aj, 162, 67,
  \dodoi{10.3847/1538-3881/ac0824}

\bibitem[{{Riess} {et~al.}(1998){Riess}, {Filippenko}, {Challis},
  {Clocchiatti}, {Diercks}, {Garnavich}, {Gilliland}, {Hogan}, {Jha},
  {Kirshner}, {Leibundgut}, {Phillips}, {Reiss}, {Schmidt}, {Schommer},
  {Smith}, {Spyromilio}, {Stubbs}, {Suntzeff}, \& {Tonry}}]{Riess1998}
{Riess}, A.~G., {Filippenko}, A.~V., {Challis}, P., {et~al.} 1998, \aj, 116,
  1009, \dodoi{10.1086/300499}

\bibitem[{{Rigault} {et~al.}(2025){Rigault}, {Smith}, {Goobar}, {Maguire},
  {Dimitriadis}, {Johansson}, {Nordin}, {Burgaz}, {Dhawan}, {Sollerman},
  {Regnault}, {Kowalski}, {Nugent}, {Andreoni}, {Amenouche}, {Aubert},
  {Barjou-Delayre}, {Bautista}, {Bellm}, {Betoule}, {Bloom}, {Carreres},
  {Chen}, {Copin}, {Deckers}, {de Jaeger}, {Feinstein}, {Fouchez}, {Fremling},
  {Galbany}, {Ginolin}, {Graham}, {Groom}, {Harvey}, {Kasliwal}, {Kenworthy},
  {Kim}, {Kuhn}, {Kulkarni}, {Lacroix}, {Laher}, {Masci}, {M{\"u}ller-Bravo},
  {Miller}, {Osman}, {Perley}, {Popovic}, {Purdum}, {Qin}, {Racine}, {Reusch},
  {Riddle}, {Rosnet}, {Rosselli}, {Ruppin}, {Senzel}, {Rusholme}, {Schweyer},
  {Terwel}, {Townsend}, {Tzanidakis}, {Wold}, \& {Yan}}]{ZTF_DR2_OVERVIEW}
{Rigault}, M., {Smith}, M., {Goobar}, A., {et~al.} 2025, \aap, 694, A1,
  \dodoi{10.1051/0004-6361/202450388}

\bibitem[{{Rodney} {et~al.}(2014){Rodney}, {Riess}, {Strolger}, {Dahlen},
  {Graur}, {Casertano}, {Dickinson}, {Ferguson}, {Garnavich}, {Hayden}, {Jha},
  {Jones}, {Kirshner}, {Koekemoer}, {McCully}, {Mobasher}, {Patel}, {Weiner},
  {Cenko}, {Clubb}, {Cooper}, {Filippenko}, {Frederiksen}, {Hjorth},
  {Leibundgut}, {Matheson}, {Nayyeri}, {Penner}, {Trump}, {Silverman}, {U},
  {Azalee Bostroem}, {Challis}, {Rajan}, {Wolff}, {Faber}, {Grogin}, \&
  {Kocevski}}]{Rodney2014}
{Rodney}, S.~A., {Riess}, A.~G., {Strolger}, L.-G., {et~al.} 2014, \aj, 148,
  13, \dodoi{10.1088/0004-6256/148/1/13}

\bibitem[{{Rose} {et~al.}(2021){Rose}, {Baltay}, {Hounsell}, {Macias}, {Rubin},
  {Scolnic}, {Aldering}, {Bohlin}, {Dai}, {Deustua}, {Foley}, {Fruchter},
  {Galbany}, {Jha}, {Jones}, {Joshi}, {Kelly}, {Kessler}, {Kirshner}, {Mandel},
  {Perlmutter}, {Pierel}, {Qu}, {Rabinowitz}, {Rest}, {Riess}, {Rodney},
  {Sako}, {Siebert}, {Strolger}, {Suzuki}, {Thorp}, {Van Dyk}, {Wang}, {Ward},
  \& {Wood-Vasey}}]{Rose2021}
{Rose}, B.~M., {Baltay}, C., {Hounsell}, R., {et~al.} 2021, arXiv e-prints,
  arXiv:2111.03081, \dodoi{10.48550/arXiv.2111.03081}

\bibitem[{{Rose} {et~al.}(2025){Rose}, {Vincenzi}, {Hounsell}, {Qu},
  {Aldoroty}, {Scolnic}, {Kessler}, {Macias}, {Brout}, {Acevedo}, {Chen},
  {Gomez}, {Peterson}, {Rubin}, {Sako}, \& {the Roman Supernova Project
  Infrastructure Team}}]{Rose2025_hourglass}
{Rose}, B.~M., {Vincenzi}, M., {Hounsell}, R., {et~al.} 2025, \apj, 988, 65,
  \dodoi{10.3847/1538-4357/ade1d6}

\bibitem[{{Rubin} {et~al.}(2025{\natexlab{a}}){Rubin}, {Aldering}, {Betoule},
  {Fruchter}, {Huang}, {Kim}, {Lidman}, {Linder}, {Perlmutter},
  {Ruiz-Lapuente}, \& {Suzuki}}]{Rubin2025_UNITY}
{Rubin}, D., {Aldering}, G., {Betoule}, M., {et~al.} 2025{\natexlab{a}}, \apj,
  986, 231, \dodoi{10.3847/1538-4357/adc0a5}

\bibitem[{{Rubin} {et~al.}(2025{\natexlab{b}}){Rubin}, {Aldering}, {Fruchter},
  {Galbany}, {Hounsell}, {Kessler}, {Perlmutter}, {Rose}, {Sako}, {Scolnic},
  {Truong}, \& {the Roman Supernova Cosmology Project Infrastructure
  Team}}]{Rubin2025_Optimize}
{Rubin}, D., {Aldering}, G., {Fruchter}, A., {et~al.} 2025{\natexlab{b}}, arXiv
  e-prints, arXiv:2506.04327, \dodoi{10.48550/arXiv.2506.04327}

\bibitem[{{S{\'a}nchez} {et~al.}(2022){S{\'a}nchez}, {Kessler}, {Scolnic},
  {Armstrong}, {Biswas}, {Bogart}, {Chiang}, {Cohen-Tanugi}, {Fouchez}, {Gris},
  {Heitmann}, {Hlo{\v{z}}ek}, {Jha}, {Kelly}, {Liu}, {Narayan}, {Racine},
  {Rykoff}, {Sullivan}, {Walter}, {Wood-Vasey}, \& {LSST Dark Energy Science
  Collaboration (DESC)}}]{Sanchez2022}
{S{\'a}nchez}, B.~O., {Kessler}, R., {Scolnic}, D., {et~al.} 2022, \apj, 934,
  96, \dodoi{10.3847/1538-4357/ac7a37}

\bibitem[{{Scolnic} {et~al.}(2018){Scolnic}, {Jones}, {Rest}, {Pan},
  {Chornock}, {Foley}, {Huber}, {Kessler}, {Narayan}, {Riess}, {Rodney},
  {Berger}, {Brout}, {Challis}, {Drout}, {Finkbeiner}, {Lunnan}, {Kirshner},
  {Sanders}, {Schlafly}, {Smartt}, {Stubbs}, {Tonry}, {Wood-Vasey}, {Foley},
  {Hand}, {Johnson}, {Burgett}, {Chambers}, {Draper}, {Hodapp}, {Kaiser},
  {Kudritzki}, {Magnier}, {Metcalfe}, {Bresolin}, {Gall}, {Kotak}, {McCrum}, \&
  {Smith}}]{Scolnic2018}
{Scolnic}, D.~M., {Jones}, D.~O., {Rest}, A., {et~al.} 2018, \apj, 859, 101,
  \dodoi{10.3847/1538-4357/aab9bb}

\bibitem[{{Shah} {et~al.}(2024){Shah}, {Davis}, {Bacon}, {Brout}, {Frieman},
  {Galbany}, {Kessler}, {Lahav}, {Lee}, {Lidman}, {Nichol}, {Sako},
  {S{\'a}nchez}, {Scolnic}, {Sullivan}, {Vincenzi}, {Wiseman}, {Allam},
  {Abbott}, {Aguena}, {Alves}, {Andrade-Oliveira}, {Annis}, {Bechtol},
  {Bertin}, {Bocquet}, {Brooks}, {Rosell}, {Carretero}, {Castander}, {da
  Costa}, {Pereira}, {Diehl}, {Doel}, {Doux}, {Everett}, {Ferrero}, {Flaugher},
  {Friedel}, {Gatti}, {Gruen}, {Gruendl}, {Gutierrez}, {Hinton}, {Hollowood},
  {Honscheid}, {Huterer}, {James}, {Kuehn}, {Lee}, {Marshall},
  {Mena-Fern{\'a}ndez}, {Miquel}, {Myles}, {Ogando}, {Palmese}, {Pieres},
  {Roodman}, {Sanchez}, {Sevilla-Noarbe}, {Smith}, {Soares-Santos}, {Suchyta},
  {Swanson}, {Tarle}, {Weaverdyck}, \& {DES Collaboration}}]{Shah2024}
{Shah}, P., {Davis}, T.~M., {Bacon}, D., {et~al.} 2024, \mnras, 532, 932,
  \dodoi{10.1093/mnras/stae1515}

\bibitem[{{Strolger} {et~al.}(2015){Strolger}, {Dahlen}, {Rodney}, {Graur},
  {Riess}, {McCully}, {Ravindranath}, {Mobasher}, \& {Shahady}}]{Strolger2015}
{Strolger}, L.-G., {Dahlen}, T., {Rodney}, S.~A., {et~al.} 2015, \apj, 813, 93,
  \dodoi{10.1088/0004-637X/813/2/93}

\bibitem[{{Suzuki} {et~al.}(2012){Suzuki}, {Rubin}, {Lidman}, {Aldering},
  {Amanullah}, {Barbary}, {Barrientos}, {Botyanszki}, {Brodwin}, {Connolly},
  {Dawson}, {Dey}, {Doi}, {Donahue}, {Deustua}, {Eisenhardt}, {Ellingson},
  {Faccioli}, {Fadeyev}, {Fakhouri}, {Fruchter}, {Gilbank}, {Gladders},
  {Goldhaber}, {Gonzalez}, {Goobar}, {Gude}, {Hattori}, {Hoekstra}, {Hsiao},
  {Huang}, {Ihara}, {Jee}, {Johnston}, {Kashikawa}, {Koester}, {Konishi},
  {Kowalski}, {Linder}, {Lubin}, {Melbourne}, {Meyers}, {Morokuma}, {Munshi},
  {Mullis}, {Oda}, {Panagia}, {Perlmutter}, {Postman}, {Pritchard}, {Rhodes},
  {Ripoche}, {Rosati}, {Schlegel}, {Spadafora}, {Stanford}, {Stanishev},
  {Stern}, {Strovink}, {Takanashi}, {Tokita}, {Wagner}, {Wang}, {Yasuda},
  {Yee}, \& {Supernova Cosmology Project}}]{Suzuki2012}
{Suzuki}, N., {Rubin}, D., {Lidman}, C., {et~al.} 2012, \apj, 746, 85,
  \dodoi{10.1088/0004-637X/746/1/85}

\bibitem[{{Vincenzi} {et~al.}(2019){Vincenzi}, {Sullivan}, {Firth},
  {Guti{\'e}rrez}, {Frohmaier}, {Smith}, {Angus}, \& {Nichol}}]{Vincenzi2019}
{Vincenzi}, M., {Sullivan}, M., {Firth}, R.~E., {et~al.} 2019, \mnras, 489,
  5802, \dodoi{10.1093/mnras/stz2448}

\bibitem[{{Vincenzi} {et~al.}(2021){Vincenzi}, {Sullivan}, {Graur}, {Brout},
  {Davis}, {Frohmaier}, {Galbany}, {Guti{\'e}rrez}, {Hinton}, {Hounsell},
  {Kelsey}, {Kessler}, {Kovacs}, {Kuhlmann}, {Lasker}, {Lidman}, {M{\"o}ller},
  {Nichol}, {Sako}, {Scolnic}, {Smith}, {Swann}, {Wiseman}, {Asorey}, {Lewis},
  {Sharp}, {Tucker}, {Aguena}, {Allam}, {Avila}, {Bertin}, {Brooks}, {Burke},
  {Carnero Rosell}, {Carrasco Kind}, {Carretero}, {Castander}, {Choi},
  {Costanzi}, {da Costa}, {Pereira}, {De Vicente}, {Desai}, {Diehl}, {Doel},
  {Everett}, {Ferrero}, {Fosalba}, {Frieman}, {Garc{\'\i}a-Bellido},
  {Gaztanaga}, {Gerdes}, {Gruen}, {Gruendl}, {Gutierrez}, {Hollowood},
  {Honscheid}, {Hoyle}, {James}, {Kuehn}, {Kuropatkin}, {Maia}, {Martini},
  {Menanteau}, {Miquel}, {Morgan}, {Palmese}, {Paz-Chinch{\'o}n}, {Plazas},
  {Romer}, {Sanchez}, {Scarpine}, {Serrano}, {Sevilla-Noarbe}, {Soares-Santos},
  {Suchyta}, {Tarle}, {Thomas}, {To}, {Varga}, {Walker}, {Wilkinson}, \& {DES
  Collaboration}}]{Vincenzi2021}
{Vincenzi}, M., {Sullivan}, M., {Graur}, O., {et~al.} 2021, \mnras, 505, 2819,
  \dodoi{10.1093/mnras/stab1353}

\bibitem[{{Vincenzi} {et~al.}(2024){Vincenzi}, {Brout}, {Armstrong}, {Popovic},
  {Taylor}, {Acevedo}, {Camilleri}, {Chen}, {Davis}, {Lee}, {Lidman}, {Hinton},
  {Kelsey}, {Kessler}, {M{\"o}ller}, {Qu}, {Sako}, {Sanchez}, {Scolnic},
  {Smith}, {Sullivan}, {Wiseman}, {Asorey}, {Bassett}, {Carollo}, {Carr},
  {Foley}, {Frohmaier}, {Galbany}, {Glazebrook}, {Graur}, {Kovacs}, {Kuehn},
  {Malik}, {Nichol}, {Rose}, {Tucker}, {Toy}, {Tucker}, {Yuan}, {Abbott},
  {Aguena}, {Alves}, {Allam}, {Andrade-Oliveira}, {Annis}, {Bacon}, {Bechtol},
  {Bernstein}, {Brooks}, {Burke}, {Carnero Rosell}, {Carretero}, {Castander},
  {Conselice}, {da Costa}, {Pereira}, {Desai}, {Diehl}, {Doel}, {Ferrero},
  {Flaugher}, {Friedel}, {Frieman}, {Garc{\'\i}a-Bellido}, {Gatti}, {Giannini},
  {Gruen}, {Gruendl}, {Hollowood}, {Honscheid}, {Huterer}, {James},
  {Kuropatkin}, {Lahav}, {Lee}, {Lin}, {Marshall}, {Mena-Fern{\'a}ndez},
  {Menanteau}, {Miquel}, {Palmese}, {Pieres}, {Plazas Malag{\'o}n}, {Porredon},
  {Romer}, {Roodman}, {Sanchez}, {Sanchez Cid}, {Schubnell}, {Sevilla-Noarbe},
  {Suchyta}, {Swanson}, {Tarle}, {To}, {Walker}, {Weaverdyck}, \&
  {Yamamoto}}]{Vincenzi2024}
{Vincenzi}, M., {Brout}, D., {Armstrong}, P., {et~al.} 2024, \apj, 975, 86,
  \dodoi{10.3847/1538-4357/ad5e6c}

\bibitem[{{Wang}(2008)}]{Wang2008_FoM}
{Wang}, Y. 2008, \prd, 77, 123525, \dodoi{10.1103/PhysRevD.77.123525}

\end{thebibliography}
\bibliographystyle{aasjournal}

\end{document}